%% file: dynamic.tex
\documentclass[11pt]{article}
\usepackage[utf8]{inputenc}
\usepackage{fullpage}
\usepackage{amsmath,amsfonts,amsthm,amssymb,multirow}
\usepackage{algorithmic}
\usepackage{floatpag}
\usepackage[ruled,vlined,commentsnumbered,titlenotnumbered]{algorithm2e}
\newenvironment{reminder}[1]{\smallskip

\noindent {\bf Reminder of #1 }\em}{\smallskip}

\newtheorem{theorem}{Theorem}[section]

\newtheorem{proposition}{Proposition}
\newtheorem{corollary}{Corollary}[section]

\newtheorem{lemma}{Lemma}[section]
\newtheorem{claim}{Claim}
\newtheorem{conjecture}{Conjecture}
\newtheorem{fconjecture}{``Conjecture''}

\newtheorem{conj}{Conjecture}
\newtheorem{fconj}{``Conjecture''}

\makeatletter
\let\c@fconjecture\c@conjecture
\makeatother

\makeatletter
\let\c@fconj\c@conj
\makeatother

\def \eps {\varepsilon}

\newcommand{\ignore}[1]{}

\def \poly { \text{\rm poly~} }
\def \polylog { \text{\rm polylog~} }
\def\th{{\text{\rm th}}}

\newcommand{\nocontentsline}[3]{}
\newcommand{\tocless}[2]{\bgroup\let\addcontentsline=\nocontentsline#1{#2}\egroup}

\def\patrascu{\text{P{\v a}tra{\c s}cu}}

\begin{document}

\title{Popular conjectures imply strong lower bounds for dynamic problems}
\author{Amir Abboud, Virginia Vassilevska Williams~\footnote{Computer Science Department, Stanford University, {\texttt{abboud@cs.stanford.edu,virgi@cs.stanford.edu}}}}

\maketitle

\input{abstract}

\thispagestyle{empty}


\newpage

\setcounter{page}{1}
\section{Introduction}
\input{intro}

\input{discussion}

\input{techniques}

\input{defstable}

\input{restable}

\input{conjecs}

\input{formalthms}

\section{Preliminaries}
\label{sec:prelim}
\input{prelims}


\section{Lower bounds from SETH}
\label{sec:seth}

\input{seth}

\section{Lower bounds from Triangle}
\label{sec:bmm}

\input{bmm}

\section{Lower bounds from APSP}
\label{sec:apsp}
\input{apsp}

\section{Lower bounds from $3$SUM}
\label{sec:3sum}

\input{3sum}

\section{Acknowledgments} 
We are grateful to Ryan Williams for suggesting the title of this manuscript, and to Liam Roditty for suggesting to us to consider the subgraph connectivity problem.
We would like to thank Liam Roditty, Ryan Williams, and Uri Zwick for valuable discussions.
This research was supported by a Stanford School of Engineering Hoover Fellowship, NSF Grant CCF-1417238 and BSF Grant BSF:2012338.
\bibliographystyle{abbrv}
\bibliography{references}

\end{document}

%% file: abstract.tex
\begin{abstract}\small
We consider several well-studied problems in dynamic algorithms and prove that sufficient progress on any of them would imply a breakthrough on one of five major open problems in the theory of algorithms:
\begin{enumerate}
\item Is the $3$SUM problem on $n$ numbers in $O(n^{2-\eps})$ time for some $\eps>0$?
\item Can one determine the satisfiability of a CNF formula on $n$ variables in $O((2-\eps)^n\poly n)$ time for some $\eps>0$? 
\item Is the All Pairs Shortest Paths problem for graphs on $n$ vertices in $O(n^{3-\eps})$ time for some $\eps>0$?
\item Is there a linear time algorithm that detects whether a given graph contains a triangle?
\item Is there an $O(n^{3-\eps})$ time combinatorial algorithm for $n\times n$ Boolean matrix multiplication?
\end{enumerate}
The problems we consider include dynamic versions of bipartite perfect matching, bipartite maximum weight matching, single source reachability, single source shortest paths, strong connectivity, subgraph connectivity, diameter approximation and some nongraph problems such as Pagh's problem defined in a recent paper by Patrascu [STOC 2010].
 \end{abstract}

%% file: intro.tex

Dynamic algorithms are a natural extension of the typical notion of an algorithm: besides computing a function on an input $x$, the algorithm needs to be able to update the computed function value as $x$ undergoes small changes, without redoing all of the computation.
Dynamic algorithms have a multitude of applications, and 
their study has evolved into a vibrant research area.
Among its many successes are efficient dynamic graph algorithms for graph connectivity ~\cite{HK99,Thorup00,PT07}, minimum spanning tree~\cite{Fr85,HolmLT01,HenzingerK01}, graph matching~\cite{sankdynmatch,BGSmatch,NSmatch,GP13} and approximate shortest paths in undirected graphs~\cite{Bernstein09,BernsteinR11,HenzingerKN13}. Graph connectivity and minimum spanning tree for instance can be supported in only polylogarithmic time per edge update or query.

Nevertheless, there are some dynamic problems that seem stubbornly difficult. For instance, consider maintaining a reachability tree from a fixed vertex under edge insertions or deletions, i.e. the so called dynamic single source reachability problem (ss-Reach). The best known dynamic ss-Reach algorithm~\cite{sank04} has update time $O(n^{1.495})$. This is only better than the trivial recomputation time for very dense graphs. Moreover, the result uses heavy machinery such as fast matrix multiplication, and is currently not practical.
There are many such problems, including dynamic shortest paths, maximum matching, strongly connected components, and some nongraph problems such as Pagh's problem~\cite{PatrDyn} supporting set intersection updates and membership queries.
For many of these problems, the only known dynamic algorithms are to recompute the answer from scratch.
(Although there has been some success when only insertions or only deletions are to be supported.)

When there are no good upper bounds, lower bounds are highly sought after. Typically, for dynamic data structure problems, one attempts to prove cell probe lower bounds. However, unfortunately, the best known cell probe lower bounds are at best logarithmic~\cite{patrasculog}, and for these hard dynamic problems we would want higher, polynomial lower bounds, i.e. of the form $\Omega(N^c)$ where $N$ is the size of the input and $c$ is an explicit constant.
Patrascu~\cite{PatrDyn} initiated the study of basing the hardness of dynamic problems on a conjecture about the hardness of the $3$SUM problem, a problem in quadratic time with no known ``truly'' subquadratic solutions ($N^{2-\eps}$ for constant $\eps>0$). He showed that one can indeed prove conditional polynomial lower bounds for some notable problems such as transitive closure and shortest paths.

Other papers have considered proving conditional lower bounds for specific problems.
Roditty and Zwick~\cite{rzesa} for instance showed tight lower bounds for decremental and incremental single source shortest paths, based on the conjecture that all pairs shortest paths (APSP) cannot be solved in truly subcubic time. Chan~\cite{Chan06} showed that a fast algorithm for subgraph connectivity would imply an unusually fast algorithm for finding a triangle in a graph. Some other works compare the complexity of their dynamic problem of study to the complexity of Boolean matrix multiplication~\cite{rzesa,HenzingerKN13}. However, the only systematic study of conditional lower bounds for a larger collection of dynamic problems is P\v{a}tra\c{s}cu's paper~\cite{PatrDyn}.

In this paper we expand on P\v{a}tra\c{s}cu's idea and prove strong conditional lower bounds for a much larger collection of dynamic problems, based on five well-known conjectures: the $3$SUM, All Pairs Shortest Paths, Triangle and Boolean Matrix Multiplication Conjectures and the Strong Exponential Time Hypothesis; we define these formally below. In section~\ref{sec:conj} we discuss the prior work on these conjectures and some potential relationships between them. As far as we know, any subset of the below conjectures could be false, and the rest could still be true. Hence it is interesting to have lower bounds based on each one of them.

\begin{conj}[No truly subquadratic $3$SUM]
In the Word RAM model with words of $O(\log n)$ bits, any algorithm requires $n^{2-o(1)}$ time in expectation to determine whether a set $S\subset \{-n^3,\ldots,n^3\}$ of $|S|=n$ integers contains three distinct elements $a,b,c\in S$ with $a+b=c$.
\label{conj:3sum}
\end{conj}

\begin{conj}[No truly subcubic APSP]
There is a constant $c$, such that in the Word RAM model with words of $O(\log n)$ bits, any algorithm requires $n^{3-o(1)}$ time in expectation to compute the distances between every pair of vertices in an $n$ node graph with edge weights in $\{1,\ldots, n^c\}$.\label{conj:apsp}
\end{conj}

\begin{conj}[Strong Exponential Time Hypothesis (SETH)]
For every $\eps>0$, there exists a $k$, such that SAT on $k$-CNF formulas on $n$ variables cannot be solved in 
$O^*(2^{(1-\eps)n})$ time\footnote{The notation $O^*(f(n))$ means $f(n)\poly n$.}.
\label{conj:seth}
\end{conj}

\begin{conj}[No almost linear time triangle]
There is a constant $\delta>0$, such that in the Word RAM model with words of $O(\log n)$ bits, any algorithm requires $m^{1+\delta-o(1)}$ time in expectation to detect whether an $m$ edge graph contains a triangle.\label{conj:tria}
\end{conj}

\begin{fconj}[No truly subcubic combinatorial BMM]
In the Word RAM model with words of $O(\log n)$ bits, any combinatorial algorithm requires $n^{3-o(1)}$ time in expectation to compute the Boolean product of two $n\times n$ matrices.\footnote{We use quotes above, mainly because the notion of a combinatorial algorithm is not well defined.}\label{conj:bmm}
\end{fconj}

This paper is the first study that relates the complexity of any dynamic problem to the exact complexity of Boolean Satisfiability (via the SETH).
Our lower bounds hold even for randomized fully dynamic algorithms with (expected) amortized update times. 
Most of our results also hold for {\em partially} dynamic (incremental and decremental) algorithms with worst-case time bounds.

Interestingly, many of our lower bounds (those based on the SETH) 
hold even when one allows {\em arbitrary} polynomial preprocessing time, and achieve essentially optimal guarantees.
These are the first lower bounds of this nature.

Most of our lower bounds also hold in the setting when one knows the list of updates and queries in advance, i.e. in the {\em lookahead} model.
This is of interest since many dynamic problems can be solved faster given sufficient lookahead, e.g. graph transitive closure~\cite{SankowskiM10} and matrix rank~\cite{Kavitha08}.


\paragraph{Organization.} In Section~\ref{sec:discussion} we discuss our results and the prior work on the problems we address. 
In Section~\ref{sec:techniques} we describe our techniques.
In Section~\ref{sec:conj} we give an overview of the prior work on the conjectures.
In Section~\ref{sec:formalthms} we give a formal statement of the theorems we prove. The problems we consider are summarized in Table~\ref{table:problems} and the results are summarized in Table~\ref{tab:res}.
In section~\ref{sec:prelim} we define some useful notation and prove reductions between dynamic problems.
In section~\ref{sec:seth} we prove lower bounds based on Conjecture~\ref{conj:seth} (SETH).
In section~\ref{sec:bmm} we prove lower bounds based on Conjectures~\ref{conj:tria} and~\ref{conj:bmm} (Triangle and BMM).
In section~\ref{sec:apsp} we prove lower bounds based on Conjecture~\ref{conj:apsp} (APSP).
And finally, in section~\ref{sec:3sum} we prove lower bounds based on Conjecture~\ref{conj:3sum} ($3$SUM).


%% file: discussion.tex
\section{Prior work and our results}
\label{sec:discussion}
Below we define each of the problems we consider and discuss the implications of our results for each problem in turn.
The problem definitions are also summarized in Table~\ref{table:problems}, and our results for each problem are summarized in Table~\ref{tab:res}.

\paragraph{Maximum cardinality bipartite matching.}
The maximum cardinality bipartite matching problem has a long history.
In a seminal paper, Hopcroft and Karp~\cite{hopkarpmatch} designed an $O(m\sqrt n)$ time algorithm for the problem in bipartite graphs with $m$ edges and $n$ nodes. Mucha and Sankowski~\cite{MS04} (and Harvey~\cite{harvey}) improved their result for dense graphs by giving an $\tilde{O}(n^\omega)$\footnote{The $\tilde{O}$ notation suppresses polylogarithmic factors.} time algorithm where $\omega<2.373$ is the matrix multiplication exponent~\cite{v12}. In a breakthrough paper earlier this year, Madry~\cite{madry} devised the first improvement over the Hopcroft-Karp algorithm for sparse bipartite graphs, with a runtime of $\tilde{O}(m^{10/7})$.

The amazing algorithms for the static case of the problem do not seem to imply efficient dynamic algorithms, however.
Since a single edge update can cause the addition of at most one augmenting path, a trivial fully dynamic algorithm algorithm for maximum bipartite matching has update time $O(m)$. The only improvement over this is a result by Sankowski~\cite{sankdynmatch} who gave a fully dynamic algorithm with an amortized update time of $O(n^{1.495})$. His result uses fast matrix multiplication and is only an improvement for sufficiently dense graphs. Two questions emerge. 

(1) Is the use of matrix multiplication inherent? 

(2) Can one get an improvement over the trivial algorithm when the graph is sparse?

We first address question (1). We show that any improvement over the trivial algorithm implies a nontrivial algorithm for Boolean matrix multiplication, thus showing that the use of matrix multiplication is indeed inherent.
We partially address question (2), by showing three interesting consequences of a dynamic algorithm for maximum bipartite matching that has (amortized) update and query time $O(m^\eps)$ for $\eps<1$. 

First, we show that an algorithm with $\eps<0.81$ would imply an improvement on the $20$-year old $O(m^{1.41})$ time bound~\cite{AlYuZw97,AlYuZw97c} for the triangle detection problem in sparse graphs. In fact, Conjecture~\ref{conj:tria} implies that there is some $\eps>0$ for which (amortized, expected) $\Omega(m^\eps)$ update or query time is necessary.
Second, we show that an algorithm with $\eps<1/3$ would imply that $3$SUM is in truly subquadratic time, thus falsifying Conjecture~\ref{conj:3sum}. 
Finally, we show that any combinatorial algorithm with any $\eps<1$ falsifies ``Conjecture''~\ref{conj:bmm}.
All of our results apply also for the bipartite perfect matching problem (BPMatch).

\paragraph{Approximately maximum matching.}
In the absence of good dynamic algorithms for maximum matching, recent research has focused on developing efficient algorithms for dynamically maintaining approximately maximum matchings. Ivkovic and Lloyd~\cite{IvLl} presented the first such algorithm, maintaining a maximal matching (and hence a $2$-approximate maximum matching) with update time $O(m^{0.71})$. Baswana, Gupta and Sen~\cite{BGSmatch} developed a randomized dynamic algorithm for maximal matching with expected amortized $O(\log n)$ update time. Neiman and Solomon~\cite{NSmatch} presented a deterministic worst case $O(\sqrt m)$ update time that maintained a $3/2$-approximate maximum matching. Finally, Gupta and Peng~\cite{GP13} showed that with the same update time one can maintain a $(1+\eps)$-approximation for any constant $\eps>0$.

All of the above papers except~\cite{GP13} obtain an approximate maximum matching by maintaining a matching that does not admit short augmenting paths.
It is well known that for any $k\geq 2$, if a matching does not admit length $2k-3$ augmenting paths, then it is a $k/(k-1)$ approximate maximum matching. The algorithms for maximal matching exclude length $1$ augmenting paths, and the $3/2$-approximation algorithm of~\cite{NSmatch} excludes length $1$ and $3$ augmenting paths. 

We show an inherent limitation to this approach for maintaining an approximately maximum matching. In particular, we show that there exists a constant $k\leq 10$ such that any dynamic algorithm that maintains a matching that excludes augmenting paths of length at most $2k-3$ can be converted into an algorithm for $3$SUM, triangle detection and Boolean matrix multiplication. 
Our results are the same as that for BPMatch:
 an $O(m^\eps)$ update time for the problem falsifies Conjecture~\ref{conj:3sum} for $\eps<1/3$, Conjecture~\ref{conj:tria} for $\eps<\delta$ and ``Conjecture''~\ref{conj:bmm} for $\eps<1$ if it is combinatorial.
In particular, the above results imply that using the augmenting paths approach for dynamic approximate matching is unlikely to yield a result such as Gupta and Peng's algorithm.
%

\paragraph{Maximum weight bipartite matching.}
There are several weighted versions of the bipartite matching problem, all equivalent to each other: find a maximum weight matching, find a maximum weight perfect matching, find a minimum weight perfect matching (also known as the assignment problem). We will refer to the weighted matching problem as MWM.
 The first polynomial time algorithm for MWM, the Hungarian algorithm, was proposed by Kuhn~\cite{kuhn}. Using Fibonacci heaps~\cite{FT} its runtime  is $O(mn+n^2\log n)$. When the edge weights are in $\{-M,\ldots, M\}$, on a word-RAM with $O(\log (Mn))$ bit words, Gabow and Tarjan~\cite{Gabow85,GTscaling} and a recent improvement by Duan and Su~\cite{DuanS12} give scaling algorithms for the problem running in $O(m\sqrt n \log M)$ time. Sankowski~\cite{sankweightmatch} gave an $\tilde{O}(Mn^\omega)$ time algorithm.
 
The dynamic case of the problem seems less studied. 
It is not hard to obtain a fully dynamic algorithm for MWM that can answer in constant time queries about the weight of the MWM, and perform edge updates in
$\tilde{O}(m)$ time.
 The algorithm is based on Edmonds-Karp's algorithm~\cite{EK72} and performs each update by searching for the shortest augmenting path.
There are no dynamic algorithms for MWM with $o(m)$ update time. The only result for the dynamic problem is an algorithm by Anand et al.~\cite{anand} that maintains an $8$-approximate MWM with expected amortized $O(\log n\log C)$ time where $C$ is the ratio between the maximum and minimum edge weight.
 
A natural question is, is it inherently hard to obtain $o(m)$ update time dynamic MWM algorithms? We address this question by showing that any dynamic MWM algorithm, even a decremental or incremental one, with amortized update time $O(n^{2-\eps})$ for constant $\eps>0$ in dense graphs would imply a truly subcubic APSP algorithm, thus explaining the lack of progress on the problem.

\paragraph{Subgraph Connectivity.}
The subgraph connectivity problem (SubConn) is as follows: given a graph $G$, maintain a subgraph $S$ where the updates are adding/removing a node of $G$ to/from $S$, and the queries are to determine whether a query node $t$ is reachable from a query node $s$ in $S$.
SubConn is a version of the graph connectivity problem, but instead of edge updates, one needs to maintain vertex updates.
As mentioned earlier, graph connectivity has extremely efficient algorithms (e.g.~\cite{Thorup00}). However, the obvious way of simulating vertex updates using edge updates is to insert/delete all incident edges to a vertex that is to be inserted/deleted. As the degree of a vertex can be linear, this type of simulation cannot give better than $O(n)$ update time for SubConn. Thus SubConn seems much more difficult than graph connectivity.

The SubConn problem was first introduced by Frigioni and Italiano~\cite{FI00} in the context of communication networks where processors may become faulty and later can come back online. They obtained an efficient dynamic algorithm for planar graphs. Later, Chan~\cite{Chan06} studied the problem in general graphs because of its applications to geometric connectivity problems.
In such problems, one is to maintain a set of $n$ axis parallel boxes in $d$ dimensions under insertions and deletions so that one can answer queries about whether there is a path between any two given points that is contained within the set of boxes.
Chan showed that for any constant $d$, the box connectivity problem can be reduced to subgraph connectivity in a graph on $\tilde{O}(n)$ edges, thus any dynamic algorithm for subgraph connectivity immediately implies an algorithm for geometric connectivity. Chan also showed that subgraph connectivity can be reduced to the 
box connectivity problem in $3$ dimensions, thus showing that subgraph connectivity and box connectivity are equivalent problems for all $d\geq 3$.
Chan, P\v{a}tra\c{s}cu, Roditty~\cite{CPR08} further showed that a variety of other geometric connectivity problems are reducible to the SubConn problem.

Chan~\cite{Chan06} obtained an algorithm for SubConn with $O(m^{1.28})$ preprocessing time, $O(m^{0.94})$ update time and $O(m^{1/3})$ query time.
Later, Chan, P\v{a}tra\c{s}cu and Roditty~\cite{CPR08} improved these bounds, obtaining an algorithm with $O(m^{4/3})$ preprocessing time, $O(m^{2/3})$ update time and $O(m^{1/3})$ query time. Duan~\cite{Duan10} presented algorithms with better space usage.

P\v{a}tra\c{s}cu~\cite{PatrDyn} showed that unless Conjecture~\ref{conj:3sum} above is false, there is some $\eps>0$ such that SubConn cannot be solved with $o(m^{1+\eps})$ preprocessing time and $o(m^\eps)$ update and query time.
Here we exhibit an explicit $\eps$, $\eps=1/3$, for which P\v{a}tra\c{s}cu's result holds. Moreover, we show that assuming Conjecture~\ref{conj:3sum}, there is a tradeoff lower bound between the query and update time for fully dynamic algorithms for SubConn. In particular, we show that unless $3$SUM has truly subquadratic algorithms, SubConn cannot be maintained with preprocessing time $O(m^{4/3-\eps})$, update time $O(m^\alpha)$ and query time $O(m^{2/3-\alpha-\eps})$, for any $\eps>0$ and $1/6<\alpha<1/3$. 

Chan~\cite{Chan06} showed that any dynamic algorithm for SubConn with preprocessing time $O(m^p)$, update and query time $O(m^u)$ would imply an $O(m^{1+u}+m^p)$ time algorithm for triangle detection. 
His result implies that if Conjecture~\ref{conj:tria} is true, then for any such algorithm either $p\geq 1+\eps$ or $u\geq \eps$ for some constant $\eps>0$, i.e. the same conclusion as P\v{a}tra\c{s}cu's assuming Conjecture~\ref{conj:3sum}.

Here we improve Chan's result slightly. In particular, we show that one can reduce the triangle detection problem on $m$ edge, $n$ node graphs to dynamic SubConn with $O(m)$ updates and only $n$ queries. This implies that 
any combinatorial dynamic algorithm with truly sublinear query time ($O(m^{1-\eps})$ for some $\eps>0$) and truly subcubic in $n$ preprocessing time, must have $\Omega(m^{1/2-\delta})$ update time for all $\delta>0$, unless ``Conjecture''~\ref{conj:bmm} is false. (Notice that it is trivial to get $O(m)$ query time and $O(1)$ update time.)
Thus, if 
the algorithm of~\cite{CPR08} can be improved to have update time $m^{0.499}$, we would have a new alternative BMM algorithm.
Our results hold even for the special case $st$-SubConn of SubConn in which we only care about whether two fixed vertices are connected in the subgraph $S$. 

\paragraph{Subgraph Connectedness.}
Chan~\cite{Chan06} identifies a problem extremely related to SubConn, that nevertheless seems much more difficult.
The problem is Subgraph Connectedness (ConnSub): similarly to SubConn, one has to maintain a subgraph $S$ of a fixed graph $G$ under vertex additions and removals, but the query one needs to be able to answer is whether $S$ is connected.

The best and only known algorithm for ConnSub is to recompute the connectivity information (via DFS in $O(m)$ time) after each update or at each query.
Here we explain this lack of progress by showing that unless the SETH (Conjecture~\ref{conj:seth}) is false, any algorithm for ConnSub with arbitrary polynomial preprocessing time, must either have essentially linear update time or essentially linear query time. Thus, the trivial algorithm is essentially optimal, under the SETH.

Our result holds even for a special case of the problem called SubUnion, originally identified by Chan: given a fixed collection of sets $X=\{X_1,\ldots, X_t\}$ of total size $m$ such that $\cup_i X_i=U$, maintain a subcollection $S\subseteq X$ under set insertions and deletions, while answering the query whether the union of the sets in $S$ is exactly $U$. That is, the query is exactly ``Is $S$ a set cover?''.

%
%
%

\paragraph{Single Source Reachability.} 
Unlike in undirected graphs where great results are possible (e.g. \cite{Thorup00}), the reachability problem in directed graphs is surprisingly tough in the dynamic setting, even when the reachability between two fixed vertices is to be maintained ($st$-Reach).
The trivial algorithm that recomputes the reachability after each update or at each query is still the best known algorithm in the case of sparse graphs.
For graphs with $\omega(n^{1.495})$ edges, Sankowski~\cite{sank04} showed that one can get a better update time. In particular, he obtained update time $O(n^{1.495})$ and $O(1)$ query time for $st$-Reach and $O(n^{1.495})$ query time for single source reachability (ss-Reach). Sankowski's result improved on the first sublinear update time result by Demetrescu and Italiano~\cite{di00} who obtained $O(n^{1.575})$ update time and $O(n^{0.575})$ query time for ss-Reach.

Both of these results heavily rely on fast matrix multiplication.
Here we show that this is inherent. In particular, any algorithm with truly subcubic (in $n$) preprocessing time and truly subquadratic query and update time can be converted without significant overhead into a truly subcubic time algorithm for Boolean matrix multiplication. Thus, any such combinatorial algorithm would falsify ``Conjecture''~\ref{conj:bmm}.

P\v{a}tra\c{s}cu~\cite{PatrDyn} showed that assuming Conjecture~\ref{conj:3sum}, there is some $\eps>0$ such that fully dynamic transitive closure requires either preprocessing time $m^{1+\eps}$, or update or query time $m^\eps$.
Here we slightly extend his result, showing that under the $3$SUM conjecture, $st$-Reach requires either preprocessing time $m^{4/3-o(1)}$ or update or query time $m^{1/3-o(1)}$.
Similar to our results on SubConn, we also exhibit a tradeoff: under the $3$SUM conjecture, 
if $st$-Reach can be solved with preprocessing time $O(m^{4/3-\eps})$ and update time $O(m^\alpha)$ for some $\eps>0$ and $1/6<\alpha<1/3$, then the query time must be at least $m^{2/3-\alpha-o(1)}$.

The single source reachability problem has been studied in the partially dynamic setting as well. 
In the incremental setting, it is not hard to obtain an algorithm for ss-Reach with $O(1)$ amortized update time and $O(1)$ query.
From the work of Even and Shiloach~\cite{evenshiloach} follows an $O(n)$ amortized update time decremental ss-Reach algorithm (with constant query).
For the special case of DAGs, Italiano~\cite{Italiano88} obtained an $O(1)$ amortized update and query time decremental ss-Reach algorithm.

In this paper we show that any combinatorial incremental or decremental algorithm for ss-Reach (and also $st$-Reach) must have $m^{1-o(1)}$ {\em worst case} update or query time, even in the special case of dense DAGs, assuming ``Conjecture''~\ref{conj:bmm}. Thus deamortizing Italiano's DAG ss-Reach algorithm, or Even and Shiloach's algorithm for general graphs, would have interesting consequences for matrix multiplication algorithms.

Finally, we consider a version of ss-Reach, $\#$SSR, in which we want to dynamically answer the query about how many nodes are reachable from the fixed source.
We note that any algorithm that explicitly maintains a reachability tree can answer this counting query.
We show strong lower bounds based on the SETH for $\#$SSR: even after polynomial preprocessing time, a dynamic algorithm cannot beat the trivial recomputation algorithm by any, however small, polynomial factor. 
Hence in particular (under the SETH) no nontrivial algorithm for ss-Reach can maintain the size of the reachability tree. 

%

\paragraph{Incremental/Decremental Single Source Shortest Paths (SSSP)}
Roditty and Zwick~\cite{rzesa} showed that any decremental or incremental algorithm for SSSP in $n$-node graphs with preprocessing time $O(n^{3-\eps})$, and update time $O(n^{2-\eps})$ and query time $O(n^{1-\eps})$ for any $\eps>0$ implies a truly subcubic time algorithm for APSP.
The trivial algorithm for the problem recomputes the shortest paths from the source in $\tilde{O}(n^2)$ time after each update, via Dijkstra's algorithm. Hence,~\cite{rzesa} showed that any tiny polynomial improvement over this result would falsify Conjecture~\ref{conj:apsp}.

Their result, however, did not exclude the possibility of an algorithm that has both $O(n^{2-\eps})$ time updates and $O(n^{2-\eps})$ time queries.
This is exactly what our result excludes, again based on the APSP conjecture. In fact, we show it for the seemingly easier problem of incremental/decremental $st$-shortest path.

\paragraph{Strongly Connected Components.}
Dynamic algorithms for maintaining the strongly connected components have many applications. One example is in compilers research, to speed up pointer analysis by finding cyclic relationships; other examples are listed in~\cite{HKMST12}.
In the partially dynamic setting, nontrivial results are known. Haeupler et al.~\cite{HKMST12} presented an incremental algorithm that maintains the strongly connected components (SCCs) in a graph with amortized update time $O(\min\{m^{1/2},n^{2/3}\})$, while being able to answer queries of the form, are $u$ and $v$ in the same SCC?
Bender et al.~\cite{benderSCC,benderjour} improved the total update time for the case of dense graphs to $O(n^2\log n)$ (thus getting amortized update time $\tilde{O}(n^2/m)$).
In the decremental setting, Roditty and Zwick~\cite{RZ02}, Lacki~\cite{Lacki13}, and Roditty~\cite{RdecSCC} obtained algorithms with amortized update time $O(n)$. The algorithm of \cite{RZ02} was randomized, whereas Lacki's was deterministic, and Roditty improved the preprocessing time to $\tilde{O}(m)$.
No nontrivial results are known for the fully dynamic setting. 

Our results are manyfold.
First, we show results similar to those for $st$-Reach. That is, under ``Conjecture''~\ref{conj:bmm} any combinatorial fully dynamic algorithm must have either preprocessing time $n^{3-o(1)}$, or update or query time $n^{2-o(1)}$.
The same bounds apply for partially dynamic algorithms, but for worst-case update times. Thus, if the known algorithmic results can be deamortized, we would have an alternative BMM algorithm.
Under Conjecture~\ref{conj:3sum}, either the preprocessing time should be at least $m^{4/3-o(1)}$, or the query or update time should be at least $m^{1/3-o(1)}$. 

The above results hold even for the special case of the problem in which we want to answer the query ``Is the graph strongly connected?''.
Next, we consider a variation of the problem which we call SC2, that maintains the graph to answer the query ``Does the graph have more than 2 strongly connected components?''.
We note that all known algorithms for dynamic SCC explicitly store the SCCs of the graph and hence can also solve SC2.
We show surprisingly that SC2 may be a much more difficult problem than SC.
In particular, any algorithm with {\em arbitrary polynomial preprocessing time} must have either $m^{1-o(1)}$ query or $m^{1-o(1)}$ update time, unless the SETH fails. That is, either Conjecture~\ref{conj:seth} is false and we have a breakthrough in the area of SAT algorithms, or the trivial algorithm for SC2 is essentially optimal.

As before, our results also hold for partially dynamic algorithms, but for worst-case update times, implying that deamortizing the results of \cite{HKMST12,benderSCC,RZ02,Lacki13,RdecSCC} is SETH-hard.

The same lower bounds under SETH hold for any of the two following variants of dynamic SCC:
\begin{itemize}
\item AppxSCC: approximate the number of SCCs within some constant factor,
\item MaxSCC: determine the size of the largest SCC.
\end{itemize}

We also consider the dynamic
 $ST$-Reach problem under edge updates: given node sets $S$ and $T$, determine whether every node in $T$ is reachable from every node in $S$.
We are able to prove much stronger update and query lower bounds for it: even after polynomial preprocessing time, the update or query time of any dynamic algorithm must be $n^{2-o(1)}$ in an $n$ node graph, even when the graph is sparse. $O(n^2)$ is the trivial update time for the problem.

%
%
%
%

\paragraph{Pagh's problem.}
P\v{a}tra\c{s}cu~\cite{PatrDyn} introduced a problem 
that he called Pagh's problem (PP) defined as follows: maintain a set $X$ of at most $k$ sets $\{X_i\}_i$ over $[n]$ under the following operation: add $X_i\cap X_j$ to $X$, while answering queries of the form ``Does $j$ belong to $X_i$?''.
(We can assume that $k\geq n$.)
The best known dynamic algorithm for PP is the trivial one: perform the set intersection explicitly in $O(n)$ time at each update and store the sets in a dictionary for which membership tests are efficient.
We introduce a natural variant of Pagh's problem, which we call $\emptyset$-PP (``Pagh's Problem with Emptiness Queries'') where the query is changed to ``Is $X_i$ empty?''.
There is also no nontrivial algorithm known for $\emptyset$-PP. 

The reductions in~\cite{PatrDyn} imply (after some work) that if Conjecture~\ref{conj:3sum} is true, then any dynamic algorithm for PP must have either $k^{4/3-o(1)}$ preprocessing time or $k^{1/3-o(1)}$ update or query time.
We prove this same type of conditional lower bound for $\emptyset$-PP.
Based on Conjecture~\ref{conj:tria} with constant $\delta$, we show that any algorithm for PP or $\emptyset$-PP must have either $k^{1+\delta-o(1)}$  preprocessing time  or $k^{\delta-o(1)}$ update or query time. (The result for PP holds only for $\delta>1/3$, however this is still interesting since as far as we know, there may be no $O(m^{4/3})$ time algorithm for Triangle detection.)
We obtain that under ``Conjecture''~\ref{conj:bmm}, any algorithm for PP or $\emptyset$-PP must have either $k^{3/2-o(1)}$  preprocessing time  or $k^{1/2-o(1)}$ update or query time.
Finally, we also relate $\emptyset$-PP to Conjecture~\ref{conj:seth}, making it the only problem for which we can prove lower bounds based on all conjectures except for~\ref{conj:apsp}.
%
We show that under the SETH, any nontrivial algorithm for $\emptyset$-PP, even assuming arbitrary polynomial time preprocessing, would violate the SETH.
Thus also, if SETH is true, any algorithm for $PP$ that beats the trivial recomputation cannot also answer emptiness queries.

\paragraph{Diameter Approximation.}
The graph diameter is the largest distance in the graph. One can compute the diameter in the same time as computing all pairs shortest paths, and no better algorithms are known.
There are faster algorithms that achieve constant factor approximations for the diameter, however. A folklore result is that in linear time one can obtain a $2$-approximation.
Aingworth et al.~\cite{ACIM99} improved the approximation factor, obtaining a $3/2$ approximation for unweighted graphs that runs in $\tilde{O}(n^2+m\sqrt n)$ time. Roditty and Vassilevska Williams~\cite{RV13}
improved the running time to $\tilde{O}(m\sqrt n)$ with randomization, and
Chechik et al.~\cite{diametersoda14} obtained deterministic $3/2$-approximation algorithms running in $\tilde{O}(\min\{m^{3/2},mn^{2/3}\})$ time that also work for weighted graphs.
Roditty and Vassilevska Williams showed that any $(3/2-\eps)$-approximation algorithm that runs in $O(n^{2-\delta})$ time in undirected unweighted graphs with $\tilde{O}(n)$ edges, for any constants $\eps,\delta>0$ would violate the SETH.

In some applications, an efficient dynamic algorithm for diameter estimation may be useful. The above result does not exclude the possibility that after some preprocessing, one can update the estimate for the diameter faster than recomputing it.
Here we show that if for some $\eps,\delta>0$, there is an algorithm that after an arbitrary polynomial time preprocessing can update a $(4/3-\eps)$-approximation to the diameter of a graph on $\tilde{O}(n)$ edges in $O(n^{2-\delta})$ amortized update time, then the SETH is false.
That is, the trivial recomputation of the diameter is essentially optimal.

%
%
%

%% file: techniques.tex
\section{Description of our techniques}
\label{sec:techniques}
%
\paragraph{Lower bounds based on the SETH.}
We begin all of our reductions with an idea used in prior reductions from the SETH in~\cite{PW10,RV13,diametersoda14}.

We assume that the strong exponential time hypothesis holds. Thus, for every $\eps$, there is some $k$, such that $k$-SAT cannot be solved faster than $O(2^{(1-\eps)n})$ time.
Using this, for each $\eps>0$, we work with a carefully chosen $k$.
Given an instance $F$ of $k$-SAT on $n$ variables, we first use the sparsification lemma of Impagliazzo, Paturi and Zane~\cite{ipz1} to
convert $F$ to a small number of $k$-CNF formulas with $n$ variables and $O(n)$ clauses each. Now we can assume that the given formula has a linear number of clauses.
After this, we can construct a graph as follows.

Split the variables $V$ into two sets $U$ and $V\setminus U$ of size $n/2$ each. We create a set $S$ on $2^{n/2}$ nodes, each corresponding to a partial assignment to the variables in $U$.
Similarly, we create a set $T$ on $2^{n/2}$ nodes, each corresponding to a partial assignment to the variables in $V\setminus U$.
We also create a set $C$ on $O(n)$ nodes, one corresponding to each clause.

Suppose now that we add a directed edge from each partial assignment $s\in S$ to a clause $c\in C$ if and only if $s$ does not satisfy $c$, and a directed edge from $c$ to a partial assignment $t\in T$ if and only if $t$ does not satisfy
$c$. Then, there is a satisfying assignment to the formula if and only if there is a pair of nodes $s\in S$ and $t\in T$ such that $t$ is not reachable from $s$.
Hence any algorithm that can solve this static $ST$-reachability problem on $N$ node, $O(N\log N)$ edge graphs in $O(N^{2-\eps})$ time for constant $\eps>0$ would imply a $O(2^{n(1-\eps')})$ time algorithm for $k$-SAT (for some $\eps'>0$ obtained from the sparsification lemma). We have chosen $k$ however so that we obtain a contradiction, and hence the SETH must be false.

Similar constructions to the above are used in prior papers~\cite{PW10,RV13,diametersoda14}.
We adapt the above argument for the case of {\em dynamic} $\#$ SSR (counting the number of nodes reachable from a source) as follows. (The reductions to the remaining problems use a similar approach with some extra work.)

Instead of having all nodes of $S$ in the above graph $G$, in our dynamic graph $G'$ we have a single node $u$. We have $2^{n/2}$ stages, one for each partial assignment $s\in S$. In each stage, we add edges from $u$ to $C$ but  only to the neighbors of $s$ in $G$, i.e. the clauses that $s$ does not satisfy. Say we have inserted $k$ edges. After the insertions, we ask the query ``Is the number of nodes reachable from $s$ less than $k+2^{n/2}$?''.
If the answer to the query is yes, then the formula is satisfiable, and we can stop. Otherwise, $s$ cannot be completed to a satisfying assignment. We then remove all the inserted edges in this stage and move on to the next partial assignment of $S$. 

The graph has $m=O(n2^{n/2})$ edges, $N=O(2^{n/2})$ vertices and we do $O(n2^{n/2})$ updates and $O(2^{n/2})$ queries. Hence any dynamic algorithm with $O(N^{2-\eps})$ preprocessing time, and $O(m^{1-\eps})$ update and query time would violate the SETH.

Now, suppose that we could achieve $O(m^{1-\eps})$ update and query time after $O(N^{t})$ preprocessing for some big constant $t$. Then we could still contradict the SETH by modifying the above construction further.
Instead of splitting the variables into two parts on $n/2$ variables each, we split them into $U$ of size $(1-\delta)n$ and $V\setminus U$ of size $\delta n$ for some constant $\delta<1/t$.
Then we apply exactly the same construction as above where $S$ is the set of $2^{(1-\delta)n}$ partial assignments to $U$ and $T$ is the set of $2^{\delta n}$ partial assignments to $V\setminus U$.

The number of vertices and edges of the graph is now $O(n 2^{\delta n}) \leq O(2^{n (1-\gamma)/t})$ for some $\gamma>0$. Hence the $O(N^t)$ preprocessing time only takes $O(2^{n(1-\gamma)})$ time.
The number of updates we do is $O(n2^{(1-\delta)n})$ but since the graph is much smaller we get that $O(m^{1-\eps})$ time updates and queries imply a runtime of
\[2^{\delta n (1-\eps)} \cdot 2^{(1-\delta)n} = 2^{n (1-\eps\delta)}\]
 (excluding polynomial factors) for solving the SAT instance. Hence we again violate the SETH.

\paragraph{Lower bounds from Triangle Detection and BMM.}
To obtain lower bounds based on ``Conjecture''~\ref{conj:bmm}, we first obtain lower bounds from Conjecture~\ref{conj:tria} that hold for arbitrary $\delta$ and an arbitrary number of edges $m$, and then apply them for $m=n^2$ and a carefully chosen $\delta$ to obtain the lower bounds from BMM. For instance if Conjecture~\ref{conj:tria} for any constant $\delta$ implies that problem $P$ cannot have a dynamic algorithm with $m^{1+\delta}$ preprocessing time, $m^\delta$ update time and $m^{2\delta}$ query time, then we get that ``Conjecture''~\ref{conj:bmm} implies that $P$ cannot have a dynamic algorithm with $n^{2+2\delta}$ preprocessing time, $n^{2\delta}$ update time and $n^{4\delta}$ query time. Then picking $\delta=(1-\eps)/2$, we get a lower bound for all $\eps>0$ of preprocessing time $n^{3-\eps}$, update time $n^{1-\eps}$ and query time $n^{2-\eps}$.

Our reductions from Triangle Detection typically begin with the following construction.
Given a graph $G=(V,E)$ on $m$ edges and $n$ vertices, we create $4$ copies of $V$, $A,B,C,D$, and for each edge $(u,v)\in E$ we add the directed edges $(u_A,v_B),(u_B,v_C),(u_C, v_D)$ where $u_X$ is the copy of $u$ in $X\in \{A,B,C,D\}$.
Now $G$ contains a triangle if and only if for some $u$, there is a path from $u_A$ to $u_D$. Since the new graph has $O(m)$ edges and $O(n)$ vertices, it suffices to simulate the $n$ reachability queries $(u_A,u_D)$ with dynamic algorithms for the problem at hand.

For $st$-Reach for instance, we add two additional nodes $s$ and $t$ to the above graph and we proceed in stages, one for each node $u\in V$.
In each stage, we add edges $(s,u_A)$ and $(u_D,t)$, and ask whether $t$ is reachable from $s$. This will be the case iff $u$ appears in a triangle in $G$. If the answer to the query is no, we remove the edges incident to $s$ and $t$ and move on to the next stage. The number of queries and updates is $O(n)$ overall, and hence any dynamic algorithm with $O(m^{1+\delta})$ preprocessing time, and $O(m^{2\delta})$ update and query time would imply an $O(m^{1+\delta}+nm^{2\delta})$ time triangle algorithm. We then apply a high-degree low-degree argument as in~\cite{AlYuZw97} to show that this also implies an $O(m^{1+\delta})$ time triangle algorithm.

To obtain the lower bounds for Strong Connectivity and Bipartite Perfect Matching, we prove general reductions from $st$-Reach to SC and BPMatch that show that if the latter two problems can be solved with preprocessing time $p(m,n)$, update time $u(m,n)$ and query time $q(m,n)$, then $st$-Reach can be solved with preprocessing time $p(O(m),O(n))$, update time $u(O(m),O(n))$ and query time $q(O(m),O(n))$.
We show a separate reduction from Triangle Detection to $st$-SubConn (similar to the one to $st$-Reach) that performs $m$ updates and $n$ queries, giving an $m^{\delta-o(1)}$ lower bound on the update time and $m^{2\delta-o(1)}$ on the query time.
%
%
%

Our lower bound for PP is more involved than the rest of the lower bounds based on Conjecture~\ref{conj:tria}.
We will explain the main ideas.
Given an $n$-node, $m$-edge graph, first let us look for triangles containing a node of high degree $\geq \Delta$.
We begin by creating for every node $j$ of high degree a set $X_j$ containing node $i$ iff $j$ is not a neighbor of $i$.
The number of such sets is $O(m/\Delta)$ and constructing them takes $O(mn/\Delta)$ time. 
Now, for each node $a$, using $d(a)$ updates, we create the intersection $Y_a$ of all sets $X_j$ for the neighbors $j$ of $a$. Then, for every edge $(a,b)$, we query whether $b\in Y_a$. 
Notice that $b\in Y_a$ if and only if $b$ is not a neighbor of any of the neighbors $j$ of $a$. Thus, if any one of the $m$ queries returns ``no'', we have detected a triangle.

Suppose now that no triangle with a node of high degree is found. Then, all nodes of any triangle have degree $<\Delta$.
We can attempt to do exactly the same reduction as above. The only problem is that the number of sets $X_j$ that we would have to create could be $n$, and thus just creating the sets would take $O(n^2)$ time.
This is sufficient for a reduction from triangle in dense graphs, however it is too costly for a reduction from sparse graphs.
Fortunately, we can avoid the high cost. Before we create the sets $X_j$, we pick a universal hash function $h$ and hash all nodes with it into a universe of size $O(\Delta^2)$. We are guaranteed that with constant probability, if we take two nodes $a$ and $b$ of low degree, then $N(a)\cup N(b)$ won't contain any two nodes hashing to the same element. Thus, we can simulate the search for a triangle with an edge $(a,b)$ where both $a$ and $b$ have low degree, just as before, except that we create a set for each {\em hash value} $v$, $X_v=\{j~|~\forall c\in N(j),~h(c)\neq v\}$. The creation time is now $O(n\Delta^2)$, and everything else works out with constant probability. We can obtain correctness with high probability by using $O(\log n)$ hash functions. Picking $\Delta=m^{1/3}$, we obtain an extra term $m^{2/3}n$ in our reduction which is negligible if we are trying to contradict Conjecture~\ref{conj:tria} for $\delta>1/3$.

\paragraph{Lower bounds from $3$SUM.}

%


P\v{a}tra\c{s}cu~\cite{PatrDyn} showed that $3$-SUM on $n$ numbers can be reduced to the problem of listing $O(n^2/R)$ triangles in a certain tripartite graph on partitions $A,B,C$ where $|A|=|B|=\sqrt{n}R$, $|C|=n$, $|E(A,B)| = O(nR)$ and $|E(A,C)|+|E(B,C)| = O(n^{1.5})$, for any $R=n^{\frac{1}{2}+\delta}$ and $0<\delta<\frac{1}{2}$, in truly subquadratic time.
Then, he reduced this triangle listing problem to ``the multiphase problem", which in turn can be reduced to several dynamic problems. 
We examine P\v{a}tra\c{s}cu's reduction in more detail and show that by directly reducing the triangle listing problem to dynamic problems like $st$-SubConn we can overcome some inefficiencies incurred by ``the multiphase problem'' and get improved lower bounds.

A first approach is to use the known reductions from triangle listing to triangle finding~\cite{focs10,JV13} to directly apply our hardness results based on triangle finding. However, 
using the currently best reductions, even a linear time algorithm for triangle finding would not be able to get us a faster than $m^{4/3}$ time algorithm for listing $m$ triangles which is what we need in order to get subquadratic $3$SUM.

Instead, we reason about P\v{a}tra\c{s}cu's construction directly.
First, we observe that to falsify Conjecture~\ref{conj:3sum}, it is enough to list in subquadratic time all pairs of nodes $(a,b)\in A \times B$ that participate in a triangle. To do this, note that
in P\v{a}tra\c{s}cu's construction, every node of $A$ has at most $O(n/R)$ neighbors in $C$.
Thus, once the $\leq O(n^2/R)$  pairs of nodes that appear in triangles are known, one can go through each one pair $(a,b)$, and check each of the at most $O(n/R)$ neighbors $c\in C$ of $a$, to find all triangles going through $(a,b)$. Thus $3$SUM would be in $O(n^2/R \cdot n/R)$ = $O(n^3/R^2)$ time which is truly subquadratic when $R=n^{\frac{1}{2}+\delta}$ for $\delta>0$.

Thus, to obtain lower bounds for our dynamic problems, we show how to list the pairs of nodes in $A\times B$ that appear in triangles using a small number of queries and updates.
We first reduce $st$-Reach to $st$-SubConn, thus also showing that $st$-SubConn is at least as hard as SC and BPMatch. Then we focus on $st$-SubConn.
%
Given P\v{a}tra\c{s}cu's graph for some choice of $R$,
we create an instance $H$ of $st$-SubConn.
$H$ is a copy of $G$ in which all the edges between parts $A,B$ are removed. Thus $H$ has only $O(n^{1.5})$ edges for any choice of $R$. 
We also add a node $s$ that is connected to all the nodes in $A$ and a node $t$ that is connected to all the nodes in $B$. Initially, $s$, $t$ and all nodes in $C$ are activated, while the nodes in $A,B$ are deactivated. 

We preprocess this graph in $p(n^{1.5})$ time which is subquadratic if $p(m)=O(m^{\frac{4}{3}-\eps})$. 
Then, we have a stage for each of the $O(nR)$ edges in $A\times B$ in $G$.
In the stage for $(a,b)$, we activate the nodes $a,b$ in $H$ and query if $s,t$ are connected. $s$ and $t$ are connected iff there is a node in $C$ that is a neighbor of both $a,b$, i.e. $(a,b)$ participates in a triangle. Then we deactivate $a$ and $b$ and move on to the next edge.
This way, we can list all the pairs that are in triangles with $O(nR)$ updates and queries to $st$-SubConn, which would be in subquadratic time if $R=n^{1+\eps/2}$ and the update and query times are $u(m),q(m) = O(m^{\frac{1}{3}-\eps})$.

This type of approach is insufficient to prove a tradeoff between the query and update time, however.
To obtain such a tradeoff, we need to be able to reduce the search for triangle edges to $st$-SubConn where the number of queries is very different from the number of updates.
To achieve this, on the same underlying graph as before,
we use $st$-SubConn to binary search for the nodes in $B$ that participate in a triangle with a given node $a$ (instead of simply trying each neighbor of $a$ as we did above). This allows us to reduce the number of queries in the reduction to $\tilde{O}(n^2/R)$, while keeping the number of updates $\tilde{O}(nR)$. This lets us pick a larger $R$ and trade-off the lower bounds for the query and the update times.

In the binary search for a fixed $a\in A$, we use the queries to check whether there is a node $b$ in a certain contiguous subset of $B$ (interval) that participates in a triangle with $a$. This can be done by activating all neighbors of $a$ in the interval at once, and asking the $s,t$ connectivity query. 
We start the search with an interval that contains all of $B$. 
If we discover that an interval $I$ contains a node $b$ that participates in a triangle with $a$, we proceed to search within both subintervals of $I$ of half the size.
(Thus, we only search in an interval if its parent interval returned ``yes''.)
%
%
Since no $b$ that appears in a triangle with $a$ appears in more than $O(\log n)$ $B$-intervals,
the number of queries to $st$-SubConn is only bigger than the number of triangles by a logarithmic factor, and is thus $\tilde{O}(n^2/R)$. 
The number of updates is no more than $O(d_B(a)\log n)$ for each $a$ where $d_B(a)$ is the number of neighbors of $a$ in $B$. Hence the total number of updates is $\tilde{O}(nR)$.

\paragraph{Lower bounds on partially dynamic algorithms.}

Notice that our reductions almost always look like this (with the exception of PP and $\emptyset$-PP). They proceed in stages, and each stage $i$ has the following form: $I_i$ insertions are performed,
then some number of queries $Q_i$ are asked. Finally the $I_i$ insertions are undone.

We can simulate this type of a reduction with an incremental algorithm as follows.
During each stage, we perform the $I_i$ insertions and $Q_i$ queries, and while we do them, we record the sequence of all changes to the data structure that the insertions (and queries) cause. This makes our reduction no longer black box (it was black box for fully dynamic algorithms). It also increases the space usage to be on the order of the time that it takes to perform the $I_i$ insertions.
However, once we have recorded all the changes, we can undo them in reverse order in roughly the same time as they originally took, and bring the data structure to the same state that it was before the beginning of the stage.
We obtain lower bounds on the preprocessing, update and query time of incremental algorithms. However, since we undo changes, the lower bounds only hold for worst case runtimes.

Simulating the above algorithms with decremental algorithms is more challenging since it would seem that we need to simulate $I_i$ insertions with roughly $I_i$ deletions, and this is not always possible.
We develop some techniques that work for many of our reductions. For instance, we are able to simulate the following with only $O(n)$ deletions (and undeletions) over all $n$ stages: in each stage $i$ a node $s$ has an edge to only the $i$th node from a set of size $n$. This is useful for our proof that efficient worst-case decremental $st$-Reach implies faster triangle algorithms.

\paragraph{Lower bounds based on APSP.}
To show our lower bounds from APSP to incremental or decremental $st$-SP and BWM, we first reduce $st$-SP to BWM, thus showing that we only have to concentrate on $st$-SP.
Then, we combine Roditty and Zwick's~\cite{rzesa} original reduction with Vassilevska Williams and Williams'~\cite{focs10} proof that negative triangle detection is equivalent to APSP.
In particular, we show that the number of shortest paths queries can be reduced to $n$ (from $n^2$) since we only need to simulate determining whether there is a path on $3$ edges from each vertex back to itself.

%% file: defstable.tex
\begin{table}\small
\thisfloatpagestyle{empty}
\begin{tabular}{|c|c|c|}
\hline
\multicolumn{3}{|c|}{Problem}\\ \hline
Maintain & Update & Query \\
\hline
\multicolumn{3}{|c|}{$(s,t)$-Subgraph Connectivity ($st$-SubConn)}\\ \hline
A fixed undirected graph, a subset $S$  &
Insert/remove a node into/from $S$ &
Are $s$ and $t$ connected in the \\
of its vertices and fixed vertices $s,t$ & &
subgraph induced by the nodes in $S$?\\
\hline

\multicolumn{3}{|c|}{Bipartite Perfect Matching (BPMatch) }\\ \hline
An undirected bipartite graph &
Edge insertions/deletions &
Does the graph have a \\
& & perfect matching?\\
\hline

\multicolumn{3}{|c|}{Bipartite Maximum Weight Matching (BWMatch)}\\  \hline
An undirected bipartite graph &
Edge insertions/deletions &
What is the weight of the \\
with integer edge weights &
&
maximum weight matching?\\
\hline

\multicolumn{3}{|c|}{Bipartite matching without length $k$ augmenting paths ($k$-BPM)}\\ \hline
An undirected bipartite graph &
Edge insertions/deletions &
What is the size of a matching \\
&
&
that does not admit \\
& &
length $k$ augmenting paths?\\
\hline

\multicolumn{3}{|c|}{Single Source Reachability (SS-Reach)}\\ \hline
A directed graph and a &
Edge insertions/deletions &
Given a vertex $t$, \\
fixed vertex $s$ &
&
is $t$ reachable from $s$?\\
\hline

\multicolumn{3}{|c|}{$(s,t)$-Reachability ($st$-Reach)}\\  \hline
A directed graph and &
Edge insertions/deletions &
Is $t$ reachable from $s$? \\
fixed vertices $s,t$ &
&
\\
\hline

\multicolumn{3}{|c|}{$(s,t)$-shortest path ($st$-SP)}\\ \hline
An undirected weighted graph and &
Edge insertions/deletions &
What is the distance \\
fixed vertices $s,t$ &
 &
between $s$ and $t$? \\
\hline

\multicolumn{3}{|c|}{Strong Connectivity (SC)}\\ \hline
A directed graph &
Edge insertions/deletions &
Is the graph strongly connected? \\
\hline

\multicolumn{3}{|c|}{2 Strong Components (SC2)}\\ \hline
A directed graph &
Edge insertions/deletions &
Are there more than $2$\\
&
&
strongly connected components? \\
\hline

\multicolumn{3}{|c|}{2 vs $k$ Strong Components (AppxSC)}\\ \hline
A directed graph &
Edge insertions/deletions &
Is the number of SCCs \\
 & &
$2$ or more than $k$? \\
\hline

\multicolumn{3}{|c|}{Maximum SCC size (MaxSCC)}\\ \hline
A directed graph &
Edge insertions/deletions &
What is the size of \\
&
&
the largest SCC?\\
\hline

\multicolumn{3}{|c|}{Single Source Reachability Count (\# SSR)}\\ \hline
A directed graph with &
Edge insertions/deletions &
Given $\ell$, is the number of\\
a fixed source $s$ &
&
nodes reachable from $s$ $<\ell$?\\
\hline

\multicolumn{3}{|c|}{Connected Subgraph (ConnSub)}\\ \hline
A fixed undirected graph  &
Insert/remove a node &
Is the subgraph induced \\
and a vertex subset $S$ &
into/from $S$ &
by $S$ connected? \\
\hline

\multicolumn{3}{|c|}{$(S,T)$-Reachability ($ST$-Reach)}\\ \hline
A directed graph and &
Edge insertions/deletions &
Are there some $s\in S$, $t\in T$ \\
fixed node subsets $S$ and $T$ &
&
s.t. $t$ is unreachable from $s$? \\
\hline

\multicolumn{3}{|c|}{$(4/3-\eps)$-Approximate Diameter ($4/3$-Diam)}\\ \hline
An undirected graph &
Edge insertions/deletions &
Is the diameter \\
&
&
$3$ or $4$?\\
\hline

\multicolumn{3}{|c|}{Chan's Subset Union Problem (SubUnion)}\\ \hline
A subset $S$ of a fixed collection &
Insert/remove a set  &
Is $\cup_{X_i\in S} X_i=U$? \\
$X=\{X_1,\ldots, X_n\}$ of subsets over a &
$X_i$ into/from $S$ &
\\
universe $U$, with $\sum_i |X_i|=m$ &
&
\\
\hline

\multicolumn{3}{|c|}{Pagh's Problem (PP)}\\ \hline
A collection $X$ of &
Given $i,j$, insert  &
Given index $i$ \\
subsets $X_1,\ldots,X_k\subseteq [n]$ &
$X_i\cap X_j$ into $X$ &
and $u\in U$, is $u\in X_i$?\\
\hline

\multicolumn{3}{|c|}{Pagh's Problem with Emptiness Queries ($\emptyset$-PP)}\\
\hline
A collection $X$ of  &
Given $i,j$, insert  &
Given index $i$, \\
subsets $X_1,\ldots,X_k\subseteq [n]$ &
$X_i\cap X_j$ into $X$ &
is $X_i=\emptyset$? \\
\hline

\end{tabular}
\caption{The problems we consider.}
\label{table:problems}
\end{table}

%% file: restable.tex
\begin{table}\small
\begin{tabular}{|c | c c c | c c c c|}
\hline 
\multirow{2}{*}{Problem} & 
\multicolumn{3}{c|}{Best Upper Bounds}  & \multicolumn{4}{c|}{Lower Bounds} \\
& $p(m,n)$ & $u(m,n)$ & $q(m,n)$ & $p(m,n)$ & $u(m,n)$ & $q(m,n)$ & Conjecture\\
\hline 
\multirow{3}{*}{$st$-Reach} &
$1$ & $m$ & $1$ (*) & $m^{4/3-\eps}$ & $m^{\alpha-\eps}$ & $m^{2/3-\alpha-\eps}$ & $3$SUM\\
& $1$ & $1$ & $m$ (*) & $m^{1+\delta-\eps}$ & $m^{2\delta-\eps}$ & $m^{2\delta-\eps}$ & Triangle\\
& $1$ & $n^{1.495}$ & $1$~\cite{sank04} & $n^{3-\eps}$ & $n^{2-\eps}$ & $n^{2-\eps}$ & BMM\\
\hline
\multirow{3}{*}{SC} &
$1$ & $m$ & $1$ (*) & $m^{4/3-\eps}$ & $m^{\alpha-\eps}$ & $m^{2/3-\alpha-\eps}$ & $3$SUM\\
& $1$ & $1$ & $m$ (*) & $m^{1+\delta-\eps}$ & $m^{2\delta-\eps}$ & $m^{2\delta-\eps}$ & Triangle\\
& &  &  & $n^{3-\eps}$ & $n^{2-\eps}$ & $n^{2-\eps}$ & BMM\\
\hline

\multirow{3}{*}{SubConn} & 
$1$ & $m$ & $1$ (*) & $n^{3-\eps}$ & $n^{1-\eps}$ & $n^{2-\eps}$ & BMM\\
& $1$ & $1$ & $m$ (*) & $m^{1+\delta-\eps}$ & $m^{\delta-\eps}$ & $m^{2\delta-\eps}$ & Triangle\\
& $m^{4/3}$ & $m^{2/3}$ & $m^{1/3}$~\cite{CPR08} & $m^{4/3-\eps}$ & $m^{\alpha-\eps}$ & $m^{2/3-\alpha-\eps}$ & $3$SUM\\
\hline

\multirow{3}{*}{BPMatch} & 
BM & $m$ & $1$ (*) &  $m^{4/3-\eps}$ & $m^{\alpha-\eps}$ & $m^{2/3-\alpha-\eps}$ & $3$SUM\\
& $1$ & $1$ & BM (*) & $m^{1+\delta-\eps}$ & $m^{2\delta-\eps}$ & $m^{2\delta-\eps}$ & Triangle\\
& $1$ & $n^{1.495}$ & $1$~\cite{sankdynmatch} & $n^{3-\eps}$ & $n^{2-\eps}$ & $n^{2-\eps}$ & BMM\\
\hline
\multirow{1}{*}{Dec/Inc BWMatch} &
WM & $m$ & $1$ (*) & $n^{3-\eps}$ & $n^{2-\eps}$ & $n^{2-\eps}$ & APSP\\
\hline

\multirow{2}{*}{Dec/Inc $st$-SP} &
$1$ & $m$ & $1$ (*) & $n^{3-\eps}$ & $n^{2-\eps}$ & $n^{2-\eps}$ & APSP\\
& $1$ & $1$ & $m$ (*) & & & & \\
\hline

\multirow{1}{*}{SC2, $\#$SSR, ConnSub,} & 
$1$ & $m$ & $1$ (*) &  &  &  &  \\
\multirow{1}{*}{AppxSCC, SubUnion} & $1$ & $1$ & $m$ (*) & $\poly(n)$ & $m^{1-\eps}$ & $m^{1-\eps}$ & SETH\\
\hline


{$\emptyset$-PP over} & 
$1$ & $n$ & $1$ (*) &  $\poly(n)$ & $n^{1-\eps}$ & $n^{1-\eps}$ & SETH\\
a universe of size $n$ & $1$ & $1$ & $kn$ (*) & $k^{1+\delta-\eps}$ & $k^{\delta-\eps}$ & $k^{\delta-\eps}$ & Triangle\\
and $k\geq n$ sets & ~ & ~ & ~  & $k^{3/2-\eps}$ & $k^{1/2-\eps}$ & $k^{1/2-\eps}$ & BMM\\
  & ~ & ~ & ~ & $k^{4/3-\eps}$ & $k^{1/3-\eps}$ & $k^{1/3-\eps}$ & $3$SUM\\
\hline

{PP over} & 
$1$ & $n$ & $1$ (*) & $k^{1+\delta-\eps}$ & $k^{\delta-\eps}$ & $k^{\delta-\eps}$ & Triangle with $\delta>1/3$\\
a universe of size $n$ & $1$ & $1$ & $kn$ (*) & $k^{3/2-\eps}$ & $k^{1/2-\eps}$ & $k^{1/2-\eps}$ & BMM\\
and $k\geq n$ sets & ~ & ~ & ~  & $k^{4/3-\eps}$ & $k^{1/3-\eps}$ & $k^{1/3-\eps}$~\cite{PatrDyn} & $3$SUM\\
\hline

\multirow{1}{*}{$ST$-Reach or $4/3$-Diam} & 
$1$ & $n^2$ & $1$ (*) &  $\poly(n)$ & $n^{2-\eps}$ & $n^{2-\eps}$ & SETH\\
in sparse graphs & $1$ & $1$ & $n^2$ (*) &  &  &  & \\
\hline
\end{tabular}
\caption{The table includes the current best upper bounds for the listed problems, together with bounds for which a listed conjecture would be falsified. In the above, 
WM refers to $\min\{Mn^\omega, m\sqrt n \log M\}$, i.e. asymptotically the fastest known time to compute a weighted matching, BM refers to $\min\{m^{10/7},n^\omega\}$, i.e. asymptotically the fastest known time to compute a bipartite perfect matching,
$\eps>0$ is an arbitrarily small constant, $\alpha\in [1/6,1/3]$ and $\delta>0$ is some constant for which Triangle detection is not in $m^{1+\delta}$ time. (*) denotes the trivial algorithm.  Dec/Inc means that the upper and lower bounds apply to fully dynamic, and also to partially dynamic, i.e. decremental and incremental, algorithms. All lower bounds can be amortized and expected. All above lower bounds also hold in the case of partially dynamic algorithms, however then the lower bounds are assumed to be worst-case (unless they are already listed in the table).}
\label{tab:res}
\end{table}

%% file: conjecs.tex
\section{The conjectures}
\label{sec:conj}
\paragraph{$3$SUM.}
The $3$SUM problem is the problem of determining whether a set of $n$ integers contains three integers $a,b,c$ so that $a+b=c$.
The problem has a simple $\tilde{O}(n^2)$ time solution: sort the integers, and for every pair $a,b$, check whether their sum is in the list using binary search.
There are faster algorithms as well. Baran, Demaine and $\patrascu$~\cite{alg3sum} showed that in the Word RAM model with $O(\log n)$ bit words, $3$SUM can be solved in $O(n^3\log^2\log n/\log^2 n)$ time.
However, there are no known $O(n^{2-\eps})$ time (so called ``truly subquadratic'') algorithms for the problem for any $\eps>0$. The lack of progress on the problem has led to the following conjecture~\cite{PatrDyn,GO95}.

\begin{conjecture}[No truly subquadratic $3$SUM]
In the Word RAM model with words of $O(\log n)$ bits, any algorithm requires $n^{2-o(1)}$ time in expectation to determine whether a set $S\subset \{-n^3,\ldots,n^3\}$ of $|S|=n$ integers contains three distinct elements $a,b,c\in S$ with $a+b=c$.
\end{conjecture}

(By standard hashing arguments, one can assume that the size of the integers in the $3$SUM instance is $O(n^3)$, and so the conjecture is not for a restricted version of the problem.)

Many researchers believe this conjecture. Besides $\patrascu$'s paper~\cite{PatrDyn} on dynamic lower bounds, $3$SUM is often used to prove conditional hardness for nondynamic problems.
Gajentaan and Overmars~\cite{GO95} formed a theory of ``$3$SUM-hard problems'' by showing that one can reduce $3$SUM to many static problems in computational geometry, showing that unless $3$SUM has a truly subquadratic time algorithm, none of them do. One example of a $3$SUM-hard problem is testing whether in a given set of $n$ points in the plane, $3$ of them are colinear.
Following~\cite{GO95} many other papers proved the $3$SUM hardness of geometric problems~\cite{BGO97,SEO02,Er99,AHLK01,AEK04,AH05,CEH04,BH99}.
Vassilevska and Williams~\cite{VW09,VW09j} showed that a certain weighted graph triangle problem cannot be found efficiently unless Conjecture~\ref{conj:3sum} is false, relating $3$SUM to problems in weighted graphs. 
Their work was recently extended~\cite{AbboudL13} for other weighted subgraph problems.


\paragraph{APSP.}
The second conjecture concerns the all pairs shortest paths problem (APSP): given a directed or undirected graph with integer edge weights, determine the distances between every pair of vertices in the graph.
Classical algorithms such as Dijkstra's or Floyd-Warshall's provide $O(n^3)$ running times for APSP in $n$-ndoe graphs. Just as with $3$SUM, there are $n^{o(1)}$ improvements over this cubic runtime.
Until 2013, the fastest such runtime was $O(n^3\log\log n / \log^2 n)$ by Han and Takaoka~\cite{HanT12}. Williams~\cite{ryanpersonal} has recently designed an algorithm that runs faster than $O(n^3/(\log n)^c)$ time for all constants $c$.
Nevertheless, no truly subcubic time ($O(n^{3-\eps})$ for $\eps>0$) algorithm for APSP is known.
This led to the following conjecture assumed in many papers, e.g.~\cite{rzesa,focs10}. 

\begin{conjecture}[No truly subcubic APSP]
There is a constant $c$, such that in the Word RAM model with words of $O(\log n)$ bits, any algorithm requires $n^{3-o(1)}$ time in expectation to compute the distances between every pair of vertices in an $n$ node graph with edge weights in $\{1,\ldots, n^c\}$.
\end{conjecture}

Vassilevska Williams and Williams~\cite{focs10} showed that many other graph problems are equivalent to APSP under subcubic reductions, and as a consequence any truly subcubic algorithm for them would violate Conjecture~\ref{conj:apsp}.
Some examples of these problems include detecting a negative weight triangle in a graph, computing replacement paths and finding the minimum cycle in the graph.

One could ask, is there a relationship between Conjectures~\ref{conj:apsp} and~\ref{conj:3sum}?
The answer is unknown. However, there is a problem that is in a sense at least as hard as both $3$SUM and APSP, and may be equivalent to either one of them. The problem, Exact Triangle, is, given a graph with integer edge weights, determine whether it contains a triangle with total weight $0$. The work of Vassilevska Williams and Williams~\cite{VW09j} based partially on~\cite{PatrDyn} shows that if Exact Triangle can be solved in truly subcubic time, then both Conjectures~\ref{conj:apsp} and~\ref{conj:3sum} are false.


\paragraph{The Strong Exponential Time Hypothesis.}
The next conjecture is about the exact complexity of an NP-hard problem, namely Boolean Satisfiability in Conjunctive Normal Form (CNF-SAT).
The best known algorithm for CNF-SAT is the $O^{*}(2^n)$ time exhaustive search algorithm which tries all possible $2^n$ assignments to the variables, and it has been a major open problem to obtain an improvement.
There are faster algorithms for $k$-SAT for constant $k$. Their running times are typically of the form $O(2^{n(1-c/k)})$ for some constant $c$ independent of $n$ and $k$ (e.g.~\cite{hirschsat,moniensat,PaturiPZ99,PPSZ05,Sch92sat,Scho99sat}). That is, as $k$ grows, the base of the exponent of the best known algorithms goes to $2$.

Impagliazzo, Paturi, and Zane~\cite{ipz1,ipz2} introduced the Strong Exponential Time
Hypothesis (SETH) to address the question of how fast one can solve $k$-SAT as $k$ grows.
They define:
\begin{eqnarray}
s_k = \inf \{\delta\mid&\exists O^{*}(2^{\delta n}) \textrm{ time algorithm solving }\nonumber \\
    & k\textrm{-SAT instances with } n \textrm{ variables}\}, \nonumber
\end{eqnarray}
The sequence $s_k$ is clearly nondecreasing. The SETH hypothesizes that $\lim_{k\rightarrow \infty} s_k=1$.


\begin{conjecture}[SETH]
For every $\eps>0$, there exists an integer $k$, such that SAT on $k$-CNF formulas on $n$ variables cannot be solved in $O(2^{(1-\eps)n}\poly n)$ time.
\end{conjecture}

The SETH is an extremely popular conjecture in the exact exponential time algorithms community. For instance, Cygan et al.~\cite{cygan} showed that the SETH is also equivalent to the assumption that several other NP-hard problems cannot be solved faster than by exhaustive search, and the best algorithms for these problems are the exhaustive search ones.
Some other work that proves conditional lower bounds based on the SETH for NP-hard problems includes~\cite{cygan,CalabroIP09,DW10,LokshtanovMS11a,cygan2011solving,RHuacheng,pilipczuk2012finding,cygan2012deterministic,cygan13,FHV13}.

Assuming the SETH, one can prove tight conditional lower bounds on the complexity of some polynomial time problems as well. P\v{a}tra\c{s}cu and Williams~\cite{PW10} give several tight lower bounds for problems such as $k$-dominating set (for any constant $k\geq 3$), $2$SAT with two extra unrestricted length clauses, and HornSAT with $k$ extra unrestricted length clauses. Roditty and Vassilevska Williams~\cite{RV13} and the follow-up work of Chechik et al.~\cite{diametersoda14} related the complexity of approximating the diameter of a graph to the SETH.
In this paper we prove the first lower bounds for dynamic problems based on the SETH. The lower bounds we obtain are surprisingly tight- any polynomial improvement over the trivial algorithm would falsify Conjecture~\ref{conj:seth}.
In addition, all lower bounds based on the SETH also hold with arbitrary polynomial preprocessing time.

\paragraph{Triangle.}
The next conjecture is on the complexity of finding a triangle in a graph. The best known algorithm for triangle detection relies on fast matrix multiplication and runs in time $O(\min \{m^{1.41},n^\omega\})$ in $m$-edge, $n$-node graphs~\cite{AlYuZw97}. Even if there were an optimal matrix multiplication algorithm, it would at best imply an $O(\min\{m^{4/3},n^2\})$ time algorithm for triangle detection. The lack of alternative algorithms leads to the conjecture that there may not be a linear time algorithm for triangle finding. (In fact, one may even conjecture that $O(m^{4/3-\eps})$ time is not possible, but we will be conservative.)

\begin{conjecture}[No almost linear time triangle]
There is a constant $\delta>0$, such that in the Word RAM model with words of $O(\log n)$ bits, any algorithm requires $m^{1+\delta-o(1)}$ time in expectation to detect whether an $m$ edge graph contains a triangle.
\end{conjecture}

One may ask whether Conjecture~\ref{conj:tria} is related to Conjectures~\ref{conj:apsp} and~\ref{conj:3sum}.
Although there is no known strong relationship between Conjectures~\ref{conj:apsp} and~\ref{conj:tria}~\footnote{One exception is in~\cite{focs10}, where the authors show that triangle detection in dense graphs and Boolean matrix multiplication (BMM) are naturally related. This gives a loose relationship between APSP and Triangle since APSP is a generalization of BMM.} the relationship between $3$SUM and Triangle detection has been explored. For instance, P\v{a}tra\c{s}cu~\cite{PatrDyn} showed that one can reduce $3$SUM on $n$ numbers to the problem of listing up to $m$ triangles in a graph on $m=O(n^{1.5})$ edges, thus any algorithm that lists $m$ triangles in an $m$-edge graph in $O(m^{4/3-\eps})$ time for $\eps>0$ would falsify Conjecture~\ref{conj:3sum}.

However, is there a relationship between triangle listing and triangle detection?
Vassilevska Williams and Williams~\cite{focs10} proved that for dense graphs, any truly subcubic algorithm for triangle detection implies a truly subcubic algorithm for listing any truly subcubic number of triangles.
Jafargholi and Viola~\cite{JV13} extended this result to the case of sparse graphs. They showed that if one can detect a triangle in $O(m^{1+\eps})$ time, then one can list $m$ triangles in $\tilde{O}(m^{4/3+2\eps/3})$ time.
Unfortunately, their reduction always produces a listing algorithm that runs in $\Omega(m^{4/3})$ time which is insufficient to falsify Conjecture~\ref{conj:3sum}. 
The authors \cite{JV13} also show that Triangle detection on a graph with $m$ edges can be reduced to $3$SUM on $O(m)$ numbers, which implies that if the Triangle Conjecture is true for some $\delta>0$ then $3$SUM requires $n^{1+\delta-o(1)}$ time.
Beyond this, the $3$SUM conjecture and the Triangle Conjecture may be unrelated.

We state our lower bounds in terms of the exponent $\delta$ in Conjecture~\ref{conj:tria}. Thus any sufficiently large improvement on the complexity of our dynamic problems would yield a new algorithm for triangle detection.

%

\paragraph{Boolean matrix multiplication (BMM).}
The Boolean product of two Boolean matrices $A$ and $B$ is the matrix $C$ with entries $C[i,j]=\vee_k (A[i,k]\land B[k,j])$. Many important problems can not only be solved using a fast BMM routine, but are also equivalent to BMM~\cite{lee,focs10}. Hence an efficient and practical BMM algorithm is highly desirable.

The Boolean product of $n\times n$ matrices can be computed using any algorithm for integer matrix multiplication, and hence the problem is in $O(n^{2.373})$ time~\cite{v12}. However, the theoretically efficient matrix multiplication algorithms (except possibly Strassen's~\cite{strassen}) use mathematical machinery that causes them to have high constant factors, and are thus currently impractical.
Because of this, alternative, so called ``combinatorial'' methods, for BMM are sought after.

The current best combinatorial algorithm for BMM by Bansal and Williams~\cite{BansalW09} runs in $O(\frac{n^3}{\log^{2.25} n})$ time, improving on the well-known Four-Russians algorithm~\cite{fourruss}.
Because it has been such a longstanding open problem to obtain an $O(n^{3-\eps})$ time (for constant $\eps>0$) algorithm for BMM, the following conjecture has been floating around the community; many papers base lower bounds for problems on it (e.g.~\cite{rzesa,lee,kavithastretch,AlNa96}). (We place ``conjecture'' in quotes, mainly because ``combinatorial'' is not a well-defined term.)

\begin{fconjecture}[No truly subcubic combinatorial BMM]
In the Word RAM model with words of $O(\log n)$ bits, any combinatorial algorithm requires $n^{3-o(1)}$ time in expectation to compute the Boolean product of two $n\times n$ matrices.
\end{fconjecture}

The only known relationship between the complexity of BMM and the rest of the conjectures in this paper is a result from~\cite{focs10} that any truly subcubic in $n$ combinatorial algorithm for finding a triangle can be converted to a truly subcubic combinatorial algorithm for BMM. Hence ``Conjecture''~\ref{conj:bmm} is equivalent to the conjecture that {\em combinatorial} triangle finding in $n$ node graphs requires $n^{3-o(1)}$ time.
However, that does not necessarily imply Conjecture~\ref{conj:tria} since it could be that there is a linear time algebraic triangle finding algorithm, but no combinatorial one. Furthermore, 
``Conjecture''~\ref{conj:bmm} could be false but Conjecture~\ref{conj:tria} might still be true. According to our current knowledge, even an optimal algorithm for BMM would at best imply an $O(m^{4/3})$ time algorithm for triangle detection.

%% file: formalthms.tex
\section{Formal statement of our results}\label{sec:formalthms}

The problems we study are defined in Table~\ref{table:problems}.
We prove the following theorems. Most of our results are summarized in Table~\ref{tab:res}.

\begin{theorem}
If for some $\eps >0$ and $t \in \mathbb{N}$, we can solve either of
\begin{itemize}

\item fully dynamic \#SSR, SC$2$, AppxSCC, MaxSCC, SubUnion, $\phi$-PP, or ConnSub, with preprocessing time $O(n^{t})$, amortized update time $O(m^{1 - \eps})$, and amortized query time $O(m^{1-\eps})$, or
\item incremental or decremental \#SSR, SC$2$, AppxSCC, MaxSCC, SubUnion, $\phi$-PP, or ConnSub, with preprocessing time $O(n^{t})$, worst case update time $O(m^{1 - \eps})$, and worst case query time $O(m^{1-\eps})$, or

\item fully dynamic $ST$-Reach or $4/3$-Diam with preprocessing time $O(n^{t})$, amortized update time $O(m^{2 - \eps})$, and amortized query time $O(m^{2-\eps})$, or
\item incremental or decremental $ST$-Reach or $4/3$-Diam with preprocessing time $O(n^{t})$, worst case update time $O(m^{2 - \eps})$, and worst case query time $O(m^{2-\eps})$,
\end{itemize}
then Conjecture~\ref{conj:seth} is false.\label{thm:seth}
\end{theorem}

\begin{theorem}\label{thm:3sum}
If for some $\eps >0$ and $1/6 \leq \alpha \leq 1/3$, we can solve either of
\begin{itemize}
\item fully dynamic $st$-SubConn, $st$-Reach, BPMatch, or SC, with preprocessing time $O(m^{\frac{4}{3}-\eps})$, amortized update time $O(m^{\alpha - \eps})$, and amortized query time $O(m^{\frac{2}{3}-\alpha-\eps})$, or
\item incremental $st$-SubConn, $st$-Reach, BPMatch, or SC, with preprocessing time $O(m^{\frac{4}{3}-\eps})$, worst case update time $O(m^{\alpha - \eps})$, and worst case query time $O(m^{\frac{2}{3}-\alpha-\eps})$, or
\item decremental $st$-Reach, BPMatch, or SC, with preprocessing time $O(m^{\frac{4}{3}-\eps})$, worst case update time $O(m^{\alpha - \eps})$, and worst case query time $O(m^{\frac{2}{3}-\alpha-\eps})$, or
\item PP or $\emptyset$-PP over $k$ sets and a universe of size $O(k)$ with preprocessing time $O(k^{\frac{4}{3}-\eps})$, amortized update and query time $O(k^{\frac{1}{3}-\eps})$,
\end{itemize}
then Conjecture~\ref{conj:3sum} is false.
\end{theorem}

\begin{theorem}\label{thm:triangle}
If for some $\delta,\eps> 0$, we can solve either of
\begin{itemize}
\item fully dynamic $st$-Reach, BPMatch, $17$-BPM, or SC, with preprocessing time $O(m^{1+\delta-\eps})$, amortized update and query times $O(m^{2\delta - \eps})$, or
\item incremental or decremental $st$-Reach, BPMatch, $17$-BPM, or SC, with preprocessing time $O(m^{1+\delta-\eps})$, worst case update and query times $O(m^{2\delta - \eps})$, or
\item fully dynamic $st$-SubConn or $5$-BPM with preprocessing time $O(m^{1+\delta-\eps})$, amortized update time $O(m^{\delta-\eps})$, and amortized query time $O(m^{2\delta - \eps})$, or
\item incremental or decremental $5$-BPM with preprocessing time $O(m^{1+\delta-\eps})$, worst case update time $O(m^{\delta - \eps})$, and worst case query time $O(m^{2\delta-\eps})$, or
\item incremental $st$-SubConn, or decremental $st$-SubConn in dense graphs, with preprocessing time $O(m^{1+\delta-\eps})$, worst case update time $O(m^{\delta - \eps})$, and worst case query time $O(m^{2\delta-\eps})$,
\item $\emptyset$-PP over $k$ sets and a universe of size $O(k)$ with preprocessing time $O(k^{1+\delta-\eps})$, amortized update and query time $O(k^{\delta-\eps})$, or
\item for $\delta>1/3$, PP over $k$ sets and a universe of size $O(k)$ with preprocessing time $O(k^{1+\delta-\eps})$, amortized update and query time $O(k^{\delta-\eps})$,
\end{itemize}
then Conjecture~\ref{conj:tria} is false for this choice of $\delta$.
\end{theorem}

Since our reductions are ``combinatorial'' and hold for arbitrary $m$ and $\delta$, we get the following hardness from Conjecture~\ref{conj:bmm}. 

\begin{corollary}
If for some $\eps>0$ we can find a ``combinatorial" algorithm for either of
\begin{itemize}
\item  fully dynamic $st$-Reach, BPMatch, $17$-BPM, or SC, with preprocessing time $O(n^{3-\eps})$, amortized update and query times $O(n^{2 - \eps})$, or
\item incremental or decremental $st$-Reach, BPMatch, $17$-BPM, or SC, with preprocessing time $O(n^{3-\eps})$, worst case update and query times $O(n^{2 - \eps})$, or
\item fully dynamic $st$-SubConn or $5$-BPM with preprocessing time $O(n^{3-\eps})$, amortized update time $O(n^{1-\eps})$, and amortized query time $O(n^{2 - \eps})$, or
\item incremental or decremental $5$-BPM with preprocessing time $O(n^{3-\eps})$, worst case update time $O(n^{1 - \eps})$, and worst case query time $O(n^{2-\eps})$, or
\item incremental or decremental $st$-SubConn with preprocessing time $O(n^{3-\eps})$, worst case update time $O(n^{1 - \eps})$, and worst case query time $O(n^{2-\eps})$,
\item PP or $\emptyset$-PP over $k$ sets and a universe of size $O(k)$ with preprocessing time $O(k^{\frac{3}{2}-\eps})$, amortized update and query time $O(k^{\frac{1}{2}-\eps})$,
\end{itemize}
then Conjecture~\ref{conj:bmm} is false.
\end{corollary}

\begin{theorem}\label{thm:apsp}
If for some $\eps >0$ we can solve decremental or incremental $st$-SP or BWMatch with preprocessing time $O(n^{3-\eps})$ and amortized update and query times $O(n^{2 - \eps})$,
then Conjecture~\ref{conj:apsp} is false.
\end{theorem}

%% file: prelims.tex
Unless otherwise noted, $n$ refers to the number of vertices and $m$ to the number of edges of a graph.
In the context of dynamic problems, $n$ and $m$ are assumed to be upper bounds on the number of nodes and edges in the input graph at any time.
The notation $\Tilde{O}(f(n))$ means $f(n)\polylog n$. The notation $O^*(f(n))$ means $f(n)\poly n$. For a positive integer $n$,
$[n]$ denotes the set $\{1,\ldots,n\}$.
For a node $v\in V$, $N(v)$ denotes the neighborhood of $v$, and $N_A(v) = N(v) \cap A$, where $A \subseteq V$ is a subset of vertices, denotes the neighborhood of $v$ in $A$. 
We denote the degree of a node $v$ by $d(v)$ and define $d_A(v)=|N_A(v)|$ to be the degree of $v$ when restricted to nodes in $A$.

Below we present several efficient reductions between dynamic problems that allow us to focus our lower bound reductions to a small set of problems. The rest of the bounds will follow from the relationships proven below.


\subsection{$st$-SubConn to $st$-Reach}

\begin{lemma}\label{lem:st}
If fully dynamic / incremental / decremental $st$-Reach can be solved with preprocessing, update, and query times $p(m,n), u(m,n), q(m,n)$, respectively, then fully dynamic / incremental / decremental $st$-SubConn can also be solved with preprocessing, update, and query times $p(O(m+n),O(n))$, $u(O(m+n),O(n))$, $q(O(m+n),O(n))$, respectively.
\end{lemma}

\begin{proof}
Let $G=(V,E)$ and $s,t \in V$ be an instance of $st$-SubConn. We create the directed graph $H=(V',E')$, where for every node $v \in V$ we create two copies $v_{in}$ and $v_{out}$ in $V'$, and for every undirected edge $(u,v) \in E$ we add the two directed edges $(u_{out},v_{in})$ and $(v_{out},u_{in})$ to $E'$.
We will maintain the invariant that a node $v\in S\subseteq V$ of $G$ is in the set of activated nodes $S$ if and only if the edge $(v_{in},v_{out})$ is in $E'$.

Let $s',t' \in V'$ be the nodes $s' = s_{out}$ and $t' = t_{in}$. To solve $st$-SubConn on $G$, we solve $s't'$-Reach on $H$.
The node updates to $G$ are handled in a straightforward manner. If a node $v$ is added to $S$, we insert the edge $(v_{in},v_{out})$ to $H$, and if $v$ is removed from $S$, we delete the edge $(v_{in},v_{out})$ from $H$.
To answer $s,t$ connectivity queries in $G$, we check whether $s'$ can reach $t'$ in $H$ and answer similarly. Note that the number of nodes in $H$ is $O(n)$ and the number of edges is $O(m+n)$.

We will show that $s$ and $t$ are connected in the subgraph of $G$ induced by $S$ if and only if there is a directed path from $s_{out}$ to $t_{in}$ in $H$.
For the first direction, let $P= (s,v_1,\ldots,v_k,t)$ be a path from $s$ to $t$ in $G$ that does not contain any nodes that are not in $S$, i.e. $v_1,\ldots, v_k \in S$. Then, 
$P' = (s_{out} \rightarrow v_{1,in} \rightarrow v_{1,out} \rightarrow v_{2,in} \rightarrow \cdots \rightarrow v_{k,out} \rightarrow t_{in})$
 is a directed path from $s'$ to $t'$ in $H$. 
For the other direction, let $P' = (s' \leadsto t')$ be a directed path from $s_{out}$ to $t_{in}$ in $H$, and note that since all edges in $H$ are either of the form $(u_{out}, v_{in})$ where $u\neq v$ and $(u,v)\in E$, or of the form $(u_{in},u_{out})$ where $u\in S$, we know that $P'$ must be of the form $P' = (s' \rightarrow v_{1,in} \rightarrow v_{1,out} \rightarrow v_{2,in} \rightarrow \cdots \rightarrow v_{k,out} \rightarrow t')$. 
Thus, we have the path $P= (s,v_1,\ldots,v_k,t)$ in $G$ where all the nodes are in $S$.

\end{proof}

\subsection{$st$-Reach to BPMatch}

\begin{lemma}\label{lem:bpm}
If fully dynamic / incremental / decremental BPMatch can be solved with preprocessing, update, and query times $p(m,n), u(m,n), q(m,n)$, respectively, then fully dynamic / incremental / decremental $st$-Reach can also be solved with preprocessing, update, and query times $p(O(m),O(n))$, $u(O(m),O(n))$, $q(O(m),O(n))$, respectively.
\end{lemma}

\begin{proof}
Given an instance of $st$-Reach, a graph $G=(V,E)$ on $n$ nodes and $m$ edges, we create an instance of BPMatch, an undirected bipartite graph $H$, in which there will exist a perfect matching whenever there is a path from $s$ to $t$ in $G$.

The nodes of $H$ will be made of two copies of the nodes of $G$, $V(H)=V_{in}\cup V_{out}$ where $V_{in}= \{ v_{in} \mid v \in V, v \neq s \}$ and $V_{out}= \{ v_{out} \mid v \in V, v \neq t \}$. For every edge $(u,v)$ in $G$ where $v\neq s$ and $u\neq t$ we will create an edge $(u_{out},v_{in})$ in $H$, then, we also add an edge $(v_{in},v_{out})$ between the two copies of each node $v\in V \setminus \{s,t\}$. Thus, $E(H) = \{ (u_{out},v_{in}) \mid (u,v)\in E \} \cup \{ (v_{in},v_{out}) \mid v \in V, v \neq s,t \}$.
Note that the number of nodes in $H$ is $O(n)$ and the number of edges is $O(m)$.

Updates are handled as follows. When an edge $(u,v)$ is removed (or added) in $G$, we remove (or add) the edge $(u_{out},v_{in})$ in $H$.

We claim that, at any point, there is a path from $s$ to $t$ in $G$ if and only if there is a perfect matching in $H$. 
For the first direction, assume there is a  path $s\rightarrow v_1 \rightarrow \ldots \rightarrow v_k \rightarrow t$ in $G$, and consider the set of edges $M = \{ (s_{out}, v_{1,in}), (v_{1,out}, v_{2,in}), \ldots, (v_{k,out}, {t}_{in}) \} \cup \{ (u_{in}, u_{out}) \mid u \in V, u \neq s,t,v_1,\ldots, v_k \}$ in $H$. Every node in $H$ appears exactly once in $M$ and therefore $M$ is a perfect matching. 
For the other direction, assume that $M$ is a perfect matching in $H$, and consider the following set of edges $S = \{ (u,v) \mid (u_{out}, v_{in}) \in M, u \neq v \}$ in $G$. We claim that $S$ must contain a path from $s$ to $t$.
For a node $v \in G$, define its in-degree and out-degree according to $S$ as $d_{in}(v) = | \{ (x,v) \in S \} |$ and $d_{out}(v) = |\{ (v,x)\in S \} |$. Since $M$ is a matching, we know that $d_{in}(v),d_{out}(v) \leq 1$ for every node $v \in V$. Moreover, we know that $d_{in}(s)=d_{out}(t)=0$ since $s_{in},t_{out}$ do not exist in $H$, and $d_{out}(s)=1$ since $M$ is a perfect matching and all neighbors of $s_{out}$ in $H$ have the form $v_{in}$ for $v\neq s$, and similarly, $d_{in}(t)=1$. While for every node $v \neq s,t$, if $d_{in}(v)=1$ then $d_{out}(v)=1$ too, since $v_{out}$ must be matched to a node other than $v_{in}$ who is already matched to another node $x_{out}$.
Therefore, $S$ must contain a path that starts at $s$ and ends at $t$, along with possibly other disjoint cycles. 

\end{proof}

\subsection{$st$-SP to BWMatch}

\begin{lemma}\label{lem:bpwm}
If fully dynamic / incremental / decremental BWMatch can be solved with preprocessing, update, and query times $p(m,n), u(m,n), q(m,n)$, respectively, then fully dynamic / incremental / decremental $st$-SP can also be solved with preprocessing, update, and query times $p(O(m),O(n))$, $u(O(m),O(n))$, and $q(O(m),O(n))$, respectively.
\end{lemma}

We will use the same construction of the bipartite graph $H$ from $G$ that was given in the proof of Lemma~\ref{lem:bpm}, but we will add edge weights this time and since $G$ is undirected now we will make it directed by bi-directing the edges. Let $w:E(G)\rightarrow [M]$ be the edge weight function of $G$, we define $w':E(H)\rightarrow [M]$ as follows. For the edges of the form $e=(v_{in},v_{out}) \in E(H)$ we set $w'(e)=M$, while for the edges of the form $e=(u_{out},v_{in}) \in E(H)$ we set $w'(e) = M - w(u,v)$.
Updates are handles the same way as before.

We claim that the value of the maximum matching in $H$ equals $n\cdot M$ minus the weight of the shortest path from $s$ to $t$ in $G$.
To see this, let $P$ be the shortest $s,t$ path $s\rightarrow v_1 \rightarrow \ldots \rightarrow v_k \rightarrow t$ in $G$, and define the perfect matching $M = \{ (s_{out}, v_{1,in}), (v_{1,out}, v_{2,in}), \ldots, (v_{k,out}, {t}_{in}) \} \cup \{ (u_{in}, u_{out}) \mid u \in V, u \neq s,t,v_1,\ldots, v_k \}$ in $H$, as before. The total weight of $M$ is 
\[ w'(M) = (M-w(s,v_1)) + (M-w(v_1,v_2)) + \ldots + (M-w(v_k,t)) + (n-k-2)\cdot M = n\cdot M -w(P).\]
Let $M'$ be a different perfect matching in $H$, we will show that $w'(M') \leq w'(M)$ which will conclude the proof of our claim.
As before, define the set of edge $S = \{ (u,v) \mid (u_{out}, v_{in}) \in M', u \neq v \}$ in $G$, and recall the definition of $d_{in}(v)$ and $d_{out}(v)$. We have shown that since $M'$ is a perfect matching, the edges in $S$ must contain a path $P'$ from $s$ to $t$ and possibly some disjoint cycles $C_1,\ldots, C_k$ in $G$. Therefore, since all weights are positive and since $w(P')\geq w(P)$ we get that
\[ w'(M') = n\cdot M - \sum_{i=1}^{k} w(C_i) - w(P') \leq n\cdot M - w(P') \leq n\cdot M- w(P) = w'(M).
\]

\subsection{$st$-Reach to SC}

\begin{lemma} \label{lem:reach-to-sc}
If fully dynamic / incremental / decremental SC can be solved with preprocessing, update, and query times $p(m,n), u(m,n), q(m,n)$, respectively, then fully dynamic / incremental / decremental $st$-Reach can also be solved with preprocessing, update, and query times $p(O(m+n),O(n))$, $u(O(m+n),O(n))$, and $q(O(m+n),O(n))$, respectively.
\end{lemma}

\begin{proof}
Given an instance of $st$-Reach, a graph $G=(V,E)$ on $n$ nodes and $m$ edges, the instance of SC we create is a directed graph $H$ which is simply $G$ with the addition of the edges $(v,s), (t,v)$ for every node $v \in V \setminus \{ s,t \}$. 
If an edge is added to $G$ we add it to $H$ too, and if an edge $(u,v)$ is removed from $G$, we remove it from $H$ only if $u \neq t$ and $v \neq s$ -- that is, we keep the additional edges in $H$ at all times.

We claim that $s$ can reach $t$ in $G$, if and only if $H$ is strongly connected. To see this, first, assume that there is a path $P$ from $s$ to $t$ in $G$, and note that any two nodes $u,v$ in $H$ will now be connected using the path $u \rightarrow s \rightarrow (P) \rightarrow t \rightarrow v$, and therefore $H$ will be strongly connected. While, for the other direction, assume that $H$ is strongly connected, and therefore there must be a simple path from $s$ to $t$ in $H$. This path cannot use any of the additional edges, and therefore it is made entirely of edges in $G$, which means that $s$ can reach $t$ in $G$.

\end{proof}

\subsection{SubUnion to ConnSub}
Here we show that there is an efficient reduction from SubUnion to ConnSub.
Chan~\cite{Chan06} conjectured that the hardness of ConnSub is due to the hardness of SubUnion and this is exactly what we will show in Section~\ref{sec:seth}.

\begin{lemma}
Suppose fully dynamic / incremental / decremental ConnSub on a graph with $n$ nodes and $m$ edges can be done with amortized/worst case update time $u(m,n)$ and query time $q(m,n)$ and $O(n^t)$ preprocessing time. Then fully dynamic / incremental / decremental SubUnion over a universe of size $n \geq |X|$ and sum of set sizes $m$ can be done with amortized/worst case update time $u(m+n,n)$ and query time $q(m+n,n)$ and $O(n^t)$ preprocessing time.\label{lemma:unioncon}
\end{lemma}

\begin{proof}
Given an instance of SubUnion over a universe $U$ we simulate it with ConnSub as follows.

Create a bipartite graph $H$ where one partition is the universe $U$ and the other is a set $A$ on $|X|$ nodes, one corresponding to each set $X_i\in X$.
For $i\in A$, add an edge to every $u\in U$ such that $u\in X_i$.
Add an additional node $a$ that all nodes of $A$ are connected to.
This node $a$ is always active; all nodes of $U$ are also always active.
When a subset $X_i$ is inserted into $S$, we turn node $i$ on, and when it is
removed from $S$, we turn node $i$ off.

The subgraph of $H$ induced on the active nodes is connected if and only if every node in $U$ has some active node $X_i$ that it is connected to, i.e. every $u\in U$ is in $\cup_{X_i\in S} X_i$, and so $\cup_{X_i\in S} X_i=U$.

The number of vertices in $H$ is $|U|+|X|+1=O(n)$, and the number of edges is $m+n$.
\end{proof}

%% file: seth.tex
In this section we prove hardness results for dynamic problems assuming SETH. The proof is divided into the lemmas below.

\begin{reminder}{Theorem~\ref{thm:seth}}
If for some $\eps >0$ and $t \in \mathbb{N}$, we can solve either of
\begin{itemize}

\item fully dynamic \#SSR, SC$2$, AppxSCC, MaxSCC, SubUnion, $\phi$-PP, or ConnSub, with preprocessing time $O(n^{t})$, amortized update time $O(m^{1 - \eps})$, and amortized query time $O(m^{1-\eps})$, or
\item incremental or decremental \#SSR, SC$2$, AppxSCC, MaxSCC, SubUnion, $\phi$-PP, or ConnSub, with preprocessing time $O(n^{t})$, worst case update time $O(m^{1 - \eps})$, and worst case query time $O(m^{1-\eps})$, or

\item fully dynamic $ST$-Reach or $4/3$-Diam with preprocessing time $O(n^{t})$, amortized update time $O(m^{2 - \eps})$, and amortized query time $O(m^{2-\eps})$, or
\item incremental or decremental $ST$-Reach or $4/3$-Diam with preprocessing time $O(n^{t})$, worst case update time $O(m^{2 - \eps})$, and worst case query time $O(m^{2-\eps})$,
\end{itemize}
then Conjecture~\ref{conj:seth} is false.
\end{reminder}




We begin with a proposition that we will use in all of our reductions.

\begin{proposition}
If the Strong Exponential Time Hypothesis is true, then for every $\gamma>0$ there exists a $k$ such that $k$-SAT instances on $O(n)$ clauses require $2^{(1-\gamma)n-o(n)}$ time.\label{prop:sparse}
\end{proposition}

\begin{proof}
Let $\gamma>0$ be given. Pick some small constant $\eps>0$, $\eps<\gamma$. 
Recall that the SETH states that for every $\gamma' > 0$ there exists an integer $k$ such that no algorithm running in time $O(2^{(1-\gamma')n}\poly n)$ solves $k$-SAT for all instances with $n$ variables.

Let $k$ be the integer as above corresponding to $\gamma'=(\gamma-\eps)$, i.e. such that $k$-SAT cannot be solved in $O^*(2^{(1-\gamma')n})$ time.
Now, given any $k$-CNF formula, we apply the sparsification lemma~\cite{ipz1}: for our choice of $\eps>0$ we obtain $2^{\eps n}$ $k$-SAT instances on $n$ variables and $O(n)$ clauses.
Now suppose that there is an $O^*(2^{(1-\gamma)n})$ time algorithm for $k$-SAT that works on instances with $O(n)$ clauses, then there is a
$O^*(2^{(1-\gamma+\eps)n})=O^*(2^{(1-\gamma')n})$ time algorithm for $k$-SAT that works on instances with an arbitrary number of clauses and that would be a contradiction to our choice of $k$. Hence assuming the SETH, $k$-SAT instances on $O(n)$ clauses require $2^{(1-\gamma)n-o(1)}$ time.
\end{proof}

\tocless\subsection{Reductions from CNF-SAT to dynamic graph problems}

\paragraph{The graph $H_\delta$.}
All of our reductions to dynamic graph problems start from the same graph $H_\delta$ (for any constant $\delta\in (0,1)$) constructed as follows. 

Let $F$ be a CNF formula on a set $V$ of $n$ variables and $O(n)$ clauses.
Let $U\subseteq V$ be a subset of the variables of size $\delta n$.
Create a node for each of the $2^{\delta n}$ partial assignments to the variables of $U$. Call these nodes $\bar{U}$. Create a node for each of the $O(n)$ clauses of $F$. Call these nodes $C$.
Create an edge between a partial assignment $\phi\in \bar{U}$ and a clause $c\in C$ iff $\phi$ does not satisfy any of the literals in $c$.

The graph $H_\delta$ above has $O(2^{\delta n})$ vertices and $O(2^{\delta n} n)$ edges.

\paragraph{Using $H_\delta$.}
The constructions in our proofs add to $H_\delta$ at most a constant number of extra nodes and $O(n)$ edges, thus keeping the size of the graph roughly the same.
If the preprocessing time of the corresponding dynamic problem is $O(N^t)$ for graphs on $N$ vertices and $\tilde{O}(N)$ edges, each of our constructions can be preprocessed in $O(2^{\delta n t} n)$ time. We set $\delta = (1-\eps)/t$ for some small $\eps>0$, so that the preprocessing time is $O(2^{(1-\eps)n} n)$ .

The reductions typically involve $2^{(1-\delta) n}$ stages. In each stage, $O(n)$ edges are inserted or deleted, $O(n)$ queries are performed, and then the edge updates of the stage are undone. Thus, if any update on a graph with $N$ nodes and $\tilde{O}(N)$ edges takes $O(N^{1-\eps})$ time for some $\eps>0$, then the $k$-SAT instance can be solved in time (modulo polynomial factors)
\[2^{(1-\delta)n}\cdot 2^{\delta n (1-\eps)} = 2^{(1-\eps\delta) n}.\]
We obtain a contradiction to the SETH by picking the clause size $k$ of the input CNF-formula using Proposition~\ref{prop:sparse} with $\gamma=1-\eps\delta$.

\paragraph{Proof of Theorem~\ref{thm:seth} for all graph problems except ConnSub.}
The lemmas below prove all bullets of Theorem~\ref{thm:seth} except those for ConnSub, SubUnion and $\emptyset$-PP. Each lemma explains how to modify the graph $H_\delta$ so that the problem at hand solves the SAT instance.

We begin by showing a lower bound for \# SSR: the problem of maintaining the size of the reachability tree of a fixed source under edge deletions and insertions.
\begin{lemma}[\# SSR]
Suppose fully dynamic \# SSR on a graph with $n$ nodes and $\tilde{O}(n)$ edges can be done with (amortized) update time $u(n)$ and query time $q(n)$ and $O(n^t)$ preprocessing time. Then assuming the SETH, $\max\{u(n),q(n)\}\geq \Omega(n^{1-\eps})$ for all $\eps>0$.
\end{lemma}

\begin{proof}
Construct $H_{\delta}$ for $\delta=(1-\eps)/t$ for some $\eps>0$, where the edges are directed from the clause nodes $C$ to the partial assignment nodes $\bar{U}$.
Add an extra node $s$.

Now, there is a stage for every partial assignment to the variables of $F$ that are not in $U$. The number of stages is hence $2^{|V\setminus U|}
=2^{(1-\delta)n}$.

In each stage we do the following. Let $\phi$ be the partial assignment to $V\setminus U$ for this stage. Add an edge from $s$ to every clause $c$ that $\phi$ does not satisfy. Let $d(s)$ be the number of edges added. Then ask the query whether the number of nodes reachable from $s$ is less than $2^{\delta n}+d(s)$.
If so, return that $F$ is satisfiable. If not, remove the edges incident to $s$ and move on to the next partial assignment to $V\setminus U$, starting the next stage.

The number of updates and queries is $O(2^{n(1-\delta)}n)$ and by the runtime analysis from before we get that the SETH is false.

To see the correctness, consider a stage corresponding to a partial assignment $\phi$ to the variables of $V\setminus U$. The number of clause nodes reachable from $s$ is exactly $d(s)$ since the only incoming edges into $C$ are those from $s$. Hence the number of nodes reachable from $s$ is $2^{\delta n}+d(s)$ only if $s$ can reach all nodes of $\bar{U}$ (and this number is less otherwise). A partial assignment $\phi'$ to the variables of $U$ is reachable from $s$ if and only if there is some clause that neither $\phi$ nor $\phi'$ satisfies. Hence the answer to the query after the stage is ``yes'' if and only if there is a $\phi'$ such that $\phi$ and $\phi'$ together satisfy all of the clauses, i.e. $F$ is satisfiable. Moreover, if $F$ is satisfiable, then let $a$ be a satisfying assignment. Let $\phi$ be the restriction of $a$ to the variables of $U$. Then the stage corresponding to $\phi$ will have a ``yes'' answer to its query, and the algorithm solves the SAT problem for $F$.
\end{proof}

A variant of the same proof shows the following. 

\begin{lemma}[Inc/Dec \# SSR]
Suppose that incremental/decremental \# SSR on a graph with $n$ nodes and $\tilde{O}(n)$ edges can be done with worst case update time $u(n)$ and query time $q(n)$ and $O(n^t)$ preprocessing time. Then assuming the SETH, $\max\{u(n),q(n)\}\geq \Omega(n^{1-\eps})$ for all $\eps>0$.
\end{lemma}

\begin{proof}
Suppose the algorithm is incremental. Then, during each stage perform all the edge insertions and the query, and while doing this, write down all changes made by the algorithm. After the query finishes, go through the changes in reverse order and undo them, going back to the state of the data structure before any stage has begun. The time to undo the changes is no more than the time to do the insertions and the query. Thus the same analysis goes through as in the proof for the fully dynamic case, except that now we can only assume that we have worst case runtime bounds.

Suppose the algorithm is decremental. Then we change the original state of the data structure to be the graph $H_\delta$ together with $s$ and edges from $s$ to all clause nodes. Now, in each stage, instead of inserting edges from $s$ to clauses, we delete the edges between $s$ and the clauses that $\phi$ satisfies. We then apply the same argument as for the incremental case.
\end{proof}

Now we prove the lower bound for SC2.

\begin{lemma}[SC2]
Suppose fully dynamic SC2 on a graph with $n$ nodes and $\tilde{O}(n)$ edges can be done with (amortized) update time $u(n)$ and query time $q(n)$ and $O(n^t)$ preprocessing time. Then assuming the SETH, $\max\{u(n),q(n)\}\geq \Omega(n^{1-\eps})$ for all $\eps>0$.

Similarly, suppose that decremental/incremental SC2 on a graph with $n$ nodes and $\tilde{O}(n)$ edges can be done with worst case update time $u(n)$ and query time $q(n)$ and $O(n^t)$ preprocessing time. Then assuming the SETH, $\max\{u(n),q(n)\}\geq \Omega(n^{1-\eps})$ for all $\eps>0$.\label{lemma:sc2}
\end{lemma}

\begin{proof}
We begin with the graph $H_\delta$ for $\delta=(1-\eps)/t$ for some $\eps>0$. We add two new vertices $s$ and $s'$ so that every partial assignment node has an edge to $s$ (and there are no edges incident to $s'$). 

In each stage we consider a partial assignment $\phi$ to the variables in $V\setminus U$. 
We add edges from $s$ to all clauses for which $\phi$ does not set any literals to true. We add bidirectional edges between $s'$ and the clauses $C_\phi$ that $\phi$ satisfies. At the end of the stage these changes are undone.
(For decremental algorithms, $s$ has edges to all clause nodes, $s'$ has bidirectional edges to all clause nodes, and during the stage the respective edges are deleted until $s$ has edges only to the clauses not in $C_\phi$ and $s'$ has bidirectional edges to the clauses in $C_\phi$.)

Notice that there is no path from $s$ to $s'$ or from $s$ to $C_\phi$. There are also no paths from the partial assignment nodes to $C_\phi$ and hence also none to  $s'$. Thus, $C_\phi$ and $s'$ form a strongly connected component.
Finally, since every partial assignment node has an edge to $s$, these nodes and $s$ form a strongly connected component if and only if for every partial assignment $\phi'$ to $U$, there is a clause node $c$ such that neither $\phi$ nor $\phi'$ satisfies $c$. Hence there are exactly $2$ strongly connected components if $\phi$ cannot be completed to a satisfying assignment, and there are more than $2$ otherwise.
\end{proof}

Modifying the above proof, we can prove the following lemma.

\begin{lemma}[AppxSCC and MaxSCC]
Suppose fully dynamic AppxSCC or MaxSCC on a graph with $n$ nodes and $\tilde{O}(n)$ edges can be done with (amortized) update time $u(n)$ and query time $q(n)$ and $O(n^t)$ preprocessing time. Then assuming the SETH, $\max\{u(n),q(n)\}\geq \Omega(n^{1-\eps})$ for all $\eps>0$.

Similarly, suppose that decremental/incremental AppxSCC or MaxSCC on a graph with $n$ nodes and $\tilde{O}(n)$ edges can be done with worst case update time $u(n)$ and query time $q(n)$ and $O(n^t)$ preprocessing time. Then assuming the SETH, $\max\{u(n),q(n)\}\geq \Omega(n^{1-\eps})$ for all $\eps>0$.
\end{lemma}

\begin{proof}
For MaxSCC, the same reduction as that in Lemma~\ref{lemma:sc2} works. 
Consider a stage for a partial assignment $\phi$. Let $d$ be the number of clauses that $\phi$ does not satisfy.
If the maximum size of an SCC equals $d+2^{\delta n}$, then $\phi$ cannot be completed to a satisfying assignment since then all $\phi'$ nodes will be in the same component as $s$. Otherwise, if the maximum SCC size is less than $d+2^{\delta n}$, then there is some $\phi'$ that completes $\phi$ to a satisfying assignment.

For AppxSCC for a constant $k$, we modify the above proof so that there are $k$ copies of each partial assignment node. The number of nodes and edges increases by a factor of $k$. The rest of the proof stays the same. If $\phi$ cannot be completed to a satisfying assignment, then the number of SCCs is still $2$, but if it can be completed to a satisfying assignment, then the number is at least $k+2$ since all $k$ copies of some assignment $\phi'$ are not in the same SCC as $s$. Moreover, all these copies are in separate components since the only possible paths between them must pass through $s$. 
\end{proof}
Next, we prove the lower bounds for $ST$-Reach.

\begin{lemma}[$ST$-Reach]
Suppose fully dynamic $ST$-Reach on a graph with $n$ nodes and $\tilde{O}(n)$ edges can be done with (amortized) update time $u(n)$ and query time $q(n)$ and $O(n^t)$ preprocessing time. Then assuming the SETH, $\max\{u(n),q(n)\}\geq \Omega(n^{2-\eps})$ for all $\eps>0$.

Similarly, suppose that decremental/incremental $ST$-Reach on a graph with $n$ nodes and $\tilde{O}(n)$ edges can be done with worst case update time $u(n)$ and query time $q(n)$ and $O(n^t)$ preprocessing time. Then assuming the SETH, $\max\{u(n),q(n)\}\geq \Omega(n^{2-\eps})$ for all $\eps>0$.
\end{lemma}

\begin{proof}
Pick $\delta<\min\{1/t,1/3\}$ and create $H_\delta$.

Let $U'$ be a subset of the variables not in $U$ of size $\delta n$. Since $\delta<1/3$, $U'$ can be picked.
Create a graph $H'_\delta$ analogous to $H_\delta$ but for $U'$ instead of $U$ and so that the edges go from the clause nodes to the assignments to $U'$ that do not satisfy them.
Let $H$ be the disjoint union of $H_\delta$ and $H'_\delta$.

$H$ has $O(2^{\delta n})$ nodes and $O(2^{\delta n}n)$ edges. The node set of $H$ is as follows: two copies of the clause nodes, $C$ and $C'$, and two sets of $2^{\delta n}$ partial assignments- a set for $U$ and a set for $U'$.

The stages correspond to the $2^{n(1-2\delta)}$ partial assignments to the variables in $V\setminus (U\cup U')$.
Consider a stage corresponding to a partial assignment $\phi$. For each clause $c$ for which $\phi$ does not satisfy any of its literals, we add an edge from the copy of $c$ in $C$ to its copy in $C'$. We then ask the query whether all nodes of $T$ are reachable from all nodes of $S$. If not, we return that the formula is satisfiable. Otherwise, we undo the updates and move on to the next stage.
(For decremental algorithms, between stages there is a perfect matching between the clause nodes $C$ and their copies in $C'$, and each stage removes some of these edges.)

Any dynamic algorithm with $O(N^{2-\eps})$ time update and query time would solve CNF-SAT asymptotically in
time (excluding polynomial factors)
\[2^{n(1-2\delta)} \cdot (2^{\delta n})^{2-\eps} = 2^{n(1-\delta\eps)}.\]
\end{proof}



Next, we prove the lower bounds for maintaining a $(4/3-\eps)$-approximation to the diameter.

\begin{lemma}
Suppose fully dynamic $4/3$-Diam on a graph with $n$ nodes and $\tilde{O}(n)$ edges can be done with (amortized) update time $u(n)$ and query time $q(n)$ and $O(n^t)$ preprocessing time. Then assuming the SETH, $\max\{u(n),q(n)\}\geq n^{2-\eps}$ for all $\eps>0$.

Similarly, suppose that decremental/incremental $4/3$-Diam on a graph with $n$ nodes and $\tilde{O}(n)$ edges can be done with worst case update time $u(n)$ and query time $q(n)$ and $O(n^t)$ preprocessing time. Then assuming the SETH, $\max\{u(n),q(n)\}\geq n^{2-\eps}$ for all $\eps>0$.
\end{lemma}

\begin{proof}
The construction is analogous to the one for $ST$-Reach. We outline the differences. All edges in the graph $H$ are made undirected. 

Let $\bar{U}$ be the nodes corresponding to the partial assignments to $U$ and let $\bar{U'}$ be the nodes corresponding to the partial assignments to $U'$.
Two nodes $s$ and $s'$ are added so that $s$ has edges to all nodes in $\bar{U}$ and $s'$ has edges to all nodes in $\bar{U'}$.
An extra node $x$ is added with edges to all nodes in $C\cup C'\cup \{s,s'\}$.


While constructing the graph, if some partial assignment (to $U$ or $U'$) satisfies all the clauses, then we can just return that the formula is satisfiable. Otherwise every node of $\bar{U}$ has an edge to $C$ and every node of $\bar{U'}$ has an edge to $C'$. Because of the addition of the node $x$, the distance between any two nodes in the graph is at most $4$, and it is exactly $4$ for all pairs of nodes from $\bar{U}\times \bar{U'}$. The distance between $x$ and all nodes, between pairs within $\bar{U}$, within $\bar{U'}$ within $C\cup \{s\}$, or within $C'\cup \{s'\}$ is $2$.

Now consider a stage for a partial assignment $\phi$ to $V\setminus \{U\cup U'\}$. If the diameter of the graph is now less than $4$, it must be that for every pair of assignments $\alpha$ to $U$ and $\alpha'$ to $U'$, there is some clause $c$ that none of $\phi,\alpha,\alpha'$ satisfy, as the only possible paths shorter than $4$ are of the form $\alpha\rightarrow c\rightarrow c'\rightarrow \alpha'$. Hence if the diameter is less than $4$, it is exactly $3$ and there is no way to complete $\phi$ to a satisfying assignment. Otherwise, if the diameter is still $4$, then there exist $\alpha\in \bar{U}$ and $\alpha'\in \bar{U'}$ such that for every clause that $\phi$ does not satisfy, one of $\alpha$ or $\alpha'$ does, so that $\phi\cdot \alpha\cdot \alpha'$ is a satisfying assignment. Hence any dynamic algorithm that can distinguish between diameter $3$ and $4$ would solve CNF-SAT.
\end{proof}

\tocless\subsection{Lower bounds for SubUnion, $\emptyset$-PP and ConnSub}
SubUnion and $\emptyset$-PP are nongraph problems, so our proofs no longer use $H_\delta$. The proof for ConnSub follows from the proof for SubUnion via Lemma~\ref{lemma:unioncon} from the preliminaries.

\begin{lemma}[SubUnion]
Suppose fully dynamic SubUnion over a universe $U$ of size $n$ and $m=\tilde{O}(n)$ can be done with (amortized) update time $u(n)$ and query time $q(n)$ and $O(n^t)$ preprocessing time. Then assuming the SETH, $\max\{u(n),q(n)\}\geq \Omega(n^{1-\eps})$ for all $\eps>0$.

Similarly, suppose that decremental/incremental SubUnion over a universe $U$ of size $n$ and $m=\tilde{O}(n)$ can be done with worst case update time $u(n)$ and query time $q(n)$ and $O(n^t)$ preprocessing time. Then assuming the SETH, $\max\{u(n),q(n)\}\geq \Omega(n^{1-\eps})$ for all $\eps>0$.\label{lemma:subunion}
\end{lemma}

\begin{proof}
Let $\delta<1/t$ be a constant.
Let $U=[2^{n\delta}]$, corresponding to the $2^{\delta n}$ partial assignments to the first $\delta n$ of the variables of the given CNF formula.
$X$ contains for every clause $c$, the set $U_c$ of assignments (with indices in $U$) that do not satisfy $c$.
We have that $\sum_c |U_c|\leq 2^{\delta n}n = \tilde{O}(2^{\delta n})=\tilde{O}(|U|)$. 

Let $S=\emptyset$ for fully dynamic and incremental algorithms. For decremental algorithms, let $S$ contain $U_c$ for every clause $c$.

Now, in each stage consider a partial assignment $\phi$ to the last $(1-\delta)n$ of the variables. 
For every clause $c$ for which $\phi$ does not satisfy any literals, add the set $U_c$ to $S$.
For decremental algorithms, remove from $S$ all sets $U_c$ for clauses $c$ that $\phi$ satisfies. Now say $S=\{X_1,\ldots, X_t\}$.
After this, query whether $\cup_i X_i=U$. If not, return that the formula is satisfiable, otherwise undo the changes and move on to the next phase.

Notice that if $\cup_i X_i\neq U$, then there is some partial assignment $\phi'$ to the first $\delta n$ variables such that $u\notin U_c$ for every $U_c\in X$, i.e. $u$ must satisfy all clauses $c$ that $\phi$ does not satisfy, and $\phi'$ completes $\phi$ to a satisfying assignment.
Otherwise, if $\cup_i X_i=U$, then for every $\phi'$ there is some clause $c$ that neither $\phi'$ nor $\phi$ satisfy, and hence $\phi$ cannot be completed to a satisfying assignment.

Suppose that $u(N),q(N)\leq N^{1-\eps}$.
Then the algorithm described solves the SAT problem for the CNF-formula in time
\[2^{(1-\delta)n}n\cdot 2^{\delta n (1-\eps)}=n 2^{n(1-\eps\delta)},\]
thus contradicting the SETH.
\end{proof}


%
%
%
%

Lemma~\ref{lemma:unioncon} from the preliminaries showed that SubUnion can be efficiently reduced to the ConnSub problem.
Thus, Lemma~\ref{lemma:subunion} immediately implies the following lemma. 

\begin{lemma}[ConnSub]
Suppose fully dynamic ConnSub on a graph with $n$ nodes and $\tilde{O}(n)$ edges can be done with (amortized) update time $u(n)$ and query time $q(n)$ and $O(n^t)$ preprocessing time. Then assuming the SETH, $\max\{u(n),q(n)\}\geq \Omega(n^{1-\eps}$) for all $\eps>0$.

Similarly, suppose that decremental/incremental ConnSub on a graph with $n$ nodes and $\tilde{O}(n)$ edges can be done with worst case update time $u(n)$ and query time $q(n)$ and $O(n^t)$ preprocessing time. Then assuming the SETH, $\max\{u(n),q(n)\}\geq \Omega(n^{1-\eps})$ for all $\eps>0$.
\end{lemma}

Finally, we prove the bounds for $\emptyset$-PP. 
\begin{lemma}[$\emptyset$-PP]
Suppose that dynamic $\emptyset$-PP over a universe of size $n$ can be done with (amortized) update time $u(n)$ and query time $q(n)$ and $O(n^t)$ preprocessing time. Then assuming the SETH, $\max\{u(n),q(n)\}\geq \Omega(n^{1-\eps})$ for all $\eps>0$.
\end{lemma}

\begin{proof}
Pick $\delta<1/t$, $\delta>0$.
As before, let $U$ be a subset of the CNF formula variables $V$ of size $\delta n$.
The universe over which we will create our sets will be $[2^{\delta n}]$.
The preprocessing time will hence be $2^{\delta n t} = O(2^{(1-\eps)n})$ for some $\eps>0$.

For every clause $c$, create a set $X_c$ containing $j$ if and only if the $j$th partial assignment to the variables in $U$ satisfies $c$.

Now we proceed in stages. In each stage we consider a partial assignment $\phi$ to the variables in $V\setminus U$.
Using $O(n)$ updates we create the set $X_\phi = \cap_{c~:~\phi \textrm{ does not satisfy } c} X_c$.
We then ask whether $X_\phi=\emptyset$. If $X_\phi\neq \emptyset$, return that the formula is satisfiable. Otherwise, move on to the next stage.

Notice that $X_\phi\neq \emptyset$ if and only if there is a partial assignment $\alpha$ to the variables in $U$ such that $\alpha$ satisfies all clauses that $\phi$ does not satisfy, i.e. $\alpha\cdot\phi$ is a satisfying assignment. Otherwise, if $X_\phi=\emptyset$, then $\phi$ cannot be completed to a satisfying assignment.

The number of queries and updates is $O(2^{n(1-\delta)} n)$. Hence if there is some $\eps>0$ such that $u(N),q(N)\leq N^{1-\eps}$, we could solve CNF-SAT asymptotically in time
\[n2^{n(1-\delta)}\cdot 2^{n\delta (1-\eps)} = n 2^{n(1-\delta\eps)},\] 
and this is $O(2^{n(1-\eps')})$ time for any constant $0<\eps'<\delta\eps$, thus violating the SETH.
\end{proof}

%% file: bmm.tex
%
%
%
%
%
%
%
%

In this section we prove our results based on the hardness of Triangle detection.

\begin{reminder}{Theorem~\ref{thm:triangle}}
If for some $\delta,\eps> 0$, we can solve either of
\begin{itemize}
\item fully dynamic $st$-Reach, BPMatch, $17$-BPM, or SC, with preprocessing time $O(m^{1+\delta-\eps})$, amortized update and query times $O(m^{2\delta - \eps})$, or
\item incremental or decremental $st$-Reach, BPMatch, $17$-BPM, or SC, with preprocessing time $O(m^{1+\delta-\eps})$, worst case update and query times $O(m^{2\delta - \eps})$, or
\item fully dynamic $st$-SubConn or $5$-BPM with preprocessing time $O(m^{1+\delta-\eps})$, amortized update time $O(m^{\delta-\eps})$, and amortized query time $O(m^{2\delta - \eps})$, or
\item incremental or decremental $5$-BPM with preprocessing time $O(m^{1+\delta-\eps})$, worst case update time $O(m^{\delta - \eps})$, and worst case query time $O(m^{2\delta-\eps})$, or
\item incremental $st$-SubConn, or decremental $st$-SubConn in dense graphs, with preprocessing time $O(m^{1+\delta-\eps})$, worst case update time $O(m^{\delta - \eps})$, and worst case query time $O(m^{2\delta-\eps})$, or
\item $\emptyset$-PP over $k$ sets and a universe of size $O(k)$ with preprocessing time $O(k^{1+\delta-\eps})$, amortized update and query time $O(k^{\delta-\eps})$, or
\item for $\delta>1/3$, PP over $k$ sets and a universe of size $O(k)$ with preprocessing time $O(k^{1+\delta-\eps})$, amortized update and query time $O(k^{\delta-\eps})$,
\end{itemize}
then Conjecture~\ref{conj:tria} is false for this choice of $\delta$.
\end{reminder}

In some of the reductions below, we will reduce Triangle to $O(n)$ updates and queries of the dynamic problem. To get a lower bound for the dynamic problem using Conjecture~\ref{conj:tria}, we show how for Triangle detection, a running time of the form $O(n\cdot T(m))$ implies a running time of the form $T'(m)$.

\begin{lemma}\label{lem:nm-to-m}
If there is an $O(nm^\eps)$ time algorithm for triangle detection for graphs on $n$ nodes and $m$ edges, then there is an $O(m^{1+\eps/2})$ time algorithm for triangle detection for graphs on $m$ edges (and arbitrary number of nodes). And similarly, if there is an $O(mn^\eps)$ time algorithm, then there is an $O(m^{1+\frac{\eps}{1+\eps}})$ time algorithm as well.
\end{lemma}

\begin{proof}
First, we assume that the graph is connected, otherwise work on each connected component separately.
Suppose there is an $O(nm^\eps)$ time algorithm and let $\Delta=m^{\eps/2}$. In $O(m\Delta)=O(m^{1+\eps/2})$ time we can find any triangle that contains a node of degree at most $\Delta$ just by going over all edges incident to low degree nodes $x$ and then going over the rest of the neighbors of $x$. If no triangle is found, then any triangle has only high degree nodes. The number of high degree nodes is $O(m/\Delta)$, thus we can find a triangle on them in $O((m/\Delta) m^\eps)=O(m^{1+\eps/2})$ time.

In case there is an $O(mn^\eps)$ time algorithm, we apply the same argument with $\Delta = m^{\frac{\eps}{1+\eps}}$.
\end{proof}

As a corollary to the above Lemma, we obtain that if there is no $O(m^{1+\delta})$ time algorithm for triangle detection, then there is also no $O(mn^{\delta/(1-\delta)})$ or $O(m^{2\delta} n)$ time algorithm either.

We start by reducing Triangle to $st$-Reach. 
In Lemmas~\ref{lem:bpm} and~\ref{lem:reach-to-sc} we have reduced $st$-Reach to both BPMatch and SC, which allows us to immediately get reductions to BPMatch and SC too.
Note that the reduction below can be used to reduce $k$-Cycle detection in directed graphs to $O(n)$ updates of $st$-Reach. Since solving $k$-Cycle detection in time $O(m^{1+\delta-\eps})$ could be a harder problem than solving Triangle detection, i.e. the $k=3$ case, in that time, this gives a lower bound for the dynamic problems based on a perhaps weaker assumption.

\begin{lemma}
[Triangle to $st$-Reach]
Suppose that $st$-Reach has
a fully dynamic algorithm with preprocessing time $p(m,n)$, update time $u(m,n)$ and query time $q(m,n)$.
Then, Triangle detection can be solved in $O(n \cdot (u(m,n)+q(m,n)) + p(m,n))$ time.
\label{lem:tri-to-streach}
\end{lemma}

\begin{proof}
Let $G=(V,E)$ be the graph for which we want to detect a triangle.
Begin by creating a 4-partite graph $H$ of partitions $A,B,C$ and $A'$, each containing a copy of each vertex of $G$.
Let the copy of vertex $x$ in partition $Y$ be denoted by $x_Y$.

For every edge $(u,v)$ in $G$, add directed edges $(u_A,v_B), (u_B,v_C), (u_C,v_{A'})$. Add two nodes $s$ and $t$.

Now we have a stage for every node $x$ of $G$. In the stage for $x$, add the directed edges $(s,x_A)$ and $(x_{A'},t)$. Ask the query to check whether $t$ is reachable from $s$.
If so, there are $y_B\in B, z_C\in C$ so that $(x,y),(y,z),(z,x)\in E$ and hence $G$ has a triangle through $x$.  Otherwise, there is no triangle in $G$ containing $x$. After the query has been answered, if no triangle containing $x$ was found, the edges $(s,x_A)$ and $(x_{A'},t)$ are removed and the next stage begins.

The total number of queries and updates is $O(n)$, and hence the time to solve Triangle Detection is $O(n\cdot (u(m,n)+q(m,n)) + p(m,n))$.
\end{proof}

The above proof immediately implies a worst case insertion time bound for incremental algorithms, but it is not immediately clear how to prove the same relationship with decremental algorithms. 

\begin{lemma}
[Triangle to decremental $st$-Reach]
Suppose that $st$-Reach has a decremental algorithm with preprocessing time $p(m,n)$, worst case update time $u(m,n)$ and query time $q(m,n)$.
Then, Triangle detection can be solved in $O(n (u(m,n)+q(m,n)) + p(m,n))$ time.
\label{lem:dec-reach}
\end{lemma}

\begin{proof}
We want to simulate the proof of the fully dynamic case using a small number of deletions (and undeletions).

To do this, assume that $n$ is a power of $2$, say $n=2^r$ and create a complete binary tree $T_s$ rooted at $s$ with leaves the nodes of $A$.
Direct the edges of $T_s$ away from $s$. Similarly, create a complete binary tree $T_{t}$ rooted at $t$ with leaves the nodes of $A'$. Direct the edges of $T_t$ towards $t$.

Associate the nodes of $G$ with the integers in $[n]$.
Without loss of generality, stage $x$ of the dynamic algorithm corresponds to node $x$.
Now, consider stage $x$, and the previous stage $y=x-1$. We will show how to update $T_s$; $T_{t}$ is analogous.

Consider the paths $P_x$ and $P_y$ in $T_s$ from $s$ to $x_A$ and $y_A$, respectively.
In order for stage $x$ to ensure that the only path from $s$ to $A$ is $P_x$, it is sufficient to ensure that for every edge $(u,v)$ on $P_x$, the edge $(u,v')$ from $u$ to the sibling of $v$ in $T_s$ is removed in stage $x$.

Now suppose that this is accomplished for $P_y$ in stage $y=x-1$.
$P_x$ and $P_y$ coincide from $s$ down to the least common ancestor $f$ of $x_A$ and $y_A$.
Also, $x_A$ is the leftmost descendent of the right child of $f$ and $y_A$ is the rightmost descendent of the left child of $f$.


We update $T_s$ as follows. 
We assume (inductively) that the edge deletions in stage $y$ proceed in order of increasing distance from $s$ on $P_y$.
Let $S_f$ be the subsequence of edge deletions starting with the edge from $f$ to its left child.
First, in stage $x$, we undo all edge deletions in $S_f$ in reverse order of their deletion in stage $y$.
The last edge undeleted is hence the edge from $f$ to its left child. Each undeletion is performed by undoing the operations which were caused by the deletion.

After all undeletions, we start deleting the edges away from $P_x$ starting with the edge from $f$ to its right child and then going down towards $x$ along $P_x$.
(I.e. for each edge $(u,v)$ on $P_x$ where $u$ is a descendent of $f$, we delete the edge $(u,v')$ to the sibling $v'$ of $v$ in $T_s$.) 

The correctness of the procedure is clear. Here we analyze the total number of deletions and undeletions over the course of all stages.
The deletion/undeletion procedure corresponds exactly to a DFS tree traversal of $T_s$. Each edge of $T_s$ is deleted and undeleted exactly once. Hence the number of update operations is $O(n)$ overall.
%
%
\end{proof}

As a corollary from the reductions and from Lemma~\ref{lem:nm-to-m}, if any of BPMatch, $st$-Reach or SC admit dynamic algorithms with $u(m,n),q(m,n)\leq O(m^\eps)$, then triangle detection can be solved in $O(m^{1+\eps/2})$ time. If $\eps<0.82$, this would imply an improvement over the fastest algorithm that we currently have for the triangle problem in sparse graphs.
The fastest known algorithm for triangle detection would run in $\Theta(m^{4/3})$ time even if the matrix multiplication exponent is $2$.
If $\eps<2/3$ above, then this would imply an entirely new approach for the triangle problem in sparse graphs.

Chan~\cite{Chan06} observed that BMM and Triangle detection can be reduced to $O(m)$ updates and queries to the Subgraph Connectivity problem. We give an alternative construction that allows us to reduce Triangle detection to $O(m)$ updates and $n$ queries of $st$-SubConn. We will use a version of this construction to reduce Triangle detection to $5$-BPM.

\begin{lemma}
[Triangle to $st$-SubConn]
Suppose that $st$-SubConn has either
\begin{itemize}
\item a fully dynamic algorithm with preprocessing time $p(m,n)$, update time $u(m,n)$, and query time $q(m,n)$, or
\item an incremental algorithm with preprocessing time $p(m,n)$, worst case update time $u(m,n)$, and query time $q(m,n)$.
\end{itemize} 
Then triangle detection can be solved in $O(m\cdot u(m,n)+ n \cdot q(m,n) + p(m,n))$ time.

If $st$-SubConn has a decremental algorithm with preprocessing time $p(m,n)$, worst case update time $u(m,n)$, and query time $q(m,n)$, then
triangle detection can be solved in $O(n^2 \cdot u(m,n)+ n\cdot q(m,n) + p(m,n))$ time.
\label{lem:tri-to-stcon}
\end{lemma}

\begin{proof}
Given a graph $G$ on $m$ edges and $n$ nodes we create a bipartite graph $H$ on partitions $A$ and $B$ so that for every vertex $v$ of $G$ we add two copies $v_A\in A$ and $v_B\in B$ and for every edge $(u,v)$ in $G$ we add the edge $(u_A,v_B)$. We add two additional vertices $s$ and $t$ so that $s$ has edges to all nodes in $A$ and $t$ has edges to all nodes of $B$. All vertices of $A$ and $B$ are originally turned off and $s$ and $t$ are turned on.

Now, the proof proceeds in stages. Each stage corresponds to a vertex $u$ in $G$.
When $u$ is considered, for every neighbor $v$ of $u$ in $G$, we activate $v_A$ and $v_B$.
After the $2d(u)$ activations, we ask the query whether $s$ and $t$ are connected. After the query, we undo the activations and move on to the next stage. If $s$ and $t$ are connected during the stage for node $u$, then there exist two nodes $x,y$ in $G$ that are neighbors of $u$ and such that $(x,y)$ is an edge, i.e. $u$ is in a triangle. Otherwise, if $s$ and $t$ are not connected, there is no edge crossing the cut between $A$ and $B$ and hence there is no pair of nodes that are both neighbors of $u$ and have an edge between them, i.e. no triangle contains $u$. 

That is, with $O(m)$ updates and $n$ queries we solve the triangle detection problem.
The algorithm works for incremental algorithms as well, but assumes that the updates are worst case.
For the decremental case, we perform $O(n^2)$ updates -- originally all nodes are active, and to simulate the activation of the neighbors of $u$, we just deactivate the nonneighbors of $u$.
\end{proof}

\begin{lemma}
[Triangle to $\emptyset$-PP]
Suppose that dynamic $\emptyset$-PP on a universe of size $n$ and up to $k=O(m)$ subsets can be solved with preprocessing time $p(n,k)$, (amortized) update time $u(n,k)$ and query time $q(n,k)$, then Triangle detection on $n$ node and $m$ edge graphs can be solved in $O(p(n,k)+m\cdot (u(n,k)+q(n,k)))$ time.
\label{lem:tri-to-phipp}
\end{lemma}

\begin{proof}
Given a graph $G=(V,E)$ on $n$ nodes and $m$ edges in which we want to find triangles, we associate each node with a number in $[n]$ and define the sets $X_u = \{ v \mid v \in N(u) \} \subseteq [n]$. Note that $u \notin X_u$ because $u \notin N(u)$. 
Note that these sets can be constructed in $\sum_v O(d(v)) = O(m)$ time.
We preprocess the sets $\{ X_u \}_{u\in V}$ in $p(n,k)$ time. 
Then, we go over the edges $(v,w) \in E$ and compute the set $X_v \cap X_w$ then ask the query to check whether this set is empty.
If one of the answers was ``no", then there is a node $x\in N(v) \cap N(w)$ and we have found a triangle and we output ``yes". This computation takes $O(m \cdot (u(n,k)+q(n,k)))$ time.
\end{proof}

As a corollary, Conjecture~\ref{conj:tria} with constant $\delta$ implies that any algorithm for $\emptyset$-PP on $k$ sets, and universe of size at most $k$ has preprocessing time $k^{1+\delta-o(1)}$ or update or query time $k^{\delta-o(1)}$. Similarly, ``Conjecture''~\ref{conj:bmm} implies that the preprocessing time cannot be $O(k^{3/2-\eps})$, and the update and query times cannot be $O(k^{1/2-\eps})$.

To get the reduction to PP we need to do some extra work, but we obtain the same conditional lower bounds whenever $\delta>1/3$.

\begin{lemma}
[Triangle to PP]
Suppose that dynamic PP on a universe of size $n$ and up to $k$ subsets can be solved with preprocessing time $p(n,k)$, (amortized) update time $u(n,k)$ and query time $q(n,k)$, then Triangle detection on $n$ node and $m$ edge graphs can be solved in $\tilde{O}(m^{2/3} n + p(n,k)+m\cdot (u(n,k)+q(n,k)))$ time, where $k=O(m)$.
\label{lem:tri-to-pp}
\end{lemma}

We note that due to Lemma~\ref{lem:nm-to-m} we can replace the $m^{2/3}n$ term above by $m^{4/3}$, and this term is negligible for any application of Conjecture~\ref{conj:tria} with $\delta>1/3$.

\begin{proof}
Let $\Delta = m^{1/3}$. Our reduction will have two phases. In the first phase we look for triangles that have two nodes with degree less than $\Delta$ and in the second phase we check if there is a triangle with a node of degree at least $\Delta$. If the graph contains a triangle, one of these cases will happen.

\paragraph{Phase 1.} Start by picking a random universal hash function $h:[n] \rightarrow [N]$, where $N= c \cdot \Delta^2 = O(m^{2/3})$ for a large enough $c$. In fact, we will need $O(\log{n})$ such hash functions.

For every $j\in [N]$ we create the set $X_j = \{ a \mid \forall c \in N(a): h(c) \neq j \} \subseteq [n]$. These sets can be computed in $O(n \cdot \Delta^2) = O(nm^{2/3})$ time, and we preprocess them in time $p(n,N)$.
Then, for every node $b\in V$ we compute the set $Y_b = \cap_{c \in N(b)} X_{h(c)}$ using $d(b)$ intersection updates. To compute all the sets $Y_b$ we need $O(m \cdot u(n,k))$ time, where $k$ is the total number of sets created over the course of the reduction.
Finally, we go over the edges $(a,b)$ of the graph for which both $d(a)< \Delta$ and $d(b)<\Delta$, and ask the query ``is $a$ in $Y_b$?". If one of the answers is ``no" we have found a triangle and we return ``yes". This final computation takes $O(m\cdot q(n,k))$ time. The total number of subsets can be bounded by $k=O(m)$.

To see the correctness of the reduction, note that $a \notin Y_b$ if and only if there exists a node $c \in N(b)$ such that $a \notin X_c$. This happens if and only if there exist two nodes $c \in N(b)$ and $c' \in N(a)$ such that $h(c)=h(c')$.
Since $d(a),d(b) <\Delta$ we know that $|N(a) \cup N(b)| < 2\Delta$ and if $N(a) \cap N(b) = \emptyset$ then the existence of such nodes $c,c'$ would be unlikely by the properties of the universal hash function. To make the probability of a false negative smaller we pick $O(\log{n})$ hash functions and run a copy of the above procedure for each one in parallel. We return ``yes" when checking a pair $(a,b)$ only if $a \notin Y_b$ using all the hash functions. Thus, with high probability, we output ``yes" if and only if $N(a)\cap N(b) \neq \emptyset$ and there is a triangle in $G$.

\paragraph{Phase 2.} To check for triangles with a high-degree node, we proceed in a similar way as above, but we do not use hashing. Instead, we create a set $X_j$ for every node $j$ with $d(j)\geq \Delta$.
Since there are at most $O(m/\Delta)$ nodes with degree $\geq \Delta$, the number of sets we start with is still $O(m^{2/3})=O(N)$ as before. 

In more detail, we enumerate the high-degree nodes $H=\{ v \in V \mid d(v) \geq \Delta \}$ and associate each one with a number in $[N]$. Then, we create the sets $X_j = \{ a \mid j \notin N(a) \}\subseteq [n]$ for $j \in [N]$. This takes time $O(n \cdot m/\Delta) = O(nm^{2/3})$, and we preprocess the sets as before in time $O(p(n,N))$. From now on we proceed exactly like in phase 1, with the exception that we do not restrict the queries to edges $(a,b)$ for which the nodes are of low degree. The analysis is similar except that there are no false negatives.
\end{proof}

\tocless\subsection{Triangle to matchings without short augmenting paths}

%
%

We now show how to reduce Triangle detection directly to dynamic matching, giving graphs in which the size of any matching $M$ that does not have length $k$ augmenting paths, for some constant $k$, can help us detect triangles.

We give two reductions. The first one uses $O(m)$ updates and $n$ queries to an algorithm for $k$-BPM where $k=5$. The second one uses fewer updates, only $O(n)$, but needs to use an algorithm for $k$-BPM where $k=17$. We start with the first reduction as a warm up for the quite longer proof of the second reduction. 

\begin{lemma}[Triangle to $5$-BPM]\label{lem:5bpm}
Given any algorithm for fully dynamic $5$-BPM with preprocessing, update, and query times $p(m,n),u(m,n)$, and $q(m,n)$, respectively, we can get an algorithm for Triangle that runs in time $O(p(m,n)+m\cdot u(m,n) + n\cdot q(m,n))$ .
\end{lemma}

\begin{proof}
Let $G=(V,E)$ be the graph for which we want to detect a triangle.
Begin by creating a $4$-layer graph $H$ with layers $I,\bar{I},J$ and $\bar{J}$, each containing a copy of each vertex of $G$.
For every edge $(u,v)$ in $G$, add the edge $(u_I,v_J)$ to $H$. Add edges $(u_{\bar{I}}, u_I)$ and $(u_{\bar{J}}, u_J)$ for each vertex of $G$.
Now we have a stage for every node $x$ of $G$. 
Go over the neighborhood of $x$ in $G$ and for every node $y\in N(x)$, remove the edges $(y_{\bar{I}},y_I)$ and $(y_{\bar{J}},y_J)$ from $H$. 
Ask the query to get the size of a matching $M$ that does not admit length $5$ augmenting paths.
If $|M| > 2(n-|N(x)|)$, then as we will show below, $G$ has triangle through $x$.
Otherwise, there is no triangle in $G$ containing $x$. After the query has been answered, if no triangle was found, we add the edges $(y_{\bar{I}},y_I)$ and $(y_{\bar{J}},y_J)$, for every node $y\in N(x)$ back to $H$ and the next stage begins.

Note that $H$ remains a bipartite graph at all times, and that it has $O(n)$ nodes and $O(m+n)$ edges.
The total number of updates is $\sum_{x\in V(G)} 2d(x) = O(m)$ and the number of queries is $O(n)$, and hence the time to solve Triangle detection is $O(p(m,n) + m\cdot u(m,n)) + n\cdot q(m,n))$.

We claim that $x$ participates in a triangle in $G$ if and only if the answer to the query at the stage of $x$ will give $|M| > 2(n-|N(x)|)$.
The key reason for this is the simple fact that $x$ participates in a triangle if and only if there are two nodes $y_1,y_2 \in N(x)$ that have an edge between them. Assume that $x$ is not in any triangle, and therefore, all edges in $H$ when we ask the query have an endpoint that is either $z_I$ or $z_J$ for some node $z \notin N(x)$. Thus, the size of any matching in $H$ is bounded by the number of these nodes, which is exactly $2(n-|N(x)|)$.
On the other hand, assume that $x$ participates in a triangle $(x,u,v)$ in $G$ and let $M$ be a matching without length $5$ augmenting paths. 
We denote unmatched edges $(u,v)\in E$ by $u \leadsto v$ and matched edges by $u \rightarrow v$, so that an augmenting path always has the form $(v_1 \leadsto v_2 \rightarrow v_3 \cdots v_{k-1} \leadsto v_k)$.
First, note that all nodes $z_I$ for $z \notin N(x)$ must be matched in $M$, since otherwise the length $1$ path $(z_{\bar{I}} \leadsto z_I)$ is an augmenting path, and moreover, $z_{I}$ cannot be matched to a node $w_{J}$ such that $w \notin N(x)$, since otherwise there is a length $3$ augmenting path, $(z_{\bar{I}} \leadsto z_I \rightarrow w_J \leadsto w_{\bar{J}})$. The same arguments apply for nodes $z_J$ in $H$, where $z\notin N(x)$. Therefore, by counting only edges in $M$ that are adjacent to nodes $z_I,z_J$ for $z \notin N(x)$ we get $2(n-|N(x)|)$, and now we argue that $M$ must contain at least one edge of the form $(u'_I,v'_J)$ for $u',v' \in N(x)$, which will imply that $|M| \geq 2(n-|N(x)|)+1$.
Since $(x,u,v)$ is a triangle, the edge $e=(u_I, v_J)$ is in $H$, and if $e\in M$ we are done. If, however, $e$ is not matched, then both $u_I$ and $v_J$ must be matched to nodes $w_J = Mate(u_I), z_I = Mate(v_J)$. If $w\in N(x)$ or $z\in N(x)$, then we have found an edge of the form $(u'_I,v'_J)$ for $u',v' \in N(x)$ and we are done. Otherwise, $w,z \notin N(x)$ and we get a contradiction since $M$ will admit the following length $5$ augmenting path $(z_{\bar{I}} \leadsto z_I \rightarrow v_J \leadsto u_I \rightarrow w_J \leadsto w_{\bar{J}})$.

\end{proof}

The construction of the graph $H$ in the proof above was an adaptation of the one in the proof of Lemma~\ref{lem:tri-to-stcon} to the matching problem. In the next proof, we adapt the construction from Lemma~\ref{lem:tri-to-streach}.

%
%
%

\begin{lemma}[Triangle to $17$-BPM]\label{lem:17bpm}
Given any algorithm for fully dynamic $17$-BPM with preprocessing, update, and query times $p(m,n),u(m,n)$, and $q(m,n)$, respectively, we can get an algorithm for Triangle that runs in time $O(p(m,n)+n\cdot (u(m,n) + q(m,n)))$ .
\end{lemma}

\begin{proof}
Let $G=(V,E)$ be the graph for which we want to detect a triangle.
Begin by creating an $8$-layer graph $H$ with layers $A_1,\bar{A_1},B,\bar{B},C,\bar{C},A_2$, and $\bar{A_2}$, each containing a copy $u_X$ of each vertex $u$ of $G$.
For every edge $(u,v)$ in $G$, add the edges $(u_{\bar{A_1}},v_B), (u_{\bar{B}},v_C), (u_{\bar{C}}, v_{A_2})$ to $H$. Add edges $(u_{A_1}, u_{\bar{A_1}}), (u_B,u_{\bar{B}}), (u_C,u_{\bar{C}})$ and $(u_{A_2}, u_{\bar{A_2}})$ for each vertex $u$ of $G$.

Now we have a stage for every node $x$ of $G$. We start the stage by removing the two edges $(x_{A_1},x_{\bar{A_1}})$ and $(x_{A_2},x_{\bar{A_2}})$, then we ask a query to get the size of a matching $M$ that does not admit length $17$ augmenting paths.
If $|M| > 4n-2$, then as we will show below, $G$ has triangle through $x$.
Otherwise, there is no triangle in $G$ containing $x$. After the query has been answered, if a triangle was not found, we add the edges $(x_{A_1},x_{\bar{A_1}})$ and $(x_{A_2},x_{\bar{A_2}})$ back to $H$ and the next stage begins.

Note that $H$ remains a bipartite graph at all times, and that it has $O(n)$ nodes and $O(m+n)$ edges.
The total number of updates and queries is $O(n)$, and hence the time to solve Triangle detection is $O(p(m+n,n) + n\cdot (u(m+n,n) + q(m+n,n)))$.

We claim that $x$ participates in a triangle in $G$ if and only if the answer to the query at the stage of $x$ is that $|M| > 4n-2$, in fact, $|M|=4n-1$. The key reason is that there is a triangle $(x,y,z)$ in $G$ if and only if there is a path $(x_{\bar{A_1}} - y_B - y_{\bar{B}} - z_{C} - z_{\bar{C}} - x_{A_2})$ in $H$.

Let us start with the easier direction, and show that if $x$ does not participate in any triangles in $G$, then $|M| \leq 4n-2$, for any matching $M$. First, note that there are two isolated vertices, $x_{A_1}$ and $x_{\bar{A_2}}$ that cannot be matched. We will show that there will be another node in $H$ that is not matched in $M$, which would show that $|M| < (8n-3)/2$, and therefore $|M|\leq 4n-2$. To see this, consider the node $x_{\bar{A_1}}$, and if it is free (i.e. not matched) we are done, we found a third unmatched node. Otherwise, let $y_B$ be $Mate(x_{\bar{A_1}})$, and consider the node $y_{\bar{B}}$. Again, if $y_{\bar{B}}$ is free, we are done, and otherwise, let $z_C = Mate(y_{\bar{B}})$. Finally, consider $z_{\bar{C}}$, and if it is not free, let $x'_{A_2} = Mate(z_{\bar{C}})$, and note that $x'\neq x$, since otherwise, we have found a triangle $(x,y,z)$, and since $x'_{\bar{A_2}}$ does not have any neighbors except for $x'_{A_2}$, it cannot be matched in $M$ and we found a third free node.

For the other direction, let $(x,y,z)$ be a triangle in $G$ and $M$ be a matching without length $17$ augmenting paths. We will show that $|M|=4n-1$ and every node except for the two isolated vertices $x_{A_1}$ and $x_{\bar{A_2}}$ will be matched. As in the proof of Lemma~\ref{lem:5bpm}, we denote unmatched edges $(u,v)\in E$ by $u \leadsto v$ and matched edges by $u \rightarrow v$, so that an augmenting path always has the form $(v_1 \leadsto v_2 \rightarrow v_3 \cdots v_{k-1} \leadsto v_k)$.

\begin{claim}
For every node $u \neq x$, $u_{\bar{A_1}}$ and $u_{A_2}$ will be matched in $M$.
\end{claim}

\begin{proof}
If $u_{\bar{A_1}}$ is free, then $u_{A_1}$ is also free, and $M$ has a length $1$ augmenting path $(u_{A_1} \leadsto u_{\bar{A_1}})$. The argument for $u_{A_2}$ is analogous.
\end{proof}

The main claim is the following.

\begin{claim} \label{claim:17main}
For every node $u \neq x$, the edges $(u_{A_1}, u_{\bar{A_1}})$ and $(u_{A_2}, u_{\bar{A_2}})$ are in $M$.
\end{claim}

\begin{proof}
Assume for contradiction that $(u_{A_1}, u_{\bar{A_1}}) \notin M$, and we will show an augmenting path for $M$ of length up to $17$. The argument for the case that $(u_{A_2}, u_{\bar{A_2}})\notin M$ is analogous.

Let $y'_B=Mate(u_{\bar{A_1}})$. If $y'_{\bar{B}}$ is free, we are done, since we found the following length $3$ augmenting path $(u_{A_1} \leadsto u_{\bar{A_1}} \rightarrow y'_B \leadsto y'_{\bar{B}})$, thus, from now on assume $y'_{\bar{B}}$ is matched. 
Let $z'_C = Mate(y'_{\bar{B}})$, and again, if $z'_{\bar{C}}$ is free we are done because of the length $5$ augmenting path $(u_{A_1} \leadsto u_{\bar{A_1}} \rightarrow y'_B \leadsto y'_{\bar{B}} \rightarrow z'_C \leadsto z'_{\bar{C}})$. 
Thus, let $x'_{A_2} = Mate(z'_{\bar{C}})$, and note that $x'$ must equal $x$, since otherwise we have found a length $7$ augmenting path $(u_{A_1} \leadsto u_{\bar{A_1}} \rightarrow y'_B \leadsto y'_{\bar{B}} \rightarrow z'_C \leadsto z'_{\bar{C}} \rightarrow x'_{A_2} \leadsto x'_{\bar{A_2}})$. 
Thus, we have reached $x_{A_2}$ in our walk from the free node $u_{A_1}$, and now we will continue the walk towards $x_{\bar{A_1}}$, using the nodes $y,z$ that make the triangle $(x,y,z)$ in $G$ under our assumption.
We will show the analysis only for the case that $y' \neq y$ and $z' \neq z$, and remark that in case $y'=y$ or $z'=z$, one can find shorter augmenting paths using the same arguments.

The node $z_{\bar{C}}$ cannot be free, since otherwise we have the augmenting path $(u_{A_1} \leadsto u_{\bar{A_1}} \rightarrow y'_B \leadsto y'_{\bar{B}} \rightarrow z'_C \leadsto z'_{\bar{C}} \rightarrow x_{A_2} \leadsto z_{\bar{C}})$. There are two options for $Mate(z_{\bar{C}})$: either it is in the $A_2$-layer, or it is $z_C$. 
In the first case, we are done, since if we let $x'_{A_2} = Mate(z_{\bar{C}})$, we get that $x' \neq x$ and we have the following augmenting path $(u_{A_1} \leadsto u_{\bar{A_1}} \rightarrow y'_B \leadsto y'_{\bar{B}} \rightarrow z'_C \leadsto z'_{\bar{C}} \rightarrow x_{A_2} \leadsto z_{\bar{C}} \rightarrow x'_{A_2} \leadsto x'_{\bar{A_2}} )$.
Thus, assume $Mate(z_{\bar{C}}) = z_C$, and similarly, we walk back to $y_{\bar{B}}$ and note that it cannot be free, since otherwise we have the augmenting path 
$(u_{A_1} \leadsto u_{\bar{A_1}} \rightarrow y'_B \leadsto y'_{\bar{B}} \rightarrow z'_C \leadsto z'_{\bar{C}} \rightarrow x_{A_2} \leadsto z_{\bar{C}} \rightarrow z_C \leadsto y_{\bar{B}} )$. 
Again, there are two cases for $Mate(y_{\bar{B}})$: either it is $z''_C$ for some $z'' \neq z',z$, or it is $y_B$. In the first case, we are done, since either $z''_{\bar{C}}$ is free and we found an augmenting path that ends in it, or it is matched to a node $x'_{A_2} = Mate(z''_{\bar{C}})$ where $x' \neq x$, in which case we have the following augmenting path
$(u_{A_1} \leadsto u_{\bar{A_1}} \rightarrow y'_B \leadsto y'_{\bar{B}} \rightarrow z'_C \leadsto z'_{\bar{C}} \rightarrow x_{A_2} \leadsto z_{\bar{C}} \rightarrow z_C \leadsto y_{\bar{B}} \rightarrow z''_{\bar{C}} \leadsto z''_C \rightarrow x'_{A_2} \leadsto x'_{\bar{A_2}} )$.
Thus, assume $Mate(y_{\bar{B}}) = y_B$, and walk back to $x_{\bar{A_1}}$.
If $x_{\bar{A_1}}$ is free, we have the following augmenting path
$(u_{A_1} \leadsto u_{\bar{A_1}} \rightarrow y'_B \leadsto y'_{\bar{B}} \rightarrow z'_C \leadsto z'_{\bar{C}} \rightarrow x_{A_2} \leadsto z_{\bar{C}} \rightarrow z_C \leadsto y_{\bar{B}} \rightarrow y_B \leadsto x_{\bar{A_1}} )$.
Otherwise, we keep walking forward towards part $A_2$ again, and this time, we cannot reach $x_{A_2}$ again, and therefore we will be able to end at a free node in part $\bar{A_2}$.
Let $y''_B = Mate(x_{\bar{A_1}})$, and consider $y''_{\bar{B}}$ which has to be matched since otherwise we found an augmenting path that ends in it. Let $z''_C = Mate(y''_{\bar{B}})$, and similarly, consider $z''_{\bar{C}}$ which has to be matched as well.
Note that $y,y'$ and $y''$ are all distinct, since each is incident to a different matched edge, and similarly $z,z'$ and $z''$ are distinct.
Finally, let $x'_{A_2} = Mate(z''_{\bar{C}})$, and since $x' \neq x$, we get the following length $17$ augmenting path
$(u_{A_1} \leadsto u_{\bar{A_1}} \rightarrow y'_B \leadsto y'_{\bar{B}} \rightarrow z'_C \leadsto z'_{\bar{C}} \rightarrow x_{A_2} \leadsto z_{\bar{C}} \rightarrow z_C \leadsto y_{\bar{B}} \rightarrow y_B \leadsto x_{\bar{A_1}} \rightarrow y''_B \leadsto y''_{\bar{B}} \rightarrow z''_C \leadsto z''_{\bar{C}} \rightarrow x'_{A_2} \leadsto x'_{\bar{A_2}})$.
 
\end{proof}

Note that by Claim~\ref{claim:17main}, we know that there can be at most one matched edge between parts $\bar{A_1}$ and $B$, and at most one matched edge between parts $\bar{C}$ and $A_2$. These edges will have to be adjacent to $x_{\bar{A_1}}$ or $x_{A_2}$.
Now we can also prove the following claim.

\begin{claim} \label{claim:17bc}
All the nodes in parts $B$ and $\bar{C}$ must be matched by $M$.
\end{claim}

\begin{proof}
We will prove the claim for part $B$, and the proof for part $\bar{C}$ is symmetric. 
Let us assume for contradiction that node $u_B$ is free. 
In this case, the node $u_{\bar{B}}$ must be matched to some node $v_C = Mate(u_{\bar{B}})$, and the node $v_{\bar{C}}$ must be matched, since otherwise we have an augmenting path $(u_B \leadsto u_{\bar{B}} \rightarrow v_C \leadsto v_{\bar{C}})$. And since, by Claim~\ref{claim:17main}, we know that every matched edge between parts $\bar{C}$ and $A_2$ must be adjacent to $x_{A_2}$, we get that $x_{A_2} = Mate(v_{\bar{C}})$.
Now we will start walking back towards $x_{\bar{A_1}}$. We will assume that $y\neq u$ and $z\neq v$, and remark that otherwise the proof is similar.

Consider $z_{\bar{C}}$ and note that either it is matched or we found the augmenting path $(u_B \leadsto u_{\bar{B}} \rightarrow v_C \leadsto v_{\bar{C}} \rightarrow x_{A_2} \leadsto z_{\bar{C}})$. Therefore, it must be the case that $z_C = Mate(z_{\bar{C}})$. For a similar reason, $y_{\bar{B}}$ must be matched, and if $Mate(y_{\bar{B}})$ is not $y_B$ but some node $z'_C$, we will have that $z'_{\bar{C}}$ is free, and we find the augmenting path $(u_B \leadsto u_{\bar{B}} \rightarrow v_C \leadsto v_{\bar{C}} \rightarrow x_{A_2} \leadsto z_{\bar{C}} \rightarrow z_C \leadsto y_{\bar{B}} \rightarrow z'_C \leadsto z'_{\bar{C}})$.
Therefore, $Mate(y_{\bar{B}})=y_B$, and if $x_{\bar{A_1}}$ was free, we could complete an augmenting path by finishing with the unmatched edge $y_{B} \leadsto x_{\bar{A_1}}$. Therefore, let $u'_{B} = Mate(x_{\bar{A_1}})$, where $u' \neq u,y$. Note that if $u'_{\bar{B}}$ is free, we would have the augmenting path $(u_B \leadsto u_{\bar{B}} \rightarrow v_C \leadsto v_{\bar{C}} \rightarrow x_{A_2} \leadsto z_{\bar{C}} \rightarrow z_C \leadsto y_{\bar{B}} \rightarrow y_B \leadsto x_{\bar{A_1}} \rightarrow u'_B \leadsto u'_{\bar{B}})$. Finally, let $v'_C = Mate(u'_{\bar{B}})$ where $v' \neq v,z$, and since $v'_{\bar{C}}$ cannot be matched to a node in part $A_2$, or we would have two matched edges between parts $\bar{C}$ and $A_2$, we conclude that $v'_{\bar{C}}$ must be free, and we found the length $15$ augmenting path
$(u_B \leadsto u_{\bar{B}} \rightarrow v_C \leadsto v_{\bar{C}} \rightarrow x_{A_2} \leadsto z_{\bar{C}} \rightarrow z_C \leadsto y_{\bar{B}} \rightarrow y_B \leadsto x_{\bar{A_1}} \rightarrow u'_B \leadsto u'_{\bar{B}} \rightarrow v'_C \leadsto v'_{\bar{C}})$ -- a contradiction.

\end{proof}

By Claims~\ref{claim:17bc} and ~\ref{claim:17main} we know that there exists at most one matched edge between parts $\bar{B}$ and $C$, since at most one node $u_B$ from $B$ can be matched to a node different than $u_{\bar{B}}$, yet all nodes in part $B$ are matched, which means that at most one node in part $\bar{B}$ can be matched to a node in $C$.
Finally, we can prove that all the nodes in $H$, except for the two isolated nodes, will be matched in $M$. 

\begin{claim}
$x_{\bar{A_1}}$ will be matched to some node $u_B$, and $x_{A_2}$ will be matched to some node $v_{\bar{C}}$. Moreover, the edge $(u_{\bar{B}}, v_C)$ will be in $M$. 
\end{claim}

\begin{proof}
We will show that each of the following cases gives a contradiction to the assumption that $M$ does not have length $17$ augmenting paths.

\paragraph{Case 1: both $x_{\bar{A_1}}$ and $x_{A_2}$ are unmatched.} In this case, $y_B$ must be matched to $y_{\bar{B}}$ and $z_C$ to $z_{\bar{C}}$, and we have the following augmenting path $(x_{\bar{A_1}} \leadsto y_B \rightarrow y_{\bar{B}} \leadsto z_C \rightarrow z_{\bar{C}} \leadsto x_{A_2})$.

\paragraph{Case 2: $x_{\bar{A_1}}$ is matched to $u_B$ but $x_{A_2}$ is free.}  Note that in this case, there are no matched edges between parts $\bar{C}$ and $A_2$, and therefore every node $v_{\bar{C}}$ has to be matched with $v_C$, and there will not be any matched edges between parts $\bar{B}$ and $C$ either.
Thus, note that $u_{\bar{B}}$ is free.
The argument will be slightly different according to whether $u = y$ or $u\neq y$. If $u =y$, then we find the augmenting path $( u_{\bar{B}} = y_{\bar{B}} \leadsto z_C \rightarrow z_{\bar{C}} \leadsto x_{A_2} )$, while if $u \neq y$, we have a slightly longer augmenting path $(u_{\bar{B}} \leadsto u_B \rightarrow x_{\bar{A_1}} \leadsto y_B \rightarrow y_{\bar{B}} \leadsto z_C \rightarrow z_{\bar{C}} \leadsto x_{A_2})$.

\paragraph{Case 3: $x_{A_2}$ is matched to $v_{\bar{C}}$ but $x_{\bar{A_1}}$ is free.} This case is symmetric to the second case.

\paragraph{Case 4: $u_B = Mate (x_{\bar{A_1}})$ and $v_C = Mate ( x_{A_2} )$, but $(u_{\bar{B}} , v_C) \notin M$.} In this case, note that there are no matched edges between parts $\bar{B}$ and $C$, and both $u_{\bar{B}}$ and $v_C$ are free. Therefore, $(u_{\bar{B}},v_C) \notin E(H)$ and $(u,v)\notin E$, which implies that either $u \neq y$ or $v \neq z$ (or both), since $(y,z)\in E$.
We show that we get a contradiction if $u \neq y$, and the argument for the case that $v \neq z$ is similar.
If $u \neq y$ then the edge $(y_B,y_{\bar{B}})$ is in $M$. If $v=z$, then $z_C$ is free and we have the augmenting path $(u_{\bar{B}} \leadsto u_B \rightarrow x_{\bar{A_1}} \leadsto y_B \rightarrow y_{\bar{B}} \leadsto z_C)$. Otherwise, $v \neq z$, and we have the augmenting path $(u_{\bar{B}} \leadsto u_B \rightarrow x_{\bar{A_1}} \leadsto y_B \rightarrow y_{\bar{B}} \leadsto z_C \rightarrow z_{\bar{C}} \leadsto x_{A_2} \rightarrow v_{\bar{C}} \leadsto v_C)$.

\end{proof}

Therefore, we proved that every node except for the two isolated nodes $x_{A_1},x_{\bar{A_2}}$ must be matched in $M$, which implies that $|M|=4n-1$.
\end{proof}

%% file: apsp.tex
In this section we prove our APSP hardness results. 

\begin{reminder}{Theorem~\ref{thm:apsp}}
If for some $\eps >0$ we can solve decremental or incremental $st$-SP or BPWMatch with preprocessing time $O(n^{3-\eps})$ and amortized update and query times $O(n^{2 - \eps})$,
then Conjecture~\ref{conj:apsp} is false.
\end{reminder}

In the Min-Weight-Triangle problem we are given an edge weighted graph where the edges are in $[M]$ and are asked to return the minimum weight of a triangle in it. 
Vassilevska Williams and Williams~\cite{focs10} showed that an $O(n^{3-\eps})$ algorithm for the Min-Weight-Triangle problem, where $M=O(n^c)$ for some $\eps>0$ and large enough $c$,  would yield a truly subcubic algorithm for the All Pairs Shortest Paths problem and violate Conjecture~\ref{conj:apsp}.

Roditty and Zwick~\cite{rzesa} gave a reduction from APSP to $O(n)$ updates and $O(n^2)$ queries of decremental or incremental single source shortest paths(SSSP). By using a similar construction, yet going through the Min-Weight-Triangle problem, we are able to strengthen their result in two ways. First, we reduce the number of queries to $O(n)$ instead of $O(n^2)$, implying a higher lower bound on the query time, and second, our reduction is to the $s,t$-shortest path problem, which is at least as easy as the SSSP problem.

In the unweighted case, Roditty and Zwick~\cite{rzesa} show that BMM can be reduced to $O(n)$ updates and $O(n^2)$ queries to the unweighted incremental or decremental SSSP problem, and using a similar modification to their proof, we can show that Triangle detection reduces to $O(n)$ updates and queries to the unweighted incremental or decremental $st$-SP problem.

\begin{lemma}
Min-Weight-Triangle on a graph with $n$ nodes and $m$ edges can be reduced to $O(n)$ updates and queries of decremental or incremental $st$-SP on a graph with $O(n)$ nodes and $O(m)$ edges. 
\end{lemma}

\begin{proof}
Let $G=(V,E)$ with weight function $w:E\rightarrow [M]$ be the edge-weighted graph for which we want to find the minimum weight triangle.
Begin by creating a $4$-partite graph $H$ of partitions $A,B,C$ and $A'$, each containing a copy of each vertex of $G$.
For every edge $(u,v)$ in $G$ of weight $w$, add edges $(u_A,v_B), (u_B,v_C), (u_C,v_{A'})$ and set the weight of each of them to $w+2M$. 
Add two nodes $s$ and $t$. Let $u_i \in V$ be the $i^{\th}$ node in $V$, and add edges $(s,u_{i,A})$ and $(u_{i,A'},t)$ both of weight $3i \cdot M$.
Now we have a stage for every node $u_i$ of $G$, where we find the minimum weight of a triangle that the node $u_i$ is a part of. 
Start from $i=1$ and set the current minimum to $t_{min}=3M$.
At stage $i \in [n]$, ask the query about the length of the shortest $s,t$-path in $H$, let it be $y$, and look at $z=y - 2\cdot3i\cdot M - 3\cdot 2M$. We show below that either $z$ equals the minimum weight of a triangle containing $u_i$ in $G$, or $z> 3M$ and $u_i$ is not a part of any triangle in $G$. 
After the query is answered, check if $z<t_{min}$ and if so, update the current minimum $t_{min} \gets z$, and remove the edges $(s,u_{i,A}), (u_{i,A'},t)$ from $H$ and go on to stage $i+1$. After the last stage, $t_{min}$ will contain the minimum weight of a triangle in $G$.

The total number of queries and updates is $O(n)$, the number of nodes in $H$ is $O(n)$ and the number of edges is $O(m)$.

For correctness, consider the shortest $s,t$-path in $H$ at stage $i$. Any $s,t$-path $P$ will have the form $s - u_{i_1,A} - (e_1,\ldots, e_k) - u_{i_2,A'} - t $, where $k\geq 3$ is odd and $i_1,i_2 \geq i$, and its weight will be
\[w_H(P) = 3i_1M+ \sum_{i=1}^k (w(e_i) + 2M) + 3i_2M
\]
Let $\Delta = (u_i, u_b, u_c)$ be the minimum weight triangle in $G$ that contains $u_i$ and compare $w_H(P)$ with the weight of the following $s,t$-path $P_{min}$ in $H$, $s - u_{i,A} - u_{b,B} - u_{c,C} - u_{i,A'} - t$,
\[ w_H(P_{min}) = 3iM+ w(\Delta) + 3\cdot2M + 3iM = (6i+6)M + w(\Delta)
\]
Assume for contradiction that $w_H(P) < w_H(P_{min})$, and note that it must be the case that $i_1=i_2 = i$, since otherwise $w_H(P) \geq (6i+3)M + \sum_{i=1}^k (w(e_i) + 2M)$, which even if $w(e_i)=1$ for all the edges in $P$ can be lower bounded by $w_H(P) \geq (6i+9)M+3$, whereas $w_H(P_{min}) \leq (6i+3)M+6M = (6i+9)M$. Furthermore, it must be the case that $k=3$, since otherwise $w_H(P) \geq 6iM + \sum_{i=1}^5 (w(e_i) + 2M) \geq (6i+10)M + 5 > (6i+9)M \geq w_H(P_{min})$. Finally, we conclude that $P$ must be of the form $s - u_{i,A} - u_{b',B} - u_{c',C} - u_{i,A'} - t$, which implies that $\Delta' = (u_i,u_{b'},u_{c'})$ is a triangle in $G$ with weight $w(\Delta')<w(\Delta)$ which is a contradiction to our choice of $\Delta$.

To get a reduction to the incremental problem, use the same reduction but perform the stages in reverse order starting from $i=n$ and adding edges every time to move down from stage $i$ to $(i-1)$.
\end{proof}

Next, we use our reduction from $st$-SP to the maximum weight perfect matching in bipartite graphs from Lemma~\ref{lem:bpwm} to prove hardness for BWMatch.

\begin{lemma}
Min-Weight-Triangle on a graph with $n$ nodes and $m$ edges can be reduced to $O(n)$ updates and queries of decremental or incremental BWMatch on a graph with $O(n)$ nodes and $O(m+n)$ edges. 
\end{lemma}

%% file: 3sum.tex
In this section we prove the following $3$SUM hardness results. The proof is divided into  the lemmas below.

\begin{reminder}{Theorem~\ref{thm:3sum}}
If for some $\eps >0$ and $1/6 \leq \alpha \leq 1/3$, we can solve either of
\begin{itemize}
\item fully dynamic $st$-SubConn, $st$-Reach, BPMatch, or SC, with preprocessing time $O(m^{\frac{4}{3}-\eps})$, amortized update time $O(m^{\alpha - \eps})$, and amortized query time $O(m^{\frac{2}{3}-\alpha-\eps})$, or
\item incremental $st$-SubConn, $st$-Reach, BPMatch, or SC, with preprocessing time $O(m^{\frac{4}{3}-\eps})$, worst case update time $O(m^{\alpha - \eps})$, and worst case query time $O(m^{\frac{2}{3}-\alpha-\eps})$, or
\item decremental $st$-Reach, BPMatch, or SC, with preprocessing time $O(m^{\frac{4}{3}-\eps})$, worst case update time $O(m^{\alpha - \eps})$, and worst case query time $O(m^{\frac{2}{3}-\alpha-\eps})$, or
\item PP or $\emptyset$-PP over $k$ sets and a universe of size $O(k)$ with preprocessing time $O(k^{\frac{4}{3}-\eps})$, amortized update and query time $O(k^{\frac{1}{3}-\eps})$,
\end{itemize}
then Conjecture~\ref{conj:3sum} is false.
\end{reminder}

P\v{a}tra\c{s}cu \cite{PatrDyn} showed that $3$-SUM can be reduced to the problem of listing triangles in a graph, and then showed how ``the multiphase problem" can be used for listing triangles. Then, by reducing ``the multiphase problem" to dynamic problems like Subgraph Connectivity, one can show conditional hardness for the dynamic problems.

Here, we reduce the triangle listing problem directly to the dynamic problems, bypassing ``the multiphase problem", which allows us to give explicit lower bounds assuming Conjecture~\ref{conj:3sum} holds. Then we show how to modify the reduction to use fewer queries, which gives a higher lower bound on the query time.

\begin{theorem}[\cite{PatrDyn}]\label{thm:listing}
If for some $\eps>0$ and $R=O(n^{1-\eps})$ but $R=\Omega(n^{\frac{1}{2}+\eps})$, we can either list all triangles in a tripartite graph $G$ or say that the number of triangles is bigger than $\Delta$ for some $\Delta=O(n^2/R)$, where:
\begin{itemize}
\item the three parts are $A,B,C$, of size $N=|A|=|B|=R\sqrt{n}$ and $|C|=n$;
\item each vertex in $A \cup B$ has $O(n/R)$ neighbors in $C$;
\item there are $O(nR)$ edges in $A\times B$,
\end{itemize}
in time $O(n^{2-\eps})$, then Conjecture~\ref{conj:3sum} is false.
\end{theorem}

The following observation gives an alternative problem that is easier to reduce from.

\begin{lemma} \label{lem:pairs}
If given a tripartite graph $G$ as described in Theorem~\ref{thm:listing}, we can either list all pairs of nodes $(a,b)\in A\times B$ that are a part of some triangle in $G$, or say that the number of such pairs is bigger than $\Delta$ for some $\Delta=O(n^2/R)$, in time $O(n^{2-\eps})$, for some $\eps>0$, then Conjecture~\ref{conj:3sum} is false.
\end{lemma}

\begin{proof}
We run the algorithm for listing all pairs $(a,b)$ in $O(n^{2-\eps})$ time, and then do the following to list all triangles.
If the number of returned pairs is more than $\Delta$ we say that there are more than $\Delta$ triangles in $G$. Otherwise, for each of the $O(n^2/R)$ pairs $(a,b)$ returned by the solution we can go over all the nodes $c\in N_C(a) \cup N_C(b)$ and check whether $(a,b,c)$ is a triangle in $O(1)$ time. Since the size of $N_C(a)\cup N_C(b)$ is $O(n/R)$ for any $a\in A, b\in B$, we can list all triangles in $G$ in time $O(n^2/R \cdot n/R) = O(n^3/R^2)$ which is $O(n^{2-\eps})$ since $R=\Omega(n^{\frac{1}{2}-\eps})$, which by Theorem~\ref{thm:listing} implies that Conjecture~\ref{conj:3sum} is false. 
\end{proof}

Using this lemma, we can immediately get lower bounds for the above dynamic problems assuming Conjecture~\ref{conj:3sum}. For example, we can reduce the problem of listing all pairs $(a,b)$ that participate in a triangle to $O(nR)$ updates and queries of the $st$-SubConn problem on a graph with $O(n^{1.5})$ edges, which by setting $R=n^{\frac{1}{2}+\delta}$ says that if we can solve $st$-SubConn with preprocessing time $O(m^{\frac{4}{3} - \eps})$ and query and update time $O(m^{\frac{1}{3}-\eps})$, for some $\eps,\delta>0$, then we get a truly sub quadratic algorithm for $3$-SUM, and Conjecture~\ref{conj:3sum} is false.

To do this, consider the graph $H$ that is equivalent to $G$ except that we remove all the edges between partitions $A$ and $B$. We also add two nodes $s,t$ to $H$ such that $s$ is connected with an edge $(s,a)$ to every node in part $A$, while $t$ is connected with an edge $(b,t)$ to every node in part $B$. Initially, only the nodes of $C$ and $s$ and $t$ are activated, i.e. in the set $S$, and note that there is no path from $s$ to $t$ in the graph induced by $S$. Then, we go over each edge $(a,b)$ in $G$, where $a \in A$ and $b \in B$ and we add the nodes $a,b$ in $H$ to $S$, then we ask a query to see if there is a path from $s$ to $t$. There will be a path if and only if there is a node $c \in C$ that has edges to both $a$ and $b$, in which case we know that $(a,b,c)$ is a triangle in $G$ and we report the pair $(a,b)$. Then, we remove $a,b$ from $S$ and go on to check the next edge. Since the number of edges in $G$ between parts $A$ and $B$ is $O(nR)$, the reduction uses $O(nR)$ updates and queries, and since the number of edge in $G$ between $A$ and $C$ and between $B$ and $C$ is $O(n^{1.5})$, it is also the number of edges in $H$.

Another example of a reduction from the problem of listing all pairs $(a,b)$ that participate in a triangle to a dynamic problem is the following. 

\begin{lemma}
Suppose that for some $\eps>0$ fully dynamic $\emptyset$-PP on a universe of size $n$ and up to $k$ subsets where $k=O(n^{1.5+\frac{\eps}{2}})$ and where each subset is of size $O(n^{\frac{1}{2}-\frac{\eps}{2}})$ can be solved with preprocessing time $p(n,k) = O(n^{2-\eps})$ and amortized update and query times $u(n),q(n)= O(n^{\frac{1}{2}-\eps})$, then Conjecture~\ref{conj:3sum} is false.
\end{lemma}

\begin{proof}
Let $R=n^{\frac{1}{2} + \frac{\eps}{2}}$ and $k= n^{1.5+\frac{\eps}{2}}$. 
Given a tripartite graph $G=(V,E)$ as described in Theorem~\ref{thm:listing}, we create the following instance of $\emptyset$-PP.
The universe will be $[n]$ and we associate every node in $C$ with an integer in $[n]$. For every node $u\in A \cup B$ we create the set $X_u = \{ c \mid c \in N_C(u) \}$. Note that the size of each of these sets is $d_C(u)=O(n/R)$, and the sets can be constructed in time $O(\sqrt{n}R \cdot n/R) = O(n^{1.5})$.
We preprocess the sets in $p(n,k)=O(n^{2-\eps})$ time.
Then we have a stage for each edge $(a,b)\in E \cap A\times B$ where we compute the set $X_a \cap X_b$ and ask a query to check whether it is empty. It is easy to see that $X_a \cap X_b \neq \phi$ if and only if $(a,b)$ participates in a triangle, and we can report the pair $(a,b)$. Note that we create at most $k=O(nR)$ subsets, and the total running time is $O(nR \cdot (u(n,k)+q(n,k))) = O(n^{1.5 + \frac{\eps}{2} + 0.5 - \eps})$  which is enough to falsify Conjecture~\ref{conj:3sum} using Lemma~\ref{lem:pairs}.
\end{proof}

As a corollary, we obtain that Conjecture~\ref{conj:3sum} implies that $\emptyset$-PP over $k$ sets and a universe of size $O(k)$ cannot be solved for any $\eps>0$ with preprocessing time $O(k^{\frac{4}{3}-\eps})$ and update and query time $O(k^{\frac{1}{3}-\eps})$.
We remark that using P\v{a}tra\c{s}cu's reduction from ``the multiphase problem'' to PP one can get a similar lower bound to the one we get for $\emptyset$-PP.

\begin{lemma}[Implicit in \cite{PatrDyn}]
Conjecture~\ref{conj:3sum} implies that PP over $k$ sets and a universe of size $O(k)$ cannot be solved for any $\eps>0$ with preprocessing time $O(k^{\frac{4}{3}-\eps})$ and update and query time $O(k^{\frac{1}{3}-\eps})$.
\end{lemma}

The above reductions allow the dynamic algorithm for $st$-SubConn, $\emptyset$-PP and PP to see the sequence of updates and queries in advance, which implies that the hardness result still holds when the updates are given ``offline''. In the rest of this section, however, the updates in our reductions will depend on the answers to the queries that we ask during the reduction, and, therefore, an algorithm with lookahead that achieves improved running times might not imply a sub quadratic algorithm for $3$-SUM.
We will use a binary search trick to decrease the number of queries we ask during the reduction. We will be able to reduce $3$-SUM, via the triangles reporting problems, to $O(nR)$ updates and only $(\sqrt{n} R +\Delta \log{n})$ queries to an $st$-SubConn problem on a graph with $O(n^{1.5})$ edges. This allows us to choose a larger $R$ and get a tradeoff -- the lower the update time is, the higher the query time needs to be.

We now present the main binary search idea that allows us to get improved lower bounds. 
Consider Algorithm~\ref{A-List-Pairs} for listing all pairs $(a,b)$ that participate in a triangle, and assume, for now, that the procedure Triangle($G,a,i,j$) returns yes if and only if there are nodes $b_k \in B$ and $c \in C$ for $k \in \{ (j-1)\cdot \frac{N}{2^{i}} +1, \ldots, j\cdot \frac{N}{2^{i}} \}$, such that $(a,b_k,c)$ is  a triangle in $G$. We will later show how to implement this procedure using the dynamic problems. 

For each node $a\in A$, we search for all nodes $b\in B$ such that the pair $(a,b)$ needs to be reported. We use calls to Triangle($\cdot$) to figure out whether there exists a node $b$ in a certain subset of $B$ that participates in a triangle with $a$. Then, if the answer was yes, we partition the subset into two halves and recurse on both halves, until we reach subsets which contain only one node of $B$ and we can output a pair. Since the subset of $B$ we start from is all of $B$, every pair $(a,b)$ that participates in a triangle will be found by our search.

To do this, we start from level $i=0$ and call the recursive function Search($G,a,i=0,j=1$). At level $i \in \{0,1,\ldots,\log{N}\}$ we partition the nodes of $B$ to $2^{i}$ sets of $N/2^{i}$ nodes each, and the index $j$ indicates which set in the partition we are currently considering, namely $B_{i,j}=\{ b_{(j-1)\cdot \frac{N}{2^{i}} +1}, \ldots, b_{j\cdot \frac{N}{2^{i}}} \}$.
We call Triangle($G,a,i,j$) to check whether there is a node $b$ in the $j^\th$ set of level $i$ such that $(a,b)$ needs to be reported, and if the answer is yes, we advance to level $(i+1)$, where we recurse on both halves of the set we were considering.  
At level $i=\log{N}$, the sets are of size $1$ and we can report a pair if Triangle($G,a,i,j$) returns yes. 
If at some point we have reported more than $\Delta$ triangles we abort and List-Pairs($G$) says that there were more than $\Delta$ pairs.

\begin{algorithm}\label{A-List-Pairs}\small
\caption{\small List-Pairs($G$)}
\ForEach{$a\in A$}{ Search($G,a,0,1$)\;
}
\end{algorithm}

\begin{algorithm}\label{A-Search}\small
\caption{\small Search($G,a,i,j$)}
    \eIf{$i = \log{N}$}
    {\If{Triangle($G,a,i,j$)}
    		{report $(a,b_j)$ \; 
		$c \gets c+1$\;
		\If{$c>\Delta$}{abort\;}}
	}
	{
	\If{Triangle($G,a,i,j$)}
	{Search($G,a,i+1,2j$)\;
	Search($G,a,i+1,2j+1$)\;
	}
	}
\end{algorithm}
 
\begin{lemma} \label{lem:bsearch}
Let $G$ be a tripartite graph as described in Theorem~\ref{thm:listing}, and denote $B_{i,j}=  \{ b_{(j-1)\cdot \frac{N}{2^{i}} +1}, \ldots, b_{j\cdot \frac{N}{2^{i}}} \}$ and $d_{a,i,j} = | N(a)\cap B_{i,j} |$ for every $a\in A, i\in\{0,\ldots,\log{N}\}$, and $j\in[2^i]$. If we can compute Triangle($G,a,i,j$) in time $O(Q + d_{a,i,j}\cdot U)$, for some $U=O((n/R)^{1-\eps})$, $Q=O(\min\{(n^{1.5}/R)^{1-\eps} , R^{1-\eps}\})$, and $\eps>0$, then Algorithm~\ref{A-List-Pairs} runs in time $\tilde{O}(n^{2-\eps})$ and Conjecture~\ref{conj:3sum} is false. 
\end{lemma}

\begin{proof}
We will first prove that the total running time of Algorithm~\ref{A-List-Pairs} can be bounded by $O(|E\cap (A\times B)| \cdot \log{N} \cdot U + (N+\Delta\cdot \log{N})\cdot Q)$, then we will show that under certain choices of parameters this implies a subquadratic algorithm for $3$-SUM.

During the runtime of Algorithm~\ref{A-List-Pairs} there will be $N=|A|$ calls of the form Triangle($G,a,0,1$), and the total running time of such calls can be bounded by
\[
\sum\limits_{a\in A} (d_{a,0,1} \cdot U + Q) = \sum\limits_{a\in A} (d_B(a) \cdot U) + N\cdot Q = |E\cap (A\times B)| \cdot U + N\cdot Q.
\]
In addition, there will be more calls to lower levels $i\geq1$. We will bound the contributions of the two terms in the runtime of Triangle($G,a,i,j$), namely the $Q$ term and the $d_{a,i,j}\cdot U$ term, separately. 
First, note that the number of calls to Triangle($\cdot$) with level $i>1$ can be bounded by $2\cdot \Delta \log{N}$, since each such call must reach a leaf $b_k$ that is a part of a triangle, while each pair $(a,b_k)$ that participates in a triangle incurs at most $2\log{N}$ extra calls, and the number of such pairs we reach is bounded by $\Delta$. Therefore, we can bound the contribution of the $Q$ term by $O(\Delta \cdot \log{N} \cdot Q)$.
Second, note that for any node $a\in A$, we never call Triangle($G,a,i,j$) with the same $i,j$ more than once, and that for any $i\in\{0,\ldots,\log{N}\}$, the sum $\sum_{j\in[2^i]}{d_{a,i,j}}$ equals $d_B(a)$. Therefore, the total contribution of the $d_{a,i,j}\cdot U$ term can be bounded by
\[
\sum\limits_{a\in A} \sum\limits_{i=0}^{\log{N}} \sum\limits_{j\in[2^i]}{d_{a,i,j} \cdot U} = \sum\limits_{a\in A} \sum\limits_{i=0}^{\log{N}} d_B(a) \cdot U 
=  |E\cap (A\times B)| \cdot \log{N} \cdot U.
\]

Thus, we have shown that, under the assumptions of the lemma, the total running time of Algorithm~\ref{A-List-Pairs} is $O(|E\cap (A\times B)| \cdot \log{N} \cdot U + (N+\Delta\cdot \log{N})\cdot Q)$, which can also be written as $\tilde{O}(nR\cdot U + (\sqrt{n}R + n^2/R)\cdot Q)$. Thus, if $U=O((n/R)^{1-\eps})$ and $Q=O(\min\{(n^{1.5}/R)^{1-\eps} , R^{1-\eps}\})$, for some $\eps>0$, then we can list all pairs $(a,b)\in A\times B$ in time $\tilde{O}(n^{2-\eps})$ which by Lemma~\ref{lem:pairs} implies that Conjecture~\ref{conj:3sum} is false.
\end{proof}

Finally, we will show how each of the dynamic problems listed above can be used to implement Triangle($G,a,i,j$), implying lower bounds for these problems under Conjecture~\ref{conj:3sum}.

\begin{lemma}\label{lem:st-sub-conn}
Let $1/6 \leq \alpha \leq 1/3$ and suppose that fully dynamic $st$-SubConn on a graph with $m$ edges can be done with (amortized) update time $u(m)$, query time $q(m)$, and $p(m)$ preprocessing time. Then assuming Conjecture~\ref{conj:3sum}, we have that, for all $\eps>0$, either $u(m)\geq m^{\alpha-\eps}$, $q(m)\geq m^{\frac{2}{3}-\alpha-\eps}$, or $p(m)\geq m^{\frac{4}{3}-\eps}$.
\end{lemma}

\begin{proof}
Given a tripartite graph $G=(V,E)$ as described in Theorem~\ref{thm:listing}, we will construct an instance of the fully dynamic SubConn problem, the graph $H$ constructed above, that will allow us to compute Triangle($G,a,i,j$). 

Recall that $H$ is a copy of $G$ in which all the edges between parts $A,B$ are removed, and there are two additional nodes $s,t$, where $s$ is connected to all the nodes of $A$ and $t$ is connected to all the nodes of $B$. Initially, only the nodes in $C\cup \{s,t\}$ are in $S$.

Note that the number of edges in $H$ is $m=O(n^{1.5})$.
We preprocess $H$ in time $p(m)$, and now we can perform updates and queries on it. 
We will assume that $p(m) = O(m^{\frac{4}{3}-\eps})$ for some $\eps>0$, so that the preprocessing takes time $O((n^{\frac{3}{2}})^{\frac{4}{3}-\eps}) = O(n^{2-\frac{3}{2}\eps})$, and therefore we can ignore the preprocessing time when using Lemma~\ref{lem:bsearch} for showing that Conjecture~\ref{conj:3sum} is false, since this subquadratic process is executed only once throughout the implementation of Algorithm~\ref{A-List-Pairs}.

To compute Triangle($G,a,i,j$) we do the following. First, (1) we add the nodes $a$ and $b'$ for every node $b' \in N(a)\cap B_{i,j}$ to $S$. Then, (2) we ask a query to determine whether $s$ and $t$ are connected with the current $S$, and (3) we remove the nodes of $S$ that were added in (1).  Finally, (4) we return yes if and only if the answer to the query in (2) was yes.

We claim that our procedure computes Triangle($G,a,i,j$) correctly, i.e. we answer yes if and only if there is a triangle $(a,b',c)$ in $G$ such that $b' \in B_{i,j}$. For the first direction, assume that we answer yes and consider a path from $s$ to $t$ in the graph induced by $S$ when we ask the query. The node $a$ must belong to this path, since it is the only neighbor of $s$ that is in the set $S$. The second node on the path must be some node $c\in C$, since $a$ only has edges to $s$ and some nodes in $C$, and the third node on the path must be a node $b' \in B$, since the nodes of $C$ have edges only to nodes in $A$ and $B$ while $a$ was the only node of $A$ that is in $S$. Since $b'$ is on the path, it must be in $S$, and therefore we know that $b' \in N(a)\cap B_{i,j}$. Thus, $(a,b'),(a,c),(b',c)\in E$ and we found a triangle $(a,b',c)$ in $G$ where $b'\in B_{i,j}$.
For the other direction, assume that $(a,b',c)$ is a triangle in $G$ where $b' \in B_{i,j}$, and consider the path $(s,a,c,b',t)$ in $H$. The nodes on this path will be in $S$ and therefore the answer to the query will be yes.

The running time of our procedure will be $q(m)+ 2\cdot d_{a,i,j}\cdot u(m)$, where $d_{a,i,j} = | N(a)\cap B_{i,j} |$. 
Therefore, if we assume that for some $0<\eps<\frac{1}{2}$, both $u(m) = O(m^{\alpha - \eps})$ and $q(m) = O(m^{\frac{2}{3}-\alpha - \eps})$, then by setting $R = n^{1-\frac{3}{2}\alpha - \frac{\eps}{2}}$, which by our choice of $\alpha$ satisfies both $R=O(n^{1-\frac{\eps}{2}})$ and $\Omega(n^{\frac{1}{2}+\frac{\eps}{2}})$, we will get that Conjecture~\ref{conj:3sum} is false by using Lemma~\ref{lem:bsearch}.
To see this, recall that $m=O(n^{1.5})$ and let $\eps' = \frac{\eps}{1+\alpha}>0$ to get that both
\[ u(m) = O((n^{1.5})^{\alpha-\eps}) = O(n^{\frac{3}{2}\alpha - \frac{3}{2}\alpha\eps'}) = O((n^{\frac{3}{2}\alpha})^{1-\eps'}) = O((n/R)^{1-\eps'})
\]
and
\[
q(m) = O((n^{1.5})^{\frac{2}{3}-\alpha-\eps}) = O(n^{1-\frac{3}{2}\alpha-\frac{\eps}{2}-\eps}) = O(n^{1-\frac{3}{2}\alpha-\frac{\eps}{2}-\eps'}) = O((n^{1-\frac{3}{2}\alpha-\frac{\eps}{2}})^{1-\eps'})
\]
since $\alpha \in [\frac{1}{6},\frac{1}{3}]$,
\[ = O((\min \{ n^{1-\frac{3}{2}\alpha-\frac{\eps}{2}} , n^{\frac{1}{2}+\frac{3}{2}\alpha + \frac{\eps}{2}} \})^{1-\eps'} ) = O((\min \{ R, \frac{n^{1.5}}{R} \})^{1-\eps'}),
\]
which is what we need in order to use Lemma~\ref{lem:bsearch} to falsify Conjecture~\ref{conj:3sum}.


\end{proof}

\begin{lemma}
Let $1/6 \leq \alpha \leq 1/3$ and suppose that fully dynamic $st$-Reach on a graph with $m$ edges can be done with (amortized) update time $u(m)$, query time $q(m)$, and $p(m)$ preprocessing time. Then assuming Conjecture~\ref{conj:3sum}, we have that, for all $\eps>0$, either $u(m)\geq m^{\alpha-\eps}$, $q(m)\geq m^{\frac{2}{3}-\alpha-\eps}$, or $p(m)\geq m^{\frac{4}{3}-\eps}$.
\end{lemma}

\begin{proof}
Follows from Lemma~\ref{lem:st} and the lower bound for $st$-SubConn.
\end{proof}

\begin{lemma}
Let $1/6 \leq \alpha \leq 1/3$ and suppose that fully dynamic BPMatch on a graph with $m$ edges can be done with (amortized) update time $u(m)$, query time $q(m)$, and $p(m)$ preprocessing time. Then assuming Conjecture~\ref{conj:3sum}, we have that, for all $\eps>0$, either $u(m)\geq m^{\alpha-\eps}$, $q(m)\geq m^{\frac{2}{3}-\alpha-\eps}$, or $p(m)\geq m^{\frac{4}{3}-\eps}$.
\end{lemma}

\begin{proof}
Follows from Lemma~\ref{lem:bpm} and the lower bound for $st$-Reach.
\end{proof}

\paragraph{Remark.} We can implement Triangle($\cdot$) using dynamic bipartite matching directly, by creating an $8$-layered graph in which, using similar arguments to those in the proofs of Lemmas~\ref{lem:5bpm} and~\ref{lem:17bpm}, the size of any matching without length $17$ augmenting paths will give us the answer to the queries we need for the implementation. Therefore, the above lower bound also applies to the $17$-BPM problem.

\begin{lemma}
Let $1/6 \leq \alpha \leq 1/3$ and suppose that fully dynamic SC on a graph with $m$ edges can be done with (amortized) update time $u(m)$, query time $q(m)$, and $p(m)$ preprocessing time. Then assuming Conjecture~\ref{conj:3sum}, we have that, for all $\eps>0$, either $u(m)\geq m^{\alpha-\eps}$, $q(m)\geq m^{\frac{2}{3}-\alpha-\eps}$, or $p(m)\geq m^{\frac{4}{3}-\eps}$.
\end{lemma}

\begin{proof}
Follows from Lemma~\ref{lem:reach-to-sc} and the lower bound for $st$-Reach.
\end{proof}

\subsection{Partially dynamic reductions}
In the reduction to $st$-SubConn in Lemma~\ref{lem:st-sub-conn}, it is easy to simulate the fully dynamic algorithm with an incremental one, since in every computation of Triangle($\cdot$) we start with a state in which all the nodes of $A$ and $B$ are deactivated, then we activate some nodes (which we can do incrementally) and then after we ask a query we deactivate the same nodes we just activated, which can be done by undoing the activations. Therefore, the lower bounds hold for worst case update times of incremental algorithms.

The decremental case is more difficult and it is not clear how we can simulate the fully dynamic $st$-SubConn algorithm with a decremental one efficiently. However, for $st$-Reach, and therefore for BPMatch and SC too, we can simulate the fully dynamic case with the decremental one with only a logarithmic overhead. The idea is very similar to the one we use in the proof of Lemma~\ref{lem:dec-reach}. 
To be able to ``activate" one node $a$ in $A$, we do exactly the same procedure we did there with the tree $T_s$.
To be able to ``activate" multiple nodes in $B$ simultaneously, we add a binary tree $T_t$  rooted at $t$ with leaves the nodes of $B$ and edges directed towards $t$. Initially, all edges of $T_t$ are in the graph $H$, and in order to ``activate" a subset of $k$ nodes from $B$, we delete $O(k \log {n})$ edges from $T_t$ so that only these $k$ nodes can reach $t$. After the query is answered, we undo the deletions to go back to the initial state. Note that the number of updates increased only by a factor of $O(\log{n})$ over the fully dynamic simulation.


%% file: dynamic.bbl
\begin{thebibliography}{10}

\bibitem{AbboudL13}
A.~Abboud and K.~Lewi.
\newblock Exact weight subgraphs and the k-sum conjecture.
\newblock In {\em ICALP (1)}, pages 1--12, 2013.

\bibitem{AHLK01}
M.~Abellanas, F.~Hurtado, C.~Icking, R.~Klein, E.~Langetepe, L.~Ma, B.~Palop,
  and V.~Sacristian.
\newblock Smallest color-spanning objects.
\newblock In {\em Proc. ESA}, pages 278 -- 289, 2001.

\bibitem{ACIM99}
D.~Aingworth, C.~Chekuri, P.~Indyk, and R.~Motwani.
\newblock Fast estimation of diameter and shortest paths (without matrix
  multiplication).
\newblock {\em SIAM Journal on Computing}, 28(4):1167--1181, 1999.

\bibitem{AlNa96}
N.~Alon and M.~Naor.
\newblock Derandomization, witnesses for boolean matrix multiplication and
  construction of perfect hash functions.
\newblock {\em Algorithmica}, 16:434--449, 1996.

\bibitem{AlYuZw97c}
N.~Alon, R.~Yuster, and U.~Zwick.
\newblock Finding and counting given length cycles.
\newblock volume 855, pages 354--364, 1994.

\bibitem{AlYuZw97}
N.~Alon, R.~Yuster, and U.~Zwick.
\newblock Finding and counting given length cycles.
\newblock {\em Algorithmica}, 17:209--223, 1997.

\bibitem{anand}
A.~Anand, S.~Baswana, M.~Gupta, and S.~Sen.
\newblock Maintaining approximate maximum weighted matching in fully dynamic
  graphs.
\newblock In {\em FSTTCS}, pages 257--266, 2012.

\bibitem{AEK04}
D.~Archambault, W.~Evans, and D.~Kirkpatrick.
\newblock Computing the set of all distant horizons of a terrain.
\newblock In {\em Proc. CCCG}, pages 76--79, 2004.

\bibitem{fourruss}
V.~L. Arlazarov, E.~A. Dinic, M.~A. Kronrod, and I.~A. Faradzev.
\newblock On economical construction of the transitive closure of an oriented
  graph.
\newblock {\em Soviet Math. Dokl.}, 11:1209--1210, 1970.

\bibitem{AH05}
B.~Aronov and S.~Har-Peled.
\newblock On approximating the depth and related problems.
\newblock In {\em Proc. SODA}, 2005.

\bibitem{BansalW09}
N.~Bansal and R.~Williams.
\newblock Regularity lemmas and combinatorial algorithms.
\newblock In {\em Proc. FOCS}, pages 745--754, 2009.

\bibitem{alg3sum}
I.~Baran, E.~Demaine, and M.~P\v{a}tra\c{s}cu.
\newblock Subquadratic algorithms for $3$sum.
\newblock {\em Algorithmica}, 50(4):584--596, 2008.

\bibitem{BH99}
G.~Barequet and S.~Har-Peled.
\newblock Some variants of polygonal containment and minimum hausdorff distance
  undertranslation are 3{SUM}-hard.
\newblock In {\em Proc. SODA}, pages 862--863, 1999.

\bibitem{BGSmatch}
S.~Baswana, M.~Gupta, and S.~Sen.
\newblock Fully dynamic maximal matching in ${O}(log n)$ update time.
\newblock In {\em FOCS}, pages 383--392, 2011.

\bibitem{benderSCC}
M.~A. Bender, J.~T. Fineman, and S.~Gilbert.
\newblock A new approach to incremental topological ordering.
\newblock In {\em SODA}, pages 1108--1115, 2009.

\bibitem{benderjour}
M.~A. Bender, J.~T. Fineman, S.~Gilbert, and R.~E. Tarjan.
\newblock A new approach to incremental cycle detection and related problems.
\newblock {\em CoRR}, abs/1112.0784, 2011.

\bibitem{Bernstein09}
A.~Bernstein.
\newblock Fully dynamic (2 + epsilon) approximate all-pairs shortest paths with
  fast query and close to linear update time.
\newblock In {\em FOCS}, pages 693--702, 2009.

\bibitem{BernsteinR11}
A.~Bernstein and L.~Roditty.
\newblock Improved dynamic algorithms for maintaining approximate shortest
  paths under deletions.
\newblock In {\em SODA}, pages 1355--1365, 2011.

\bibitem{CalabroIP09}
C.~Calabro, R.~Impagliazzo, and R.~Paturi.
\newblock The complexity of satisfiability of small depth circuits.
\newblock In {\em Parameterized and Exact Computation}, pages 75--85. Springer,
  2009.

\bibitem{Chan06}
T.~M. Chan.
\newblock Dynamic subgraph connectivity with geometric applications.
\newblock {\em SIAM J. Comput.}, 36(3):681--694, 2006.

\bibitem{CPR08}
T.~M. Chan, M.~P\v{a}tra\c{s}cu, and L.~Roditty.
\newblock Dynamic connectivity: Connecting to networks and geometry.
\newblock In {\em FOCS}, pages 95--104, 2008.

\bibitem{diametersoda14}
S.~Chechik, D.~Larkin, L.~Roditty, G.~Schoenebeck, R.~E. Tarjan, and
  V.~{Vassilevska Williams}.
\newblock Better approximation algorithms for the graph diameter.
\newblock In {\em Proc. SODA}, 2014.

\bibitem{CEH04}
O.~Cheong, A.~Efrat, and S.~Har-Peled.
\newblock On finding a guard that sees most and a shop that sells most.
\newblock In {\em Proc. SODA}, pages 1098--1107, 2004.

\bibitem{cygan2012deterministic}
M.~Cygan.
\newblock Deterministic parameterized connected vertex cover.
\newblock In {\em Algorithm Theory--SWAT 2012}, pages 95--106, 2012.

\bibitem{cygan}
M.~Cygan, H.~Dell, D.~Lokshtanov, D.~Marx, J.~Nederlof, Y.~Okamoto, R.~Paturi,
  S.~Saurabh, and M.~Wahlstrom.
\newblock On problems as hard as {CNFSAT}.
\newblock In {\em Proc. CCC}, pages 74--84, 2012.

\bibitem{cygan13}
M.~Cygan, S.~Kratsch, and J.~Nederlof.
\newblock Fast {H}amiltonicity checking via bases of perfect matchings.
\newblock In {\em STOC}, pages 301--310, 2013.

\bibitem{cygan2011solving}
M.~Cygan, J.~Nederlof, M.~Pilipczuk, J.~van Rooij, and J.~Wojtaszczyk.
\newblock Solving connectivity problems parameterized by treewidth in single
  exponential time.
\newblock In {\em FOCS}, pages 150--159, 2011.

\bibitem{DW10}
E.~Dantsin and A.~Wolpert.
\newblock On moderately exponential time for {SAT}.
\newblock In {\em Proc. 13th International Conference on Theory and
  Applications of Satisfiability Testing}, pages 313--325, 2010.

\bibitem{BGO97}
M.~{de Berg}, M.~{de Groot}, and M.~H. Overmars.
\newblock Perfect binary space partitions.
\newblock {\em Computational Geometry: Theory and Applications}, 7(81):81--91,
  1997.

\bibitem{di00}
C.~Demetrescu and G.~F. Italiano.
\newblock Fully dynamic transitive closure: Breaking through the $o(n^2)$
  barrier.
\newblock In {\em Proc. FOCS}, volume~41, pages 381--389, 2000.

\bibitem{Duan10}
R.~Duan.
\newblock New data structures for subgraph connectivity.
\newblock In {\em ICALP (1)}, pages 201--212, 2010.

\bibitem{DuanS12}
R.~Duan and H.-H. Su.
\newblock A scaling algorithm for maximum weight matching in bipartite graphs.
\newblock In {\em Proceedings of the Twenty-Third Annual ACM-SIAM Symposium on
  Discrete Algorithms}, SODA '12, pages 1413--1424. SIAM, 2012.

\bibitem{EK72}
J.~Edmonds and R.~M. Karp.
\newblock Theoretical improvements in algorithmic efficiency for network flow
  problems.
\newblock {\em Journal of the ACM}, 19(2):248--264, 1972.

\bibitem{Er99}
J.~Erickson.
\newblock New lower bounds for convex hull problems in odd dimensions.
\newblock {\em SIAM Journal on Computing}, 28(4):1198--1214, 1999.

\bibitem{evenshiloach}
S.~Even and Y.~Shiloach.
\newblock An on-line edge-deletion problem.
\newblock {\em J. ACM}, 28(1):1--4, 1981.

\bibitem{FHV13}
H.~Fernau, P.~Heggernes, and Y.~Villanger.
\newblock A multivariate analysis of some {DFA} problems.
\newblock In {\em Proceedings of LATA}, pages 275--286, 2013.

\bibitem{Fr85}
G.~N. Frederickson.
\newblock Data structures for on-line updating of minimum spanning trees, with
  applications.
\newblock {\em SIAM J. Comput.}, 14(4):781--798, 1985.

\bibitem{FT}
M.~L. Fredman and R.~E. Tarjan.
\newblock Fibonacci heaps and their uses in improved network optimization
  algorithms.
\newblock {\em J. ACM}, 34(3):596--615, 1987.

\bibitem{FI00}
D.~Frigioni and G.~F. Italiano.
\newblock Dynamically switching vertices in planar graphs.
\newblock {\em Algorithmica}, 28(1):76--103, 2000.

\bibitem{Gabow85}
H.~Gabow.
\newblock A scaling algorithm for weighted matching on general graphs.
\newblock In {\em Prof. FOCS}, pages 90--100, 1985.

\bibitem{GTscaling}
H.~N. Gabow and R.~E. Tarjan.
\newblock Faster scaling algorithms for general graph-matching problems.
\newblock {\em J. ACM}, 38(4):815--853, 1991.

\bibitem{GO95}
A.~Gajentaan and M.~Overmars.
\newblock On a class of $o(n^2)$ problems in computational geometry.
\newblock {\em Computational Geometry}, 5(3):165--185, 1995.

\bibitem{GP13}
M.~Gupta and R.~Peng.
\newblock Fully dynamic $(1+\eps)$-approximate matchings.
\newblock In {\em FOCS}, 2013.

\bibitem{HKMST12}
B.~Haeupler, T.~Kavitha, R.~Mathew, S.~Sen, and R.~E. Tarjan.
\newblock Incremental cycle detection, topological ordering, and strong
  component maintenance.
\newblock {\em ACM Transactions on Algorithms}, 8(1):3, 2012.

\bibitem{HanT12}
Y.~Han and T.~Takaoka.
\newblock An ${O}(n^3 \log\log n/\log^2 n)$ time algorithm for all pairs
  shortest paths.
\newblock In {\em SWAT}, pages 131--141, 2012.

\bibitem{harvey}
N.~J.~A. Harvey.
\newblock Algebraic structures and algorithms for matching and matroid
  problems.
\newblock In {\em Proc. FOCS}, volume~47, pages 531--542, 2006.

\bibitem{HenzingerKN13}
M.~Henzinger, S.~Krinninger, and D.~Nanongkai.
\newblock Dynamic approximate all-pairs shortest paths: Breaking the {O}(mn)
  barrier and derandomization.
\newblock In {\em Proc. FOCS}, 2013.

\bibitem{HK99}
M.~R. Henzinger and V.~King.
\newblock Randomized fully dynamic graph algorithms with polylogarithmic time
  per operation.
\newblock {\em J. ACM}, 46(4):502--516, 1999.

\bibitem{HenzingerK01}
M.~R. Henzinger and V.~King.
\newblock Maintaining minimum spanning forests in dynamic graphs.
\newblock {\em SIAM J. Comput.}, 31(2):364--374, 2001.

\bibitem{hirschsat}
E.~A. Hirsch.
\newblock Two new upper bounds for {SAT}.
\newblock In {\em Proc. SODA}, pages 521--530, 1998.

\bibitem{HolmLT01}
J.~Holm, K.~de~Lichtenberg, and M.~Thorup.
\newblock Poly-logarithmic deterministic fully-dynamic algorithms for
  connectivity, minimum spanning tree, 2-edge, and biconnectivity.
\newblock {\em J. ACM}, 48(4):723--760, 2001.

\bibitem{hopkarpmatch}
J.~Hopcroft and R.~Karp.
\newblock An $n^{5/2} $ algorithm for maximum matchings in bipartite graphs.
\newblock {\em SIAM Journal on Computing}, 2(4):225--231, 1973.

\bibitem{ipz1}
R.~Impagliazzo and R.~Paturi.
\newblock On the complexity of k-sat.
\newblock {\em J. Comput. Syst. Sci.}, 62(2):367--375, 2001.

\bibitem{ipz2}
R.~Impagliazzo, R.~Paturi, and F.~Zane.
\newblock Which problems have strongly exponential complexity?
\newblock {\em J. Comput. Syst. Sci.}, 63(4):512--530, 2001.

\bibitem{Italiano88}
G.~F. Italiano.
\newblock Finding paths and deleting edges in directed acyclic graphs.
\newblock {\em Inf. Process. Lett.}, 28(1):5--11, 1988.

\bibitem{IvLl}
Z.~Ivkovic and E.~L. Lloyd.
\newblock Fully dynamic maintenance of vertex cover.
\newblock In {\em WG}, pages 99--111, 1993.

\bibitem{JV13}
Z.~Jafargholi and E.~Viola.
\newblock 3sum, 3xor, triangles.
\newblock {\em Electronic Colloquium on Computational Complexity (ECCC)}, 20:9,
  2013.

\bibitem{kavithastretch}
T.~Kavitha.
\newblock Faster algorithms for all-pairs small stretch distances in weighted
  graphs.
\newblock In {\em Proc. FSTTCS}, pages 328--339, 2007.

\bibitem{Kavitha08}
T.~Kavitha.
\newblock Dynamic matrix rank with partial lookahead.
\newblock In {\em FSTTCS}, pages 268--279, 2008.

\bibitem{kuhn}
H.~W. Kuhn.
\newblock The hungarian method for the assignment problem.
\newblock {\em Naval Research Logistics Quarterly}, 2:83--97, 1955.

\bibitem{Lacki13}
J.~Lacki.
\newblock Improved deterministic algorithms for decremental reachability and
  strongly connected components.
\newblock {\em ACM Transactions on Algorithms}, 9(3):27, 2013.

\bibitem{lee}
L.~Lee.
\newblock Fast context-free grammar parsing requires fast boolean matrix
  multiplication.
\newblock {\em J. ACM}, 49(1):1--15, 2002.

\bibitem{LokshtanovMS11a}
D.~Lokshtanov, D.~Marx, and S.~Saurabh.
\newblock Known algorithms on graphs on bounded treewidth are probably optimal.
\newblock In {\em SODA}, pages 777--789, 2011.

\bibitem{SEO02}
J.~E. M.~Soss and M.~H. Overmars.
\newblock Preprocessing chains for fast dihedral rotations is hard or even
  impossible.
\newblock {\em Computational Geometry: Theory and Applications},
  26(3):235--246, 2002.

\bibitem{madry}
A.~Madry.
\newblock Navigating central path with electrical flows: from flows to
  matchings, and back.
\newblock In {\em Proc. FOCS}, 2013.

\bibitem{moniensat}
B.~Monien and E.~Speckenmeyer.
\newblock Solving satisfiability in less than $2^n$ steps.
\newblock {\em Discrete Applied Mathematics}, 10(3):287 -- 295, 1985.

\bibitem{MS04}
M.~Mucha and P.~Sankowski.
\newblock Maximum matchings via gaussian elimination.
\newblock In {\em Proc. FOCS}, volume~45, pages 248--255, 2004.

\bibitem{NSmatch}
O.~Neiman and S.~Solomon.
\newblock Simple deterministic algorithms for fully dynamic maximal matching.
\newblock In {\em STOC}, pages 745--754, 2013.

\bibitem{PPSZ05}
R.~Paturi, P.~Pudl\'{a}k, M.~E. Saks, and F.~Zane.
\newblock An improved exponential-time algorithm for $k$-{SAT}.
\newblock {\em J. ACM}, 52(3):337--364, 2005.

\bibitem{PaturiPZ99}
R.~Paturi, P.~Pudl{\'a}k, and F.~Zane.
\newblock Satisfiability coding lemma.
\newblock {\em Chicago J. Theor. Comput. Sci.}, 1999, 1999.

\bibitem{pilipczuk2012finding}
M.~Pilipczuk and M.~Pilipczuk.
\newblock Finding a maximum induced degenerate subgraph faster than $2^n$.
\newblock In {\em Parameterized and Exact Computation}, pages 3--12. 2012.

\bibitem{PatrDyn}
M.~P\v{a}tra\c{s}cu.
\newblock Towards polynomial lower bounds for dynamic problems.
\newblock In {\em STOC}, pages 603--610, 2010.

\bibitem{patrasculog}
M.~P\v{a}tra\c{s}cu and E.~D. Demaine.
\newblock Logarithmic lower bounds in the cell-probe model.
\newblock {\em SIAM J. Comput.}, 35(4):932--963, 2006.

\bibitem{PT07}
M.~P\v{a}tra\c{s}cu and M.~Thorup.
\newblock Planning for fast connectivity updates.
\newblock In {\em FOCS}, pages 263--271, 2007.

\bibitem{PW10}
M.~P\v{a}tra\c{s}cu and R.~Williams.
\newblock On the possibility of faster {SAT} algorithms.
\newblock In {\em Proc. SODA}, pages 1065--1075, 2010.

\bibitem{RdecSCC}
L.~Roditty.
\newblock Decremental maintenance of strongly connected components.
\newblock In {\em SODA}, pages 1143--1150, 2013.

\bibitem{RV13}
L.~Roditty and V.~{Vassilevska Williams}.
\newblock Fast approximation algorithms for the diameter and radius of sparse
  graphs.
\newblock In {\em Proceedings of the 45th annual ACM symposium on Symposium on
  theory of computing}, STOC '13, pages 515--524, New York, NY, USA, 2013. ACM.

\bibitem{RZ02}
L.~Roditty and U.~Zwick.
\newblock Improved dynamic reachability algorithms for directed graphs.
\newblock In {\em FOCS}, pages 679--689, 2002.

\bibitem{rzesa}
L.~Roditty and U.~Zwick.
\newblock On dynamic shortest paths problems.
\newblock In {\em ESA}, pages 580--591, 2004.

\bibitem{sank04}
P.~Sankowski.
\newblock Dynamic transitive closure via dynamic matrix inverse.
\newblock In {\em Proc. FOCS}, volume~45, pages 509--517, 2004.

\bibitem{sankdynmatch}
P.~Sankowski.
\newblock Faster dynamic matchings and vertex connectivity.
\newblock In {\em Proc. SODA}, pages 118--126, 2007.

\bibitem{sankweightmatch}
P.~Sankowski.
\newblock Maximum weight bipartite matching in matrix multiplication time.
\newblock {\em Theor. Comput. Sci.}, 410(44):4480--4488, 2009.

\bibitem{SankowskiM10}
P.~Sankowski and M.~Mucha.
\newblock Fast dynamic transitive closure with lookahead.
\newblock {\em Algorithmica}, 56(2):180--197, 2010.

\bibitem{Sch92sat}
I.~Schiermeyer.
\newblock Solving 3-satisfiability in less then $1.579^{n}$ steps.
\newblock In {\em CSL}, pages 379--394, 1992.

\bibitem{Scho99sat}
U.~Sch\"{o}ning.
\newblock A probabilistic algorithm for $k$-{SAT} and constraint satisfaction
  problems.
\newblock In {\em Proc. FOCS}, pages 410--414, 1999.

\bibitem{strassen}
V.~Strassen.
\newblock Gaussian elimination is not optimal.
\newblock {\em Numer. Math.}, 13:354--356, 1969.

\bibitem{Thorup00}
M.~Thorup.
\newblock Near-optimal fully-dynamic graph connectivity.
\newblock In {\em STOC}, pages 343--350, 2000.

\bibitem{VW09}
V.~Vassilevska and R.~Williams.
\newblock Finding, minimizing, and counting weighted subgraphs.
\newblock In {\em Proc. STOC}, pages 455--464, 2009.

\bibitem{v12}
V.~{Vassilevska~Williams}.
\newblock Multiplying matrices faster than {C}oppersmith-{W}inograd.
\newblock In {\em Proc. STOC}, pages 887--898, 2012.

\bibitem{VW09j}
V.~{Vassilevska Williams} and R.~Williams.
\newblock Finding, minimizing, and counting weighted subgraphs.
\newblock {\em SIAM J. Comput.}, 42(3):831--854, 2013.

\bibitem{ryanpersonal}
R.~Williams.
\newblock Faster all-pairs shortest paths via circuit complexity.
\newblock {\em Submitted}, 2013.

\bibitem{RHuacheng}
R.~Williams and H.~Yu.
\newblock Finding orthogonal vectors in discrete structures.
\newblock In {\em SODA}, 2014.
\newblock to appear.

\bibitem{focs10}
V.~V. Williams and R.~Williams.
\newblock Subcubic equivalences between path, matrix and triangle problems.
\newblock In {\em Proc. FOCS}, pages 645--654, 2010.

\end{thebibliography}
